\documentclass[pdflatex,prd,twocolumn,showpacs,superscriptaddress,nofootinbib]{revtex4-1}
\usepackage{bm,amsmath,amssymb}
\usepackage{graphicx}
\usepackage{subfigure}
\usepackage{color}
\usepackage{soul}
\usepackage{hyperref}
\usepackage{multirow}
\usepackage{lineno}
\usepackage{ulem}
\setulcolor{blue}
\setstcolor{red}
\sethlcolor{yellow}

\newcommand \beq {\begin{equation}}
\newcommand \eeq {\end{equation}}
\newcommand \beqa {\begin{eqnarray}}
\newcommand \eeqa {\end{eqnarray}}

\newcommand \hmu {\hat{\mu}}

\def\lsim{\raise0.3ex\hbox{$<$\kern-0.75em\raise-1.1ex\hbox{$\sim$}}}
\def\gsim{\raise0.3ex\hbox{$>$\kern-0.75em\raise-1.1ex\hbox{$\sim$}}}

\begin{document}

\title{Second order cumulants of conserved charge fluctuations revisited \\
I. Vanishing chemical potentials}

\author{D. Bollweg}
\affiliation{Fakult\"at f\"ur Physik, Universit\"at Bielefeld, D-33615 Bielefeld, Germany}
\author{J. Goswami}
\affiliation{Fakult\"at f\"ur Physik, Universit\"at Bielefeld, D-33615 Bielefeld, Germany}
\author{O. Kaczmarek}
\affiliation{Fakult\"at f\"ur Physik, Universit\"at Bielefeld, D-33615 Bielefeld, Germany}
\author{F. Karsch}
\affiliation{Fakult\"at f\"ur Physik, Universit\"at Bielefeld, D-33615 Bielefeld, Germany}
\author{Swagato Mukherjee}
\affiliation{Physics Department, Brookhaven National Laboratory, Upton, New York 11973, USA}
\author{P. Petreczky}
\affiliation{Physics Department, Brookhaven National Laboratory, Upton, New York 11973, USA}

\author{C. Schmidt}
\affiliation{Fakult\"at f\"ur Physik, Universit\"at Bielefeld, D-33615 Bielefeld, Germany}
\author{P. Scior}
\affiliation{Physics Department, Brookhaven National Laboratory, Upton, New York 11973, USA}

\collaboration{HotQCD collaboration}
\date{\today}
\begin{abstract}

We update lattice QCD results for second order cumulants of conserved charge fluctuations and correlations at 
non-zero temperature and vanishing values of the conserved charge chemical potentials. We compare these results to hadron resonance gas calculations with and without excluded volume terms as well as S-matrix results in the hadronic phase of QCD, and comment on their current limitations. We, furthermore, use these results to characterize  thermal conditions in the vicinity of the pseudo-critical line of the chiral transition in QCD. We argue that the ratio of strange to baryon
chemical potentials is a robust observable that, on the one hand, deviates only little from  hadron resonance gas results, but, on the other hand, is very sensitive to the spectrum of strange baryon resonances.

\end{abstract}


\maketitle


\section{Introduction}
The thermal state of strong-interaction matter described by Quantum Chromodynamics (QCD)
with two light up and down quarks and a heavier strange quark
is fixed by four external control parameters, the temperature ($T$) and three chemical
potentials ($\mu_B, \mu_Q, \mu_S$) that couple to the conserved currents for 
net baryon-number ($B$), electric-charge ($Q$) and strangeness ($S$), respectively.
In a grand canonical description of equilibrium thermodynamics these external control
parameters are Lagrange multiplier, appearing in the statistical operator, that 
characterize the thermodynamic conditions of, e.g., matter described by the strong force. 
The set of external parameters, ($T, \mu_B, \mu_Q, \mu_S$) may be used to 
calculate various bulk thermodynamic observables, e.g. energy and number densities
as well as generalized susceptibilities that are obtained from higher order 
derivatives of the grand canonical partition function. 
Such observables are, in principle, measurable in experiments, while the set of
Lagrange multiplier, ($T, \mu_B, \mu_Q, \mu_S$), is not. They are specific to a given model 
to the extent that the value of a certain bulk thermodynamic observable, e.g. the net baryon-number density, which is related to a certain temperature value
$T$ in QCD (or nature) will correspond to another
temperature value in a model calculation. The latter 
will be close to the former only when the model approximations provide a realistic description of QCD.

Lattice QCD calculations provide continuum extrapolated results for bulk thermodynamic 
observables at small values of the chemical potential with percent-level accuracy. 
This provides, in particular, the pseudo-critical temperature for the chiral transition
at vanishing values of the baryon chemical potential with better than 1\% accuracy
\cite{Bazavov:2018mes} for a specific set of observables. This result differs by less than 2\%
from calculations using different observables and discretization schemes to characterize the
crossover transition in QCD \cite{Bazavov:2018mes,Borsanyi:2020fev}. 
Similarly the temperature dependence
of the pseudo-critical line, $T_{pc}(\mu_B)$, is known with better than 2\% 
accuracy at least for baryon chemical potentials $\mu_B\le T_{pc,0}\equiv T_{pc}(0)$
\cite{Bazavov:2018mes,Borsanyi:2020fev,Bonati:2015bha}. This also includes 
a determination of constraints on the strangeness chemical potential, 
$\mu_S(\mu_B)$, required to insure strangeness neutrality in strongly 
interacting matter. 
At fixed value of temperature this constraint is known to better than 
5\% for a range of baryon chemical potentials $\mu_B \le T_{pc}(\mu_B)$.

In heavy-ion experiments at the Large Hadron Collider (LHC) and the Relativistic Heavy Ion Collider (RHIC) the
thermal properties of strongly interacting matter as it
existed at the time of hadronization (freeze-out of
various hadrons) is probed. The underlying
thermal parameters, ($T, \mu_B, \mu_Q, \mu_S$), are extracted from particle yields
or higher order cumulants characterizing the distributions of these hadrons. 
These parameters, however, generally are not obtained through a direct comparison with QCD
but with statistical models, e.g. models that use the statistical operator of hadron 
resonance gas (HRG) models \cite{BraunMunzinger:2003zd} 
for point-like non-interacting hadrons, which are variants of the Hagedorn 
resonance gas model
\cite{Hagedorn:1965st,Karsch:2010ck,Andronic:2017pug}.
This too leads to a determination of thermal parameters with errors that are on the percent 
level. The approach, however, may suffer from systematic uncertainties,
as the statistical operator used in model calculations obviously will differ 
from that of QCD in certain ranges of the temperature and baryon chemical
potential. Differences between the statistical operator of QCD and that of
HRG models based on a spectrum of point-like, non-interacting
hadron resonances become apparent in properties of higher order
cumulants, e.g. deviations from (generalized) Skellam distributions
that are inherent to HRG models using point-like, non-interacting
hadron resonances. A proper inclusion of interactions in a hadronic medium,
using a relativistic virial expansion \cite{Dashen:1969ep,Venugopalan:1992hy,lo_probing_2013,Fernandez-Ramirez:2018vzu,Andronic:2018qqt} or more phenomenology oriented excluded volume
\cite{Hagedorn:1980kb,Hagedorn:1980cv,Gorenstein:1981fa,Vovchenko:2016rkn,Vovchenko:2017xad,Taradiy:2019taz}
as well as repulsive mean approaches
\cite{Olive:1980dy,Olive:1982we,Huovinen:2017ogf} thus may be necessary. These 
approaches generally need to adjust parameters which is done by 
comparing to lattice QCD calculations of thermodynamic observables.

Second order cumulants
can be used as important benchmark observables that 
allow to establish the range of validity of hadronic
models that are needed to provide an interface between
experimental observables, e.g. hadron yields and fluctuations, and thermal observables obtained in 
field theoretic calculations, e.g. QCD. 
In this paper we will provide high precision, continuum extrapolated
lattice QCD results for
second order cumulants of conserved charge 
fluctuations and their correlations at vanishing values of the 
baryon chemical potential. We will extend these calculations in a follow-up paper to non-vanishing 
values of the chemical potential.

The calculations
presented here at vanishing values of the chemical potential allow to constrain the basic parameters that
enter in model calculations, e.g.  the excluded volume parameters introduced in hadron resonance gas  (HRG) models to mimic the effect of repulsive interactions or  
the modeling of higher order corrections in S-matrix
calculations that are needed to go beyond the calculation of second order coefficients in a virial expansion.
Preliminary results from this study have been presented
by us previously \cite{Goswami:2020yez,Goswami:2021opr}. A similar compar- ison of lattice QCD results on cumulants of conserved charge fluctuations with excluded volume HRG models also appeared recently\cite{Karthein:2021cmb}.

This paper is organized as follows. In the next section we describe the computational set-up for our calculations and describe the scale setting used by us to define temperature scales. Section III gives a short description of the basic observables that we analyze. Section IV describes our determination of continuum extrapolated results for all second order cumulants. Section V is devoted to a comparison of the QCD results for these cumulants with hadron resonance gas model calculations. Finally we give our conclusions in Section VI. Three appendices are devoted to the presentation of further details on the scale setting, the fits to second order cumulants and a description of differences between two lists of hadron resonances used in HRG model calculations. 

\section{Computational set-up}
The framework for our calculations with the Highly Improved
Staggered Quark (HISQ) \cite{Follana:2006rc} discretization scheme for $(2+1)$-flavor 
QCD with a physical strange quark mass and two degenerate, physical light quark masses
is well established and has been used by us for several studies of higher order
cumulants of conserved charge fluctuations and correlations. The specific set-up
used in our current study has been described in \cite{Bazavov:2020bjn}. 

In addition to the data-sets used for that study we increased the statistics
on lattices with temporal extent $N_\tau=12$ and $16$. 
This allows us to present continuum extrapolated results for all second order cumulants, for which the statistical
error on the data themselves amounts to less than 50\%, the
reminder coming from systematic errors related to the
fit ans\"atze used for continuum extrapolations and 
uncertainties on the zero temperature observables used
to set the scale for the temperature.

\subsection{Data sets and statistics}
We follow here the notation and conventions used in \cite{Bazavov:2017dus} for the
calculation of the equation of state in $(2+1)$-flavor QCD
with non-vanishing chemical potentials. In that work first
continuum extrapolated results for second order cumulants, obtained with the HISQ action, had been presented. 
The statistics collected for our current analysis is more
than a factor 10 larger than that used in 
\cite{Bazavov:2017dus} for the calculation of the equation of state on lattices with temporal extent
$N_\tau = 8$ and $12$. Moreover, compared to 
the statistics used previously for a determination of the pseudo-critical temperature $T_{pc}$
of the chiral transition in $(2+1)$-flavor QCD \cite{Bazavov:2018mes}, we increased the
statistics by a factor ($3$-$4$) for existing data-sets 
on lattices with temporal extent 
$N_\tau=16$ in the temperature range $T\in [135~{\rm MeV}:178~{\rm MeV}]$. 
We also make use of data-sets obtained in earlier calculations on lattices with temporal extent $N_\tau=6$
\cite{Bazavov:2018mes,Bazavov:2017dus}.
Details on the simulation parameters and our current statistics are given in Table~\ref{tab:statistics}.
The gauge field configurations stored in our data-sets are separated by 10 time units in a simulation with the rational Hybrid Monte Carlo (RHMC)
algorithm. 

\begin{table*}[htb]
\centering
\begin{tabular}{|cccr||cccr||cccr|}
\hline
\hline
\multicolumn{4}{|c||}{ $N_\tau=8$}&\multicolumn{4}{c|}{ $N_\tau=12$}&\multicolumn{4}{c|}{ $N_\tau=16$} \\
\hline
$\beta$ & $m_l$ & T[MeV] & \#conf. & $\beta$ & $m_l$ & T[MeV] & \#conf. & $\beta$ & $m_l$ & T[MeV] & \#conf. \\
\hline
6.175 & 0.003307 & 125.28 & 1,471,861 & & & & & & & &  \\
6.245 & 0.00307  & 134.84 & 1,275,380 &6.640 & 0.00196 & 135.24 & 330,447& 6.935 & 0.00145 & 135.80 &  17671 \\
6.285 & 0.00293 &  140.62 & 1,598,555 &6.680 & 0.00187 & 140.80 & 441,115& 6.973 & 0.00139 & 140.86 &  23855\\
6.315 & 0.00281 &  145.11 & 1,559,003 &6.712 & 0.00181 & 145.40 & 416,703& 7.010 & 0.00132 & 145.95 &  26122\\
6.354 & 0.00270 &  151.14 & 1,286,603 &6.754 & 0.00173 & 151.62 & 323,738& 7.054 & 0.00129 & 152.19 &  26965\\
6.390 & 0.00257 &  156.92 & 1,602,684 &6.794 & 0.00167 & 157.75 & 299,029& 7.095 & 0.00124 & 158.21 &  21656 \\
6.423 & 0.00248 &  162.39 & 1,437,436 &6.825 & 0.00161 & 162.65 & 214,671& 7.130 & 0.00119 & 163.50 & 18173\\
6.445 & 0.00241 &  166.14 & 1,186,523 &6.850 & 0.00157 & 166.69 & 156,111& 7.156 & 0.00116 & 167.53 & 19926\\
6.474 & 0.00234 &  171.19 &   373,644 &6.880 & 0.00153 & 171.65 & 144,633& 7.188 & 0.00113 & 172.60 & 17163\\
6.500 & 0.00228 &  175.84 &   294,311 &6.910 & 0.00148 & 176.73 & 131,248& 7.220 & 0.00110 & 177.80 &  3282\\
\hline
\hline
\end{tabular}
\caption{Simulation parameters and statistics collected on lattices of size $N_\sigma^3\times N_\tau$ with $N_\sigma= 4N_\tau$
	in calculations with light to strange quark mass ratio $m_l/m_s=1/27$.}
\label{tab:statistics}
\end{table*}

All our calculations have been performed on lattices 
with spatial extent $N_\sigma = 4 N_\tau$. 
Additional data on second order cumulants on lattices with
temporal extent $N_\tau=6$
are  taken from \cite{Bazavov:2017dus}. 
In order to stabilize the asymptotic behavior of our fits
to second order cumulants we also use data at higher temperatures, which also are taken from \cite{Bazavov:2017dus}.

In this new analysis we focus on a temperature
range of $20$~MeV above and below the pseudo-critical temperature for the chiral transition at vanishing value of the chemical potentials,
$T_{pc,0}=(156.5\pm 1.5)$~MeV \cite{Bazavov:2018mes}.

\subsection{Scale setting}
In order to set the scale for the temperature used in
lattice QCD calculations we closely follow the
strategy laid down for the definition of a line of constant physics and the calculation of the equation of state in \cite{Bazavov:2014pvz}. There we introduced temperature scales based on a calculation of $r_1$, which characterizes the short distance part of the heavy quark potential, and the kaon decay constant $f_K$. We use parametrizations of both
observables to set the scale for the temperature,
\begin{eqnarray}
T_{fK} &=& \frac{1}{N_\tau af_K}\ f_K\; , \label{eq:TfK}\\
    T_{r1}&=& \left( \frac{1}{N_\tau} \frac{r_1}{a}\right)   \frac{1}{r_1} \; . \label{eq:Tr1}
\end{eqnarray}
The two temperature scales are related through the value of $r_1f_K$, which is solely determined
through a lattice calculation, and the physical value of $r_1$, which requires input from 
experiment, e.g. the pion decay constant $f_\pi$. 
At non-zero lattice spacing
the temperature scales derived from different observables differ. At finite values of the gauge
coupling $\beta$ they are related through
\begin{equation}
\frac{T_{fK}}{T_{r1}} =
   \left( \frac{1}{af_K}   \frac{a}{r_1} \right)
   r_1 f_K 
   \label{eq:fKr1} \; .
\end{equation}
The parametrizations of $a/r_1$ and $af_K$ used by 
us at non-vanishing lattice spacing are given in Appendix~\ref{app:scale-setting}. In all figures, 
that show lattice QCD results obtained at non-vanishing
lattice spacing, we use, for definiteness, the temperature scale based on calculations of $af_K$, as has been done by us also in the past \cite{Bazavov:2019www}. For these figures and the
continuum extrapolations at fixed temperature values we 
use as basic input the central value of the MILC results for $r_1$, i.e. $r_1=0.3106$~fm. The fits presented in 
Appendix~\ref{app:scale-setting} then fix the central 
value of the kaon decay constant to $f_K=155.7/\sqrt{2}$~MeV.

As discussed in more detail in Section~\ref{sec:continuum}, we treat the error on $r_1$ and $f_K$ as an overall systematic error on the temperature scale that will enter the final error budget in our analysis. 

\section{First and second order moments of conserved net charge fluctuations}
\label{sec:notation}

We focus here on a discussion of second order cumulants
of conserved charge fluctuations at vanishing chemical 
potentials for the conserved charges of $(2+1)$-flavor QCD, 
i.e. net baryon-number ($B$),
electric charge ($Q$) and strangeness ($S$).
They are obtained from the QCD partition function,
$\mathcal{Z}(T,V,\vec{\mu})$, as derivatives with
respect to the associated chemical potentials
$\vec{\mu}=(\mu_B, \mu_Q, \mu_S)$,
\begin{equation}
\chi_{ijk}^{BQS} =\left. 
\frac{1}{VT^3}\frac{\partial \ln\mathcal{Z}(T,V,\vec{\mu}) }{\partial\hmu_B^i \partial\hmu_Q^j \partial\hmu_S^k}\right|_{\hmu=0} \; ,
\; i+j+k=2 \; .
\label{suscept}
\end{equation}
This set of six second order cumulants are leading order
terms in Taylor expansion of various thermodynamic 
quantities derived from Taylor expansions of the 
pressure of $(2+1)$-flavor QCD,
\begin{equation}
	\frac{P}{T^4} = \frac{1}{VT^3}\ln\mathcal{Z}(T,V,\vec{\mu}) = \sum_{i,j,k=0}^\infty%
\frac{\chi_{ijk}^{BQS}}{i!j!\,k!} \hmu_B^i \hmu_Q^j \hmu_S^k \; ,
\label{Pdefinition}
\end{equation}
with $\hat{\mu}_X\equiv \mu_X/T$ and arbitrary natural numbers $i,j,k$. In particular, they provide the leading
order expansion coefficients of
mean ($n_X/T^3\equiv \chi_1^X$) and variance ($\sigma_X^2/T^2\equiv \chi_2^X$) 
of conserved charge distributions. 
At leading order in the chemical potentials the former
are given by
\begin{eqnarray}
	n_B/T^3 &=& \chi_2^B \hmu_B + \chi_{11}^{BS} \hmu_S
	+\chi_{11}^{BQ} \hmu_Q  \; ,  \\
	n_Q/T^3 &=& \chi_{11}^{BQ}\hmu_B  + \chi_{11}^{QS} \hmu_S
        +\chi_{2}^{Q} \hmu_Q \; , \\
	n_S/T^3 &=& \chi_{11}^{BS}\hmu_B + \chi_{2}^{S} \hmu_S
        +\chi_{11}^{QS} \hmu_Q \; .
	\label{densities}
\end{eqnarray}
Of particular interest, for a discussion of properties
of strongly interacting matter, created in heavy ion
collisions, is the case of strangeness neutral matter 
($n_S\equiv 0$). 
The second order cumulants provide important information 
on the relation between strangeness and baryon chemical potentials in strongly interacting matter \cite{Bazavov:2012vg}. 
To leading order this is given by,
\begin{equation}
    \frac{\mu_S}{\mu_B}\equiv s_1(T) = - \frac{\chi_{11}^{BS}}{\chi_{2}^{S}} 
    -\frac{\chi_{11}^{QS}}{\chi_{2}^{S}} q_1
    +{\cal O} (\hmu_B^2)
    \; ,
    \label{muSmuB}
\end{equation}
with 
\begin{equation}
q_1 =
\frac{
r \left( \chi_2^B\chi_2^S - \chi_{11}^{BS}\chi_{11}^{BS} \right)
-\left( \chi_{11}^{BQ}\chi_2^S -\chi_{11}^{BS} \chi_{11}^{QS} \right)
}{
\left( \chi_2^Q\chi_2^S  - \chi_{11}^{QS} \chi_{11}^{QS} \right)
- r \left(\chi_{11}^{BQ}\chi_2^S - \chi_{11}^{BS}\chi_{11}^{QS} \right)
} 
    \label{muSmuBq1}
\end{equation}
Here $r=n_Q/n_S$. Note that 
in the isospin symmetric case, $r=1/2$, one has
$q_1=0$.
We will discuss Section \ref{sec:neutral} the sensitivity of this ratio of chemical potentials
to the hadron resonance spectrum contributing to the thermodynamics of strongly interacting matter.

\section{Second order cumulants at vanishing chemical potential}
\label{sec:second0}

\subsection{Continuum extrapolation of second order cumulants}
\label{sec:continuum}
Second order cumulants of conserved charge fluctuations and cross-correlations
among them have been calculated in lattice QCD using various discretization
schemes. This led to continuum extrapolated results in $(2+1)$-flavor
QCD with physical (degenerate) light quark masses ($m_u=m_d$) and
a physical strange quark mass ($m_s$) that have been presented, previously \cite{Bazavov:2012jq,Bellwied:2019pxh}.
The HotQCD Collaboration presented such results obtained in calculations with the HISQ action. Making use of recently obtained high statistics data
we update here the continuum extrapolation of the second order
cumulants and present fits to a set of four independent second order
cumulants that are suitable for analyzing all six $2^{nd}$ order cumulants in the conserved charge ($B, Q, S$) basis as well as the ($u, d, s$) flavor basis \cite{Bazavov:2017dus} for the chemical potentials.

Among the six second order cumulants 
only four are independent at vanishing values of the chemical potential. The other two
are constrained by isospin symmetry as our calculations, as well as most other lattice QCD calculations,
are performed for two degenerate light up and down quarks. This leads to two constraints \cite{Bazavov:2017dus},
\begin{eqnarray}
  \chi_2^S &=& 2 \chi_{11}^{QS} - \chi_{11}^{BS}\;\; , \label{constraintS} \\ 
  \chi_2^B &=& 2 \chi_{11}^{BQ} - \chi_{11}^{BS}\; .
\label{constraintB}    
\end{eqnarray}
In the quark flavor basis the four independent observables and the two constraints relate 
to the three diagonal and three off-diagonal cumulants for the light and strange quark 
number fluctuations and correlations. The two constraints merely reflect that $u$ and 
$d$-quark fluctuations are identical as are their correlations with the strange quarks, 
{\it i.e.} $\chi_2^u=\chi_2^d$ and $\chi_{11}^{us}=\chi_{11}^{ds}$. We also note that the 
above constraints are fulfilled to better than 1\% also in HRG model calculations
that utilize the experimentally determined, physical hadron resonance spectrum 
\cite{Zyla:2020zbs}. Here small mass differences among 
isospin partners arise from differences in the 
up and down quark masses as 
well as electromagnetic interactions. 
The above constraints, of course, also
hold for HRG model calculations using hadron spectra 
calculated e.g. within relativistic quark models
\cite{Capstick:1986bmxx,Ebert:2009ub,Faustov:2015eba} 
as well as for spectra obtained in lattice QCD calculations \cite{Edwards:2011jj,Edwards:2020rbo}.

All possible second order cumulants and ratios thereof, calculated either in the 
($B,Q,S)$ or $(u,d,s)$ basis for the chemical potentials \cite{Allton:2005gk}, thus can be  obtained from 
four independent observables; we focus on
the set of observables
$\chi_2^{Q}$, $\chi_{11}^{QS}$, $\chi_{11}^{BQ}$, $\chi_{11}^{BS}$.
While the first two cumulants are dominated by contributions from
the non-strange ($\chi_2^{Q}$) and strange ($\chi_{11}^{QS}$) meson 
spectrum the latter two are dominated by the non-strange ($\chi_{11}^{BQ}$) 
and strange ($\chi_{11}^{BS}$) baryon sectors of the 
hadron spectrum.

The second order cumulants have been obtained in recent
high statistics calculations on lattices with temporal
extent $N_\tau = 8,\ 12,\ 16$ that focused on a calculation
of higher order cumulants
\cite{Bazavov:2020bjn,HotQCD:2017qwq}. 
These calculations have been extended, focusing on the 
temperature range in the vicinity of the pseudo-critical temperature, 
$T_{pc,0}$, for the chiral transition. The current statistics,
accumulated in these calculations, is given in Table~\ref{tab:statistics}. 
Moreover, we used data sets obtained earlier
on lattices with temporal extent $N_\tau=6$
\cite{Bazavov:2018mes,Bazavov:2017dus}.
For each $N_\tau$ we used cubic spline interpolations of our data in the interval $T\in[134~{\rm MeV}: 178~{\rm MeV}]$
(see Appendix~\ref{app:fits}). These interpolations allow us to obtain for each of the four different lattice sizes
cumulants at the same temperature. This has been done for
each of the two schemes used to define a temperature scale
at non-vanishing values of the lattice spacing.

\begin{figure*}[ht]
\includegraphics[scale=0.44]{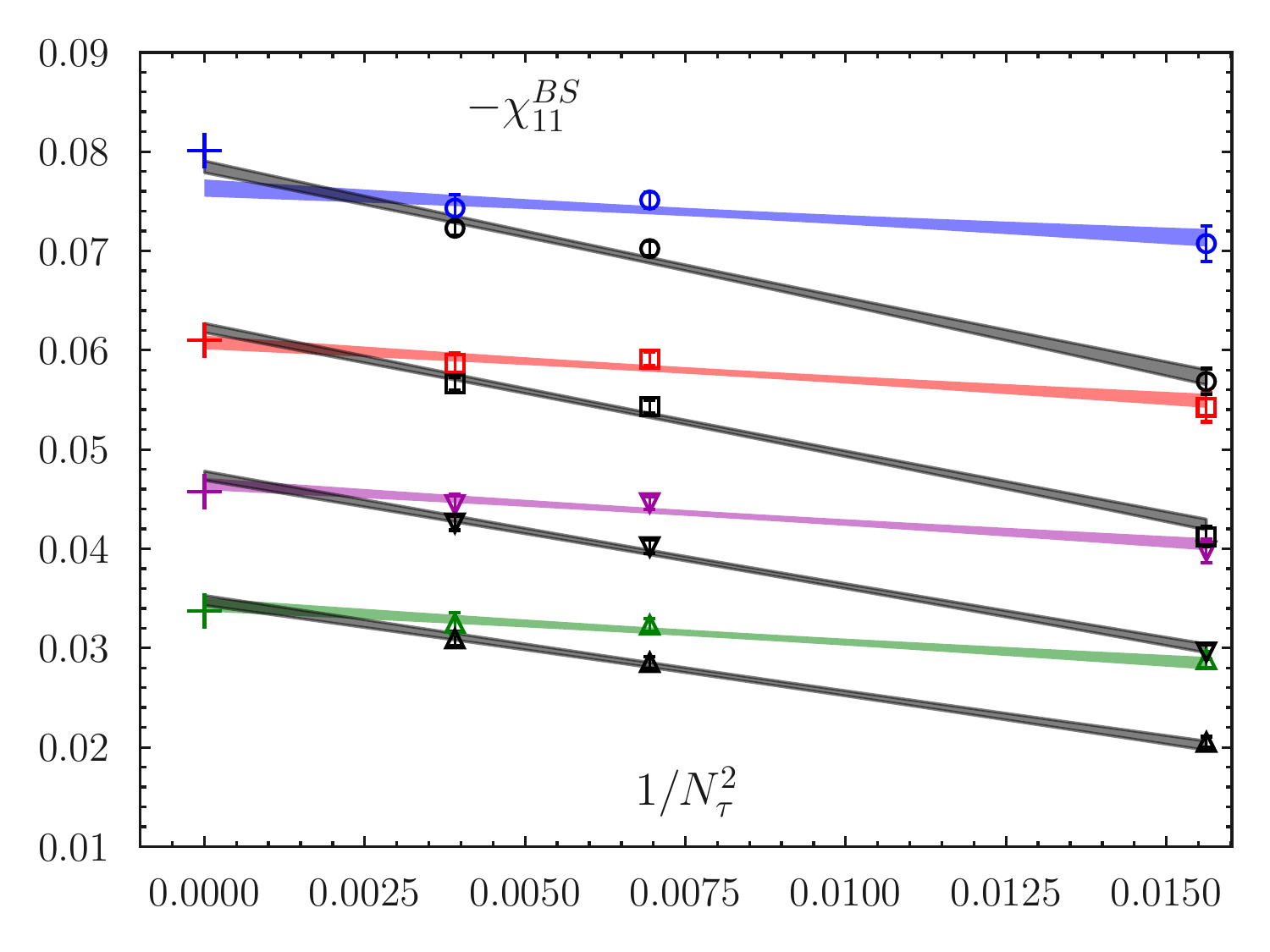}
\includegraphics[scale=0.44]{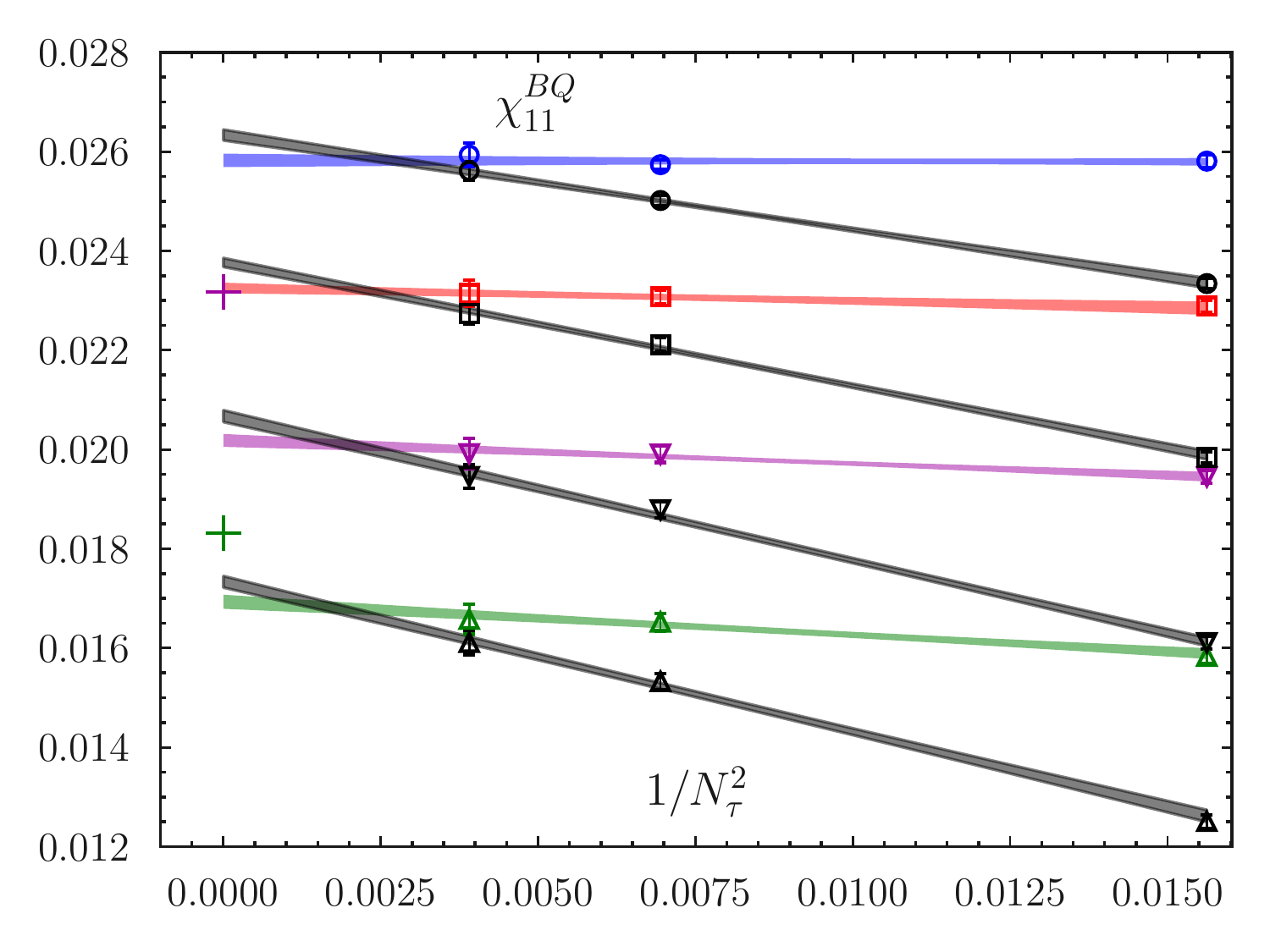}
\includegraphics[scale=0.44]{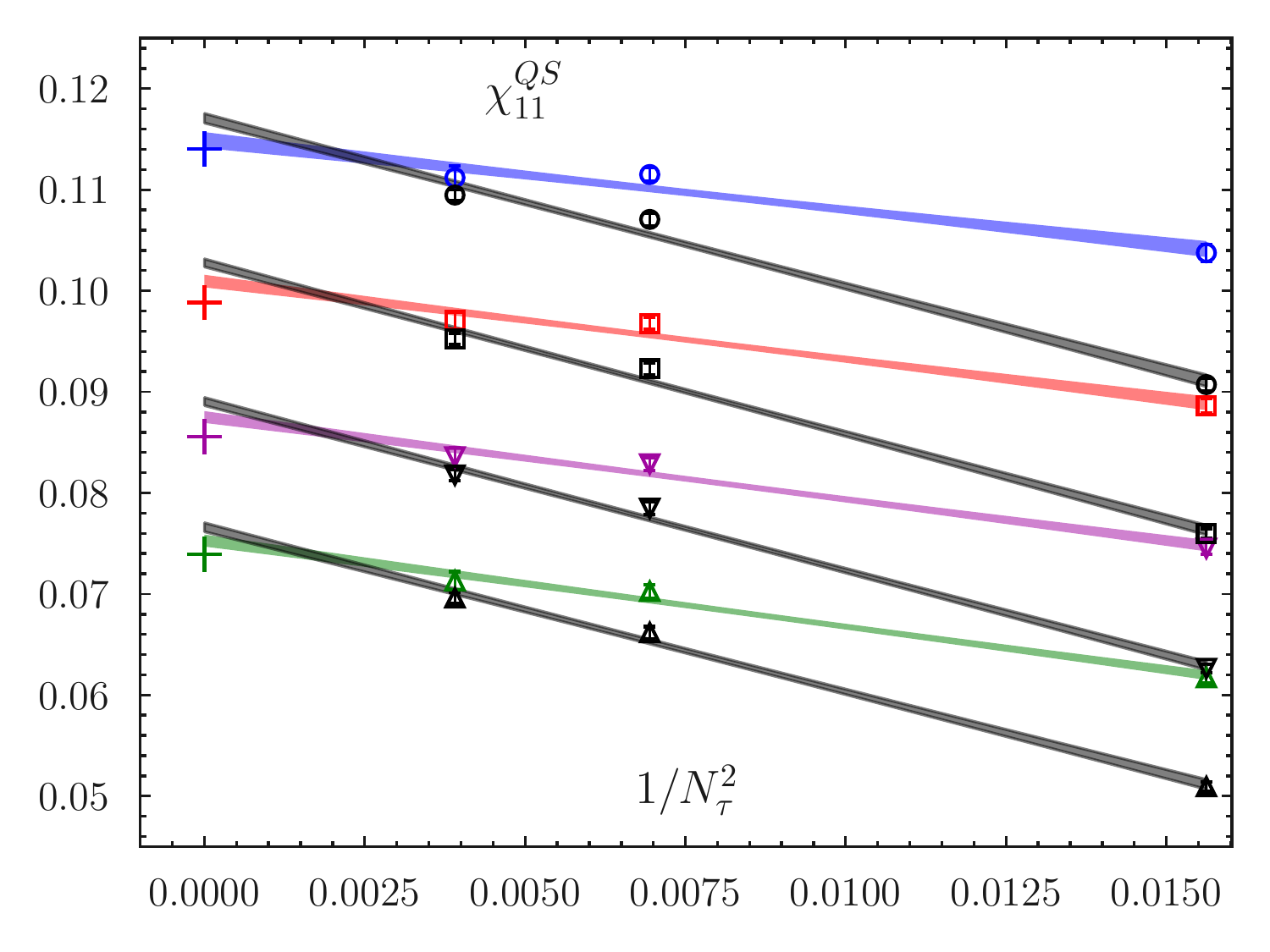}
\includegraphics[scale=0.44]{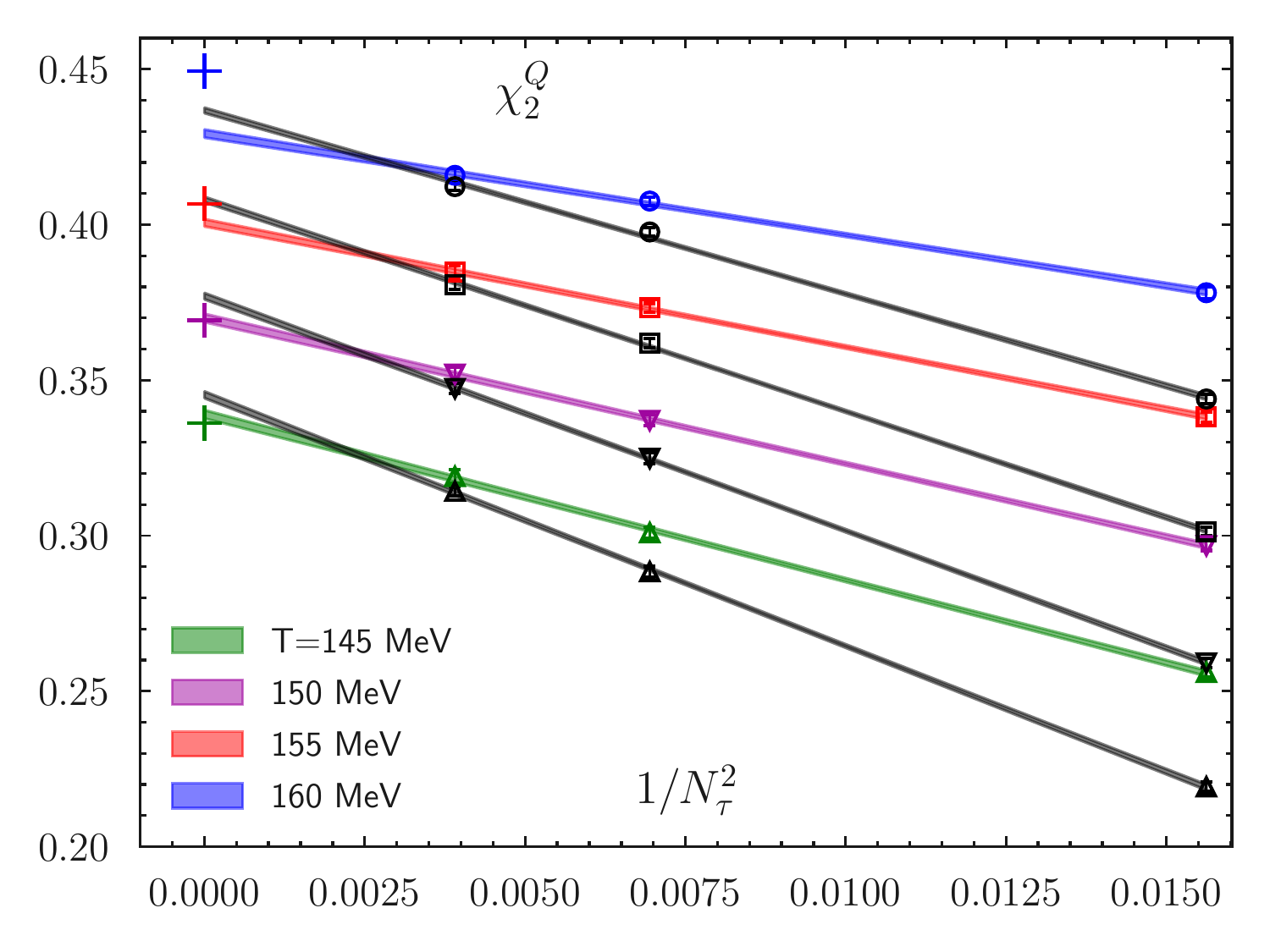}
\caption{Continuum limit extrapolations for the cumulants  $\chi_{11}^{BS}$ (top, left),
$\chi_{11}^{BQ}$ (top, right),  $\chi_{11}^{QS}$ (bottom, left) and $\chi_{2}^{Q}$ (bottom, right) at several values of the temperature,
        which has been obtained from $af_K$ (colored lines) and $a/r_1$ (black lines), respectively. Extrapolations are linear
        in $1/N_\tau^2$ and for data obtained on lattices with temporal extent $N_\tau=8,\ 12,\ 16$. Crosses indicate the corresponding QMHRG2020 value at that temperature. Note that for
        $\chi_{11}^{BQ}$ and $\chi_2^Q$ these 
        HRG results are not shown for all temperature values as the deviations from the corresponding QCD results are too large.
        For $\chi_2^Q$  QMHRG2020 with the finite-volume corrected contributions for pions and kaons in a volume
        $LT\equiv N_\sigma / N_\tau =4$ has been used (see Section~\ref{chiQ2}).
}
\label{fig:r1fKcompare}
\end{figure*}

Continuum extrapolations of these observables have been
performed in both schemes, using linear and quadratic extrapolations in $1/N_\tau^2$, 
\begin{eqnarray}
f_2(T,N_\tau) &=& f_2(T)  +  \frac{a_2}{N_\tau^2}
\label{fit-linear} \; ,\\
    f_4(T,N_\tau) &=& f_4(T)  +  \frac{b_2}{N_\tau^2} +
    \frac{b_4}{N_\tau^4}
    \; .
    \label{fit-quadratic}
\end{eqnarray}
For some temperature values the linear and quadratic extrapolations are shown 
in Appendix~\ref{app:fits}. Results obtained with
linear extrapolations on the $N_\tau=8,\ 12,\ 16$ data sets
based on the $r_1$ and $f_K$ temperature scales are compared
in Fig.~\ref{fig:r1fKcompare}.

Differences in the continuum extrapolation
shown in Fig.~\ref{fig:r1fKcompare}, that arise from the usage of two different temperature scales, reflect systematic errors arising from the parametrization of $a/r_1$ and
$af_K$ at finite values of the gauge coupling. Moreover, the restriction to linear fit ans\"atze
reacts differently to truncated higher order corrections
in both schemes. Our final continuum extrapolation result for second order cumulants is obtained by averaging over
the two different linear fit results.
The difference in the two extrapolations at each temperature is taken as a systematic error and has been added linearly to the statistical errors of the two extrapolations based on the $r_1$ and $f_K$ temperature scales, respectively.
Further details on our continuum extrapolations are given in
Appendix~\ref{app:fits}.

The presentation of the continuum extrapolations in Fig.~\ref{fig:r1fKcompare}  makes use of a specific 
choice for the value of $r_1$ to define the 
temperature scale, i.e. we used\footnote{Note that we could have shown Fig.~\ref{fig:r1fKcompare}  also for fixed
$Tr_1$ or $T/f_K$. We used a scale in MeV only for 
clarity and better orientation.} $r_1=0.3106$~fm, which is the central value for $r_1$ quoted by the MILC Collaboration \cite{MILC:2010hzw}, $r_1=0.3106(8)(14)(4)$~fm. 
Here the errors are statistical, systematic and experimental, respectively.
The statistical error and part of the systematic error on the value for $r_1$, quoted by the MILC collaboration, corresponds to errors also arising in our calculation when analyzing results obtained with two different
$T$-scales, different fit ans\"atze and fit ranges
used for the continuum extrapolations.

In addition to this error we obviously need to add the systematic uncertainty in the $T$-scale arising from the error on the experimental value of the pion decay constant, $f_\pi$, quoted by
MILC as $\Delta r_1=0.0004$~fm. This gives rise to
an uncertainty of the $T$-scale of about $0.2$~MeV in the $T$-range considered by us. There also is a substantial spread in values for $r_1$ obtained by several groups and quoted by FLAG \cite{Aoki:2019cca}, for instance in Table~56. 
We estimate this overall systematic uncertainty on $r_1$ as
$\Delta r_1=0.001$~fm, which for the temperature range
considered by us amounts to a scale uncertainty of 
$\Delta T \simeq 0.6$~MeV. The resulting systematic error 
is shown separately for all our continuum extrapolated results on second order cumulants. 

\begin{figure*}[t]
\includegraphics[scale=0.6]{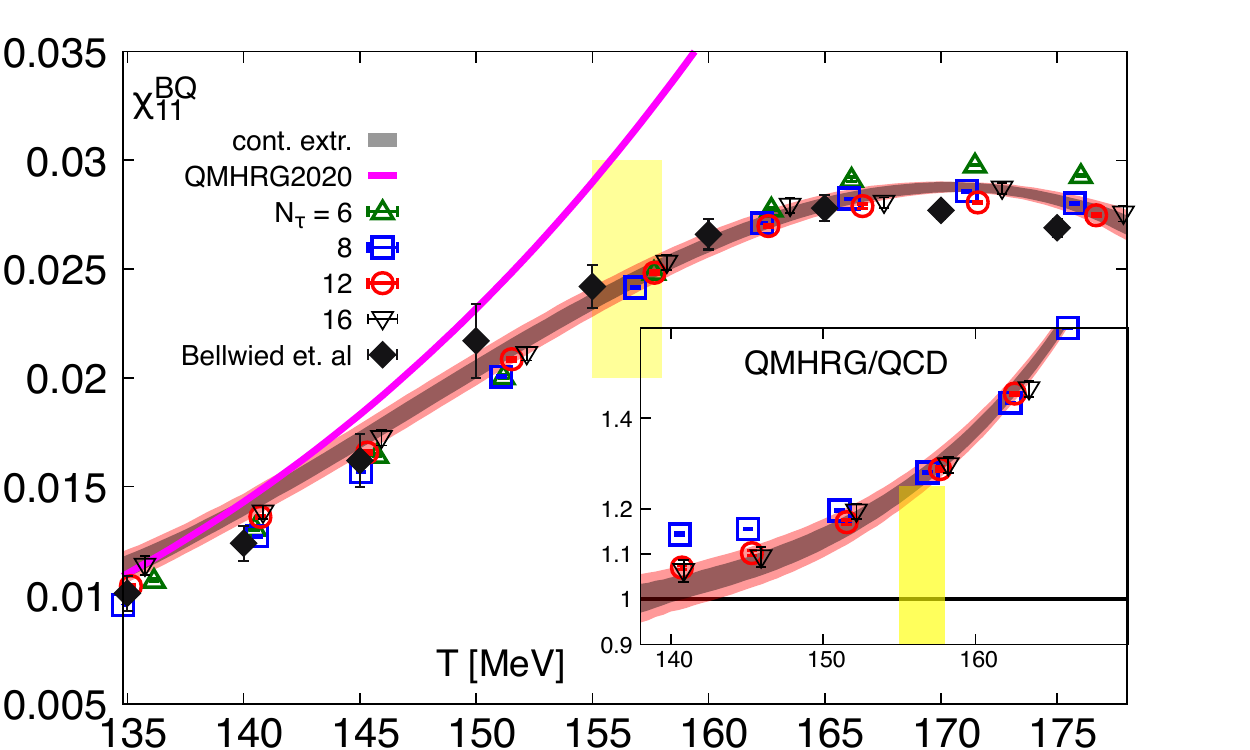}
\includegraphics[scale=0.6]{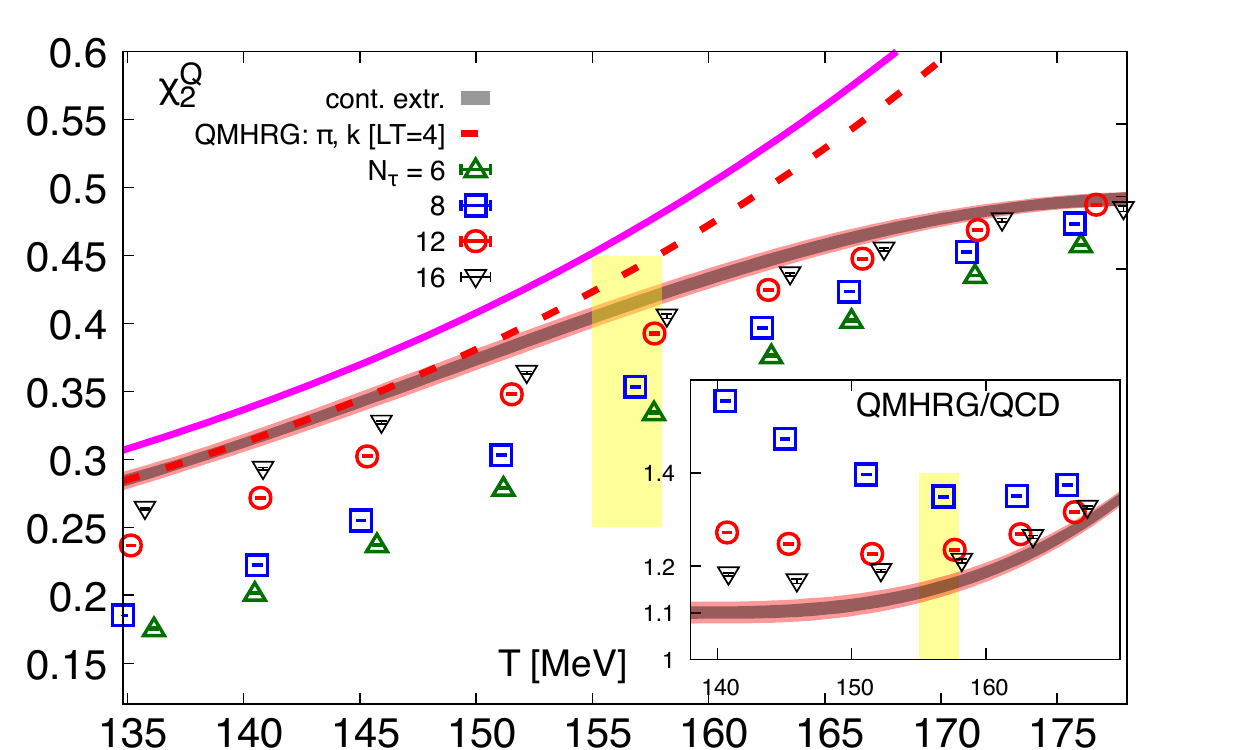}
\includegraphics[scale=0.6]{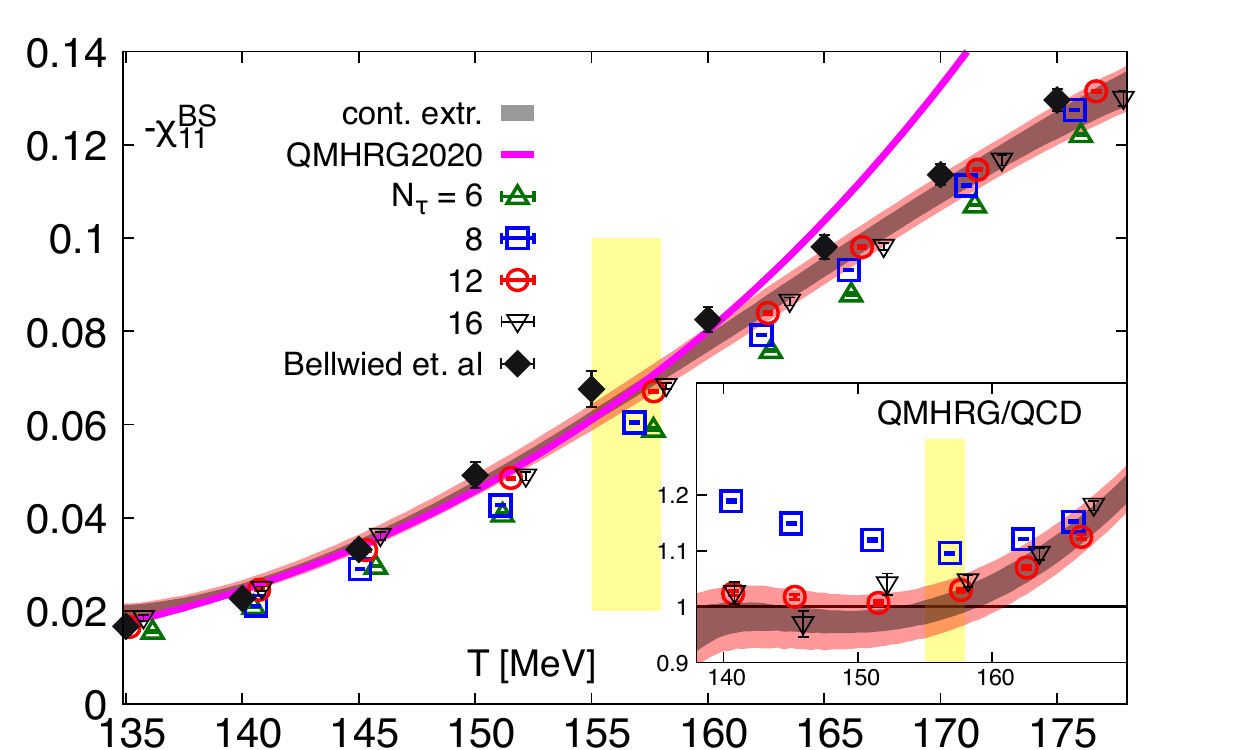}
\includegraphics[scale=0.6]{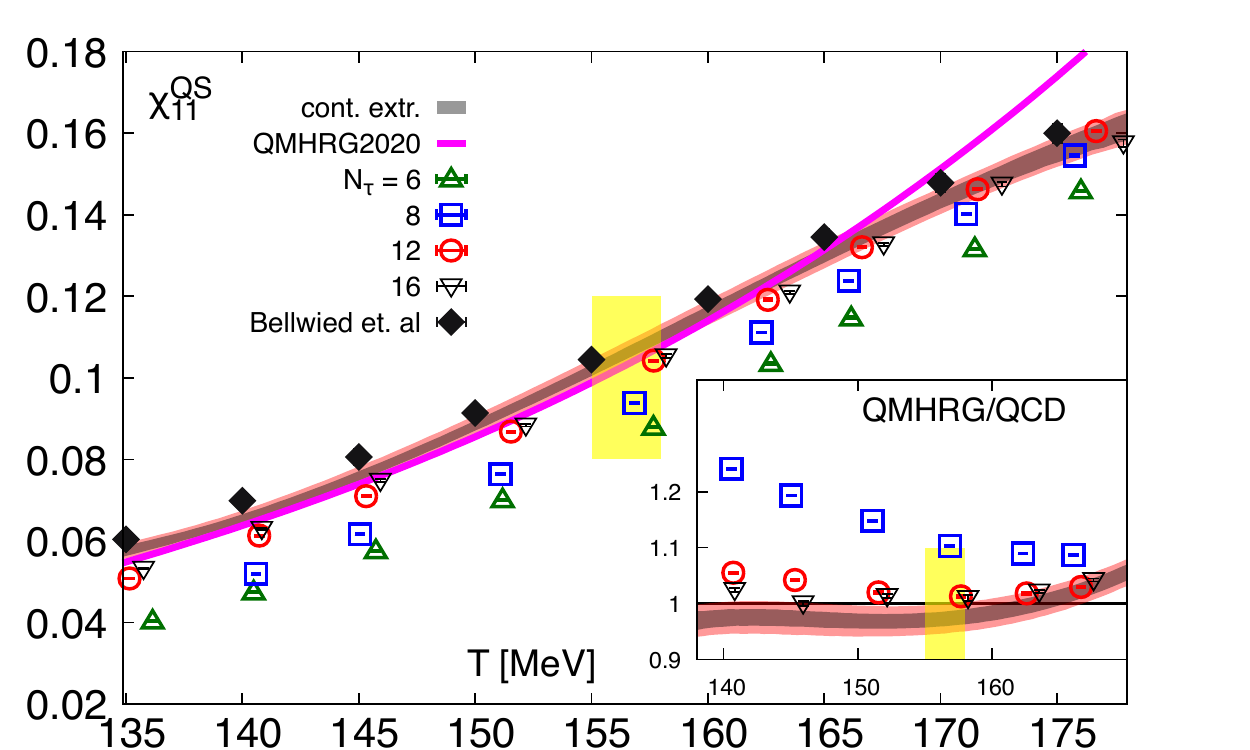}
\caption{Four independent second order cumulants versus temperature calculated
on lattices with different temporal extent $N_\tau$. Lines show results from
HRG model calculations using point-like, non-interacting resonances 
and the hadron spectrum list QMHRG2020 discussed in Section~\ref{QMHRG2020}.
The insets show the ratio of continuum extrapolated lattice QCD results
and HRG model calculations based on QMHRG2020.
Also shown are results from Bellwied et al. \cite{Bellwied:2019pxh}. For the presentation
of data at finite values of $N_\tau$ the temperature scale based on calculations of 
$af_K$ has been used.
Here and in all other figures the yellow band corresponds to the crossover temperature,
$T_{pc,0}$.
}
\label{fig:second}
\end{figure*}

Continuum extrapolated results for all four independent
second order cumulants, calculated on lattices with temporal extent $N_\tau =6,\ 8,\ 12$ and $16$ are shown in Fig.~\ref{fig:second} in the temperature range 
$130~{\rm MeV} < T< 180~{\rm MeV}$.
In that figure we also show separately the systematic error arising from the scale uncertainties discussed in the previous paragraph (red band), and the
combined statistical and systematic error arising from 
the continuum extrapolation using two different $T$-scales (grey band). 
The insets in these figures show a comparison between the lattice QCD results and 
a specific HRG model calculation that is based on the 
QMHRG2020 list of resonances (see Section~\ref{QMHRG2020}). This will be discussed further in the next section.

The  uncertainty of the $T$-scale determination has been
propagated into a systematic error for our continuum extrapolated observables at temperature $T$.
This is quoted as a systematic error on the observables at
temperature $T$. Results for some values of the 
temperature are given in Table~\ref{tab:HW-Hot-results} 
and compared with corresponding results obtained in Ref.~\cite{Bellwied:2019pxh}. As can be seen agreement between these two analyses is quite good.

\begin{table*}[ht]
\begin{tabular}{|c||c|c||c|c||c|c||c|}
\hline
$T [\rm MeV] $ & \multicolumn{2}{c||}{$\chi_{11}^{BQ}$} & \multicolumn{2}{c||}{$\chi_{11}^{BS}$} & \multicolumn{2}{c||}{$\chi_{11}^{QS}$} & $(\chi_2^Q)_{LT=4}$\\
\hline
& this work & \cite{Bellwied:2019pxh} &  this work & \cite{Bellwied:2019pxh} & this work & \cite{Bellwied:2019pxh} & this work\\
\hline
135&  0.0114(5)(2)   & 0.0101(8)   & -0.0197(15)(3)    & -0.0167(17) & 0.0576(13)(6)  & 0.0604(20)  & 0.285(4)(2) \\
140&  0.0140(4)(3)   & 0.0124(8)   & -0.0251(7)(7)     & -0.0227(13) & 0.0655(10)(9)  & 0.0699(18)  & 0.312(4)(3) \\
145&  0.0172(4)(3)   & 0.0162(12)  & -0.0345(8)(11)    & -0.0332(18) & 0.0760(12)(12)  & 0.0806(20)  & 0.342(4)(3) \\
150&  0.0204(4)(3)   & 0.0217(17)  & -0.0469(10)(14)   & -0.0491(28) & 0.0883(12)(13) & 0.0914(12)  & 0.374(5)(3) \\
155&  0.0235(4)(3)   &  0.0242(10) & -0.0616(14)(15)   & -0.0676(38) & 0.1018(14)(14) & 0.1045(9)   & 0.404(5)(3) \\
160&  0.0261(4)(2)   &  0.0266(7)  & -0.0775(20)(16)   & -0.0825(27) & 0.1160(16)(14) & 0.1193(15)  & 0.433(5)(3) \\
165&  0.0280(3)(1)   &  0.0278(6)  & -0.0938(22)(15)   & -0.0981(26) & 0.1300(18)(14) & 0.1345(20)  & 0.458(5)(2) \\
170&  0.0288(2)(0)   &  0.0277(4)  & -0.1097(24)(14)   & -0.1136(23) & 0.1434(20)(12) & 0.1478(22)  & 0.476(4)(1) \\
175&  0.0281(3)(1)   &  0.0269(4)  & -0.1244(30)(12)   & -0.1296(24) & 0.1553(26)(12) & 0.1600(23)  & 0.489(4)(1) \\
\hline
\end{tabular}
\caption{\label{tab:HW-Hot-results}
Continuum extrapolated results for the set of four independent second order cumulants obtained at some values of $T$. The second error given for
our data corresponds to the uncertainty in the
physical value of $r_1$ as discussed in the text. In the case of $\chi_2^Q$ we explicitly indicate that these results are sensitive to 
finite volume effects and have been obtained
on lattices with aspect ratio $LT\equiv N_\sigma/N_\tau=4$ (see discussion in Section~\ref{chiQ2}).
Also shown are results from Bellwied et al. \cite{Bellwied:2019pxh}. 
}
\end{table*}

Further details on our continuum extrapolated results and
comparisons with various HRG model calculations as well as results obtained from virial expansions will be presented in the next sections. 

\subsection{Parametrization of the continuum extrapolated
second order cumulants}

When performing the continuum extrapolations of second order cumulants we did not assume any  specific ansatz for the temperature 
dependence of the cumulants. It, however, will be useful
also for comparisons with other observables and model 
calculations to have at hand a parametrization of the four independent second order
cumulants as well as the two dependent observables, $\chi_2^B$ and $\chi_2^S$. 
For this purpose we provide
a parametrization in terms of rational polynomials,
\begin{equation}
	\chi_{11}^{XY}(T)=\frac{\sum\limits_{k=0}^3 n_{k}^{XY} \bar{t}^k}{1+
		\sum\limits_{k=1}^3 d_{k}^{XY} \bar{t}^k}
		\; , \; \bar{t}=\left( 1-\frac{T_{pc,0}}{T} \right) ~\; ,
	\label{cumulantfit}
\end{equation}
where $X, Y\in {B,Q,S}$, and it is
understood that $\chi_{11}^{XY}$ should be replaced by $\chi_2^X$ 
for $X=Y$. 
This parametrization corresponds to the center of the 
error bands shown in Fig.~\ref{fig:second}. The coefficients
for these parametrizations are  given in Table~\ref{tab1:fit_params}.

\begin{table}
	\begin{tabular}{ |c||c|c|c|c|c|c|}
		\hline
		   & $\chi^{BQ}_{11}$ & $\chi^{QS}_{11}$ & $\chi^{BS}_{11}$ & $\chi^{Q}_{2}$ & $\chi^{B}_{2}$ &  $\chi^{S}_{2}$ \\
		\hline
		 $n^{XY}_0$ &  0.0243 & 0.106 & -0.066  & 0.413  & 0.115 & 0.279  \\
		 $n^{XY}_1$ &  0.0122 &  0.0629 & -0.327  & -0.159 & 0.328  & -1.172 \\
		 $n^{XY}_2$ & -0.376  & -0.6097 & 0.0290 & -2.099 & -0.933 & -6.661 \\
		 $n^{XY}_3$ & -1.219  & -3.896 & 3.834  & -6.362 & -6.522 &  -28.378 \\
		 \hline
		 \hline
		 $d^{XY}_1$  & -3.036  &  -3.572 & -2.505 & -2.605 & -2.922 & -9.135 \\
		 $d^{XY}_2$  &  3.006  &  3.166 & 2.952 & 1.677 & 3.189 & 13.624 \\
		 $d^{XY}_3$  & -2.133  & -5.080 & 3.0973 & 3.892 & -0.245 & -66.402 \\
		\hline
		\end{tabular}
	\caption{Parametrization of second order cumulants representing lines for central values of the fits shown in Fig.~\ref{fig:second} in the interval
	$T\in [135~{\rm MeV}:175~{\rm MeV}]$.}
	\label{tab1:fit_params}
\end{table}

\begin{figure*}[t]
\includegraphics[scale=0.6]{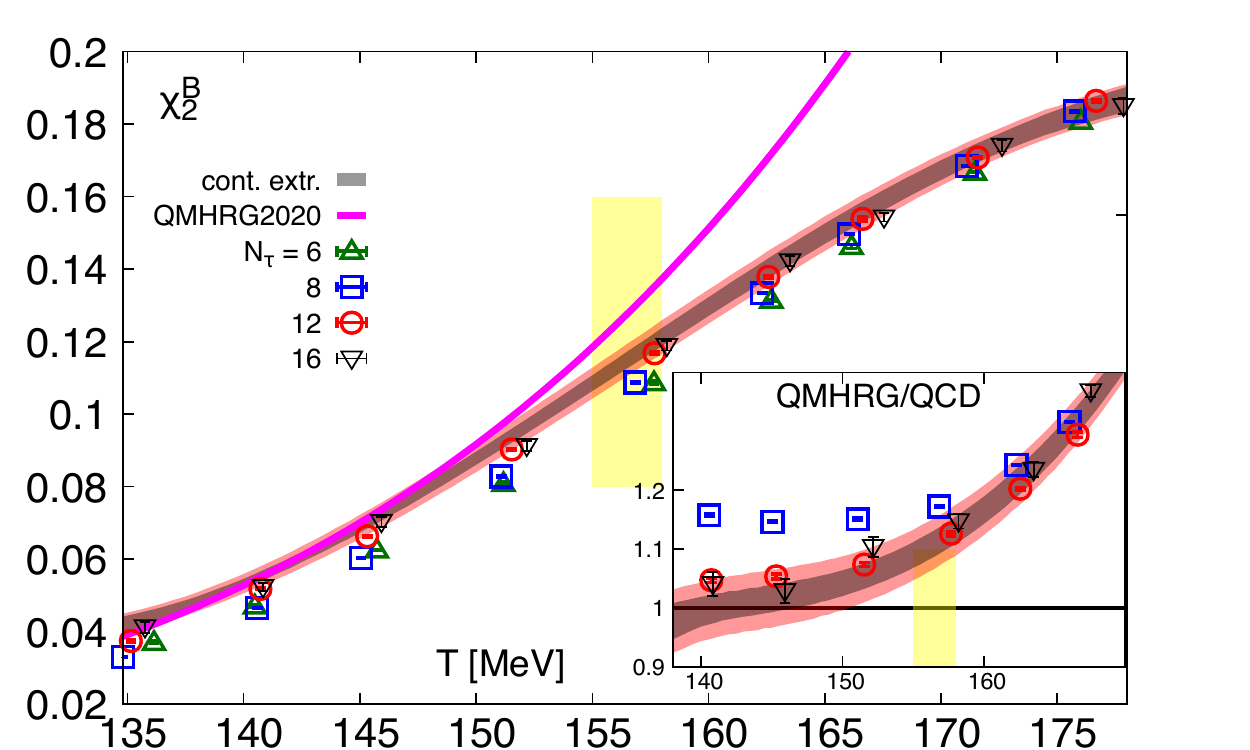}
\includegraphics[scale=0.6]{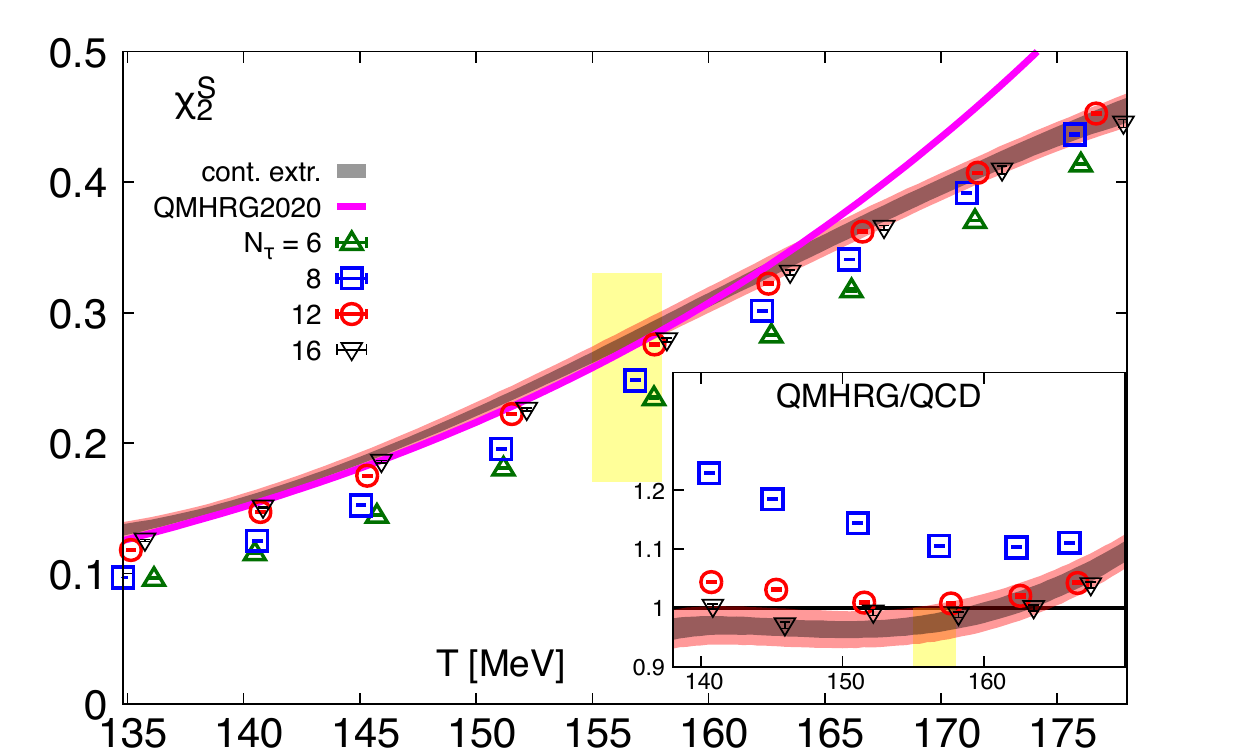}
\caption{The diagonal second order cumulants $\chi_2^B$ (left) and $\chi_2^S$ (right) versus temperature calculated
on lattices with different temporal extent $N_\tau$. Lines show results from
HRG model calculations using point-like, non-interacting resonances 
and the hadron spectrum list QMHRG202,
introduced in Section~\ref{QMHRG2020}.
The insets show the ratio of continuum extrapolated lattice QCD results
and the HRG model calculations based on QMHRG2020.
}
\label{fig:diag}
\end{figure*}

As a consistency check we use the parametrizations of the four independent cumulants and
compare with the diagonal susceptibilities $\chi_2^S$
and $\chi_2^B$, respectively. Note that for fixed $N_\tau$ 
these data fulfill exactly
the constraints given in Eqs.~\ref{constraintS}
and \ref{constraintB}. The data for  $\chi_{11}^{BQ}$, $\chi_{11}^{BS}$ and $\chi_{11}^{QS}$, however, are 
correlated, which may lead to slight differences in the
continuum extrapolated results. We show the continuum extrapolations for $\chi_2^B$ and $\chi_2^S$ in Fig.~\ref{fig:diag}. Here the error band has been obtained in the 
same way as for the off-diagonal cumulants. 
The continuum extrapolated results for these 
diagonal second order cumulants are given in Table~\ref{tab:HW-Hot-results-diag} for the same 
temperature values as those given in Table~\ref{tab:HW-Hot-results} for the set of four 
independent second order cumulants. Although the 
continuum extrapolated results for $\chi_2^B$ and 
$\chi_2^S$ have been obtained without imposing
explicitly the relations given in Eqs.~\ref{constraintS} and \ref{constraintB} we find that the continuum
extrapolated cumulants are consistent with these constraints within errors and also the parametrization
of second order cumulants given in Table~\ref{tab1:fit_params} do so to better than 1\%.

\begin{table}[ht]
\begin{tabular}{|c|c|c|}
\hline
$T [\rm MeV] $ & $\chi_{2}^{B}$ & $\chi_{2}^{S}$ \\
\hline
135 & 0.0422(22)(10) & 0.134(4)(2) \\
140 & 0.0532(14)(14) & 0.156(2)(2) \\
145 & 0.0689(15)(18) & 0.187(3)(3) \\
150 & 0.0878(18)(20) & 0.224(3)(4) \\
155 & 0.1085(22)(21) & 0.266(4)(4) \\
160 & 0.1296(26)(21) & 0.310(5)(4) \\
165 & 0.1497(28)(19) & 0.354(6)(4) \\
170 & 0.1673(27)(15) & 0.396(7)(4) \\
175 & 0.1809(29)(11) & 0.435(8)(4) \\
\hline
\end{tabular}
\caption{\label{tab:HW-Hot-results-diag}
Continuum extrapolated results for $\chi_{2}^{B}$ and $\chi_{2}^{S}$ 
obtained at some values of $T$. The second error reflects the uncertainty in the
value of $r_1$ used for setting the temperature scale as discussed in the text.
}
\end{table}

\section{Second order cumulants in QCD and the hadron resonance gas model} 

In this section we compare QCD results for
second order cumulants with HRG model calculations. We will discuss
the importance of additional resonances contributing to correlations
of net strangeness fluctuations with
net baryon-number and net electric-charge fluctuations, respectively.
We also discuss in detail the quite different behavior 
of correlations between net baryon-number and net electric-charge fluctuations on the one hand and 
net baryon-number and net strangeness fluctuation
on the other hand, which is apparent when comparing
QCD and HRG model calculations. Furthermore, we comment on the significance of finite volume effects
visible in the second order cumulant of net electric-charge
fluctuations.

\subsection{Sensitivity of second order cumulants
to details of the hadron resonance gas spectrum}

Obviously, conclusions drawn from a comparison of HRG model calculations with QCD results crucially depend 
on the hadron spectrum which is input to the HRG  model calculations. Although such models can rely on a lot of
information from experimentally determined resonances \cite{Zyla:2020zbs},
it has been noted that this information is not 
sufficient to constrain interactions in a 
hadron resonance gas to such an extent that these models do provide satisfactory 
comparisons with QCD results.
As noted earlier \cite{Bazavov:2014xya}, in particular in the baryon sector
of the spectrum additional strange hadron resonances seem to be needed to 
obtain reasonable agreement between HRG model calculations and QCD results 
for strangeness fluctuations and their correlations with electric charge 
and baryon number fluctuations, respectively. 

\begin{figure}[htb]
    \includegraphics[scale=0.67]{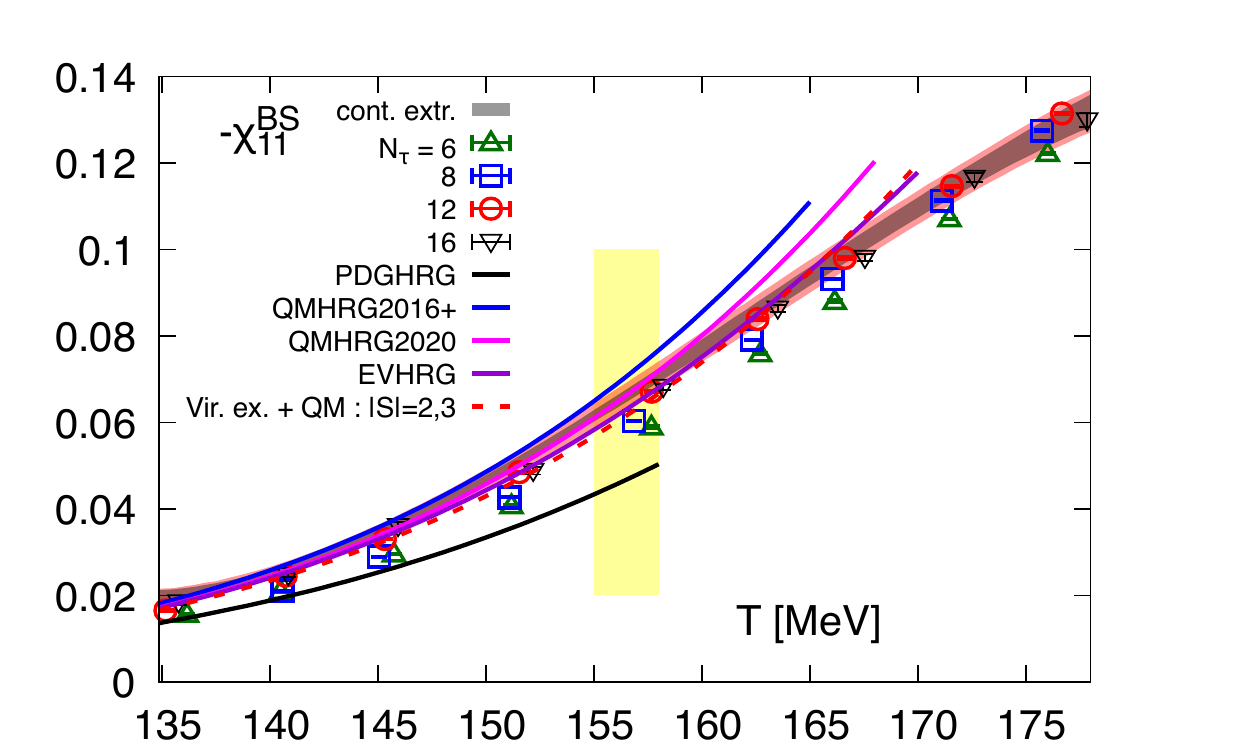}
    \includegraphics[scale=0.67]{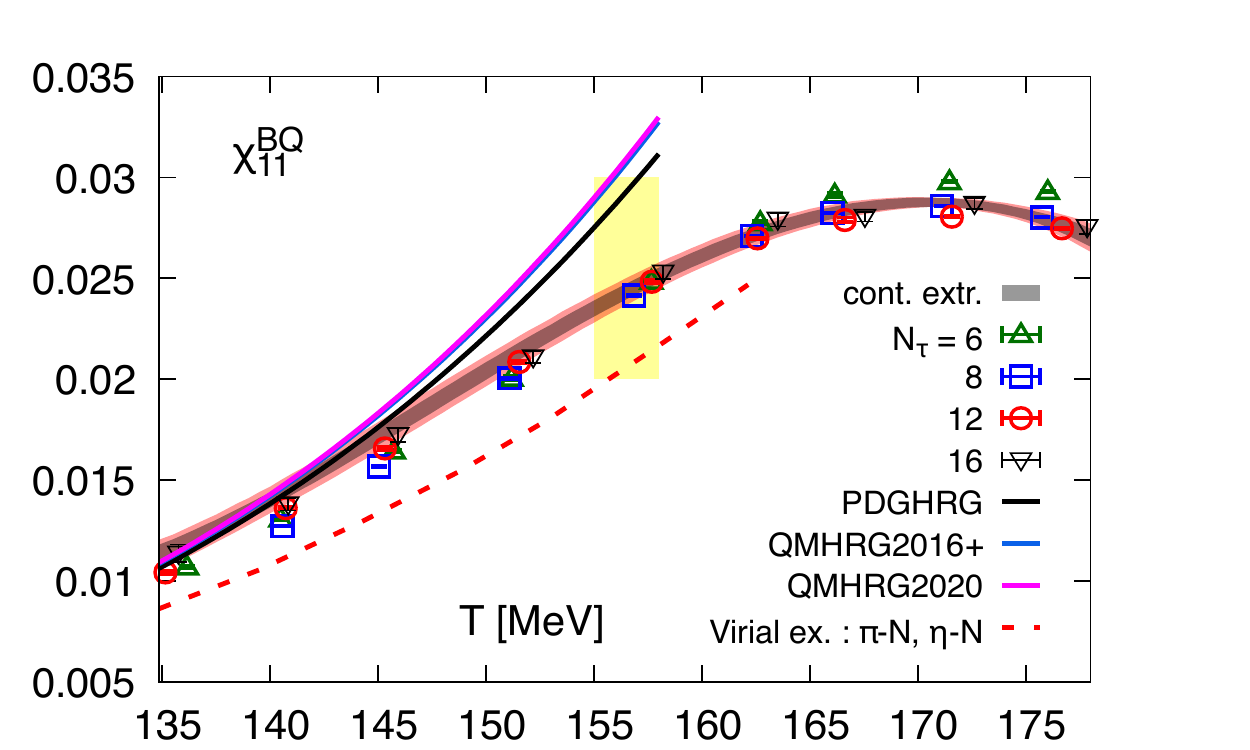}
\caption{Continuum extrapolated results for 
$\chi_{11}^{BS}$ (top) and $\chi_{11}^{BQ}$ (bottom).
Shown is a comparison to HRG model calculations
based on different lists for hadron resonances as 
discussed in the text. Also shown are results obtained
with excluded volume HRG models, using an excluded volume parameter $b=0.4$~fm$^3$, and virial expansions
\cite{Lo:2017lym,Fernandez-Ramirez:2018vzu},
respectively.
}
\label{fig:HRG}
\end{figure}

In particular, when using in HRG model calculations with 
point-like, non-interacting resonances
only well established mesons and 3- and 4-star baryon resonances listed 
in the summary tables of the particle data group (PDG) \cite{Zyla:2020zbs}, 
the comparison with QCD results yields only poor agreement in the 
strangeness sector (see Fig.~\ref{fig:HRG}~(top)).
Including additional resonances that
have been predicted in QCD based quark model (QM) calculations, e.g. in 
\cite{Capstick:1986bmxx,Ebert:2009ub,Faustov:2015eba}, as well as lattice QCD calculations 
\cite{Edwards:2011jj,Edwards:2020rbo}, significantly improves the comparison 
between HRG model and QCD calculations. However, such an approach is not 
unique. It depends on details of the relativistic quark model calculations (for a recent compilation
of results see \cite{Menapara:2021dzi})
as well as the treatment of the not well established resonances listed by the PDG.
Moreover, it is well understood since the early work
of Hagedorn \cite{Hagedorn:1965st} and 
Dashen, Ma and Bernstein \cite{Dashen:1969ep}
that a modeling of strong interactions in terms of 
point-like, interacting resonances is not sufficient
to account for all interactions in strongly interacting
matter. The need for a proper treatment
of additional repulsive 
interactions, for instance by assigning an intrinsic
volume to each hadron, has been pointed out early on
\cite{Hagedorn:1980kb,Hagedorn:1980cv}. However,
it also has been noted that at the same time 
the interplay between repulsive and attractive interaction needs to be taken into account.

A more rigorous treatment of interactions among
hadrons in medium can be achieved in the S-matrix
formalism \cite{Dashen:1969ep} which is the starting point
for a systematic virial expansion of relativistic quantum gases \cite{Fiore:1977pb,Venugopalan:1992hy}. 
The subtle interplay of attractive and repulsive
interactions is, at least in principle, taken
care of in the virial expansion. 
In the case of the strange meson $K_0^*(700)$, for 
instance, an analysis of the effect of attractive 
and repulsive contributions in the S-matrix, partial
wave analysis led to the conclusion that 
the contribution of this resonance 
to the thermodynamics of strong interaction matter is strongly suppressed. Effectively
it does not contribute at all and the resonance
thus should not be included in point-like, non-interacting HRG models \cite{Broniowski:2015oha,Friman:2015zua},
despite the fact that it is listed as a well established resonance in the PDG tables.

Setting up a HRG model for the description of 
the thermodynamics of strong interaction matter in
the low temperature, hadronic regime thus is subject
to a certain amount of ambiguity. Nonetheless, such 
models are a good starting point for the comparison
to QCD calculations and can serve as a mediator between
QCD calculations and experimental observations.

\subsubsection{QMHRG2020}
\label{QMHRG2020}
Previously we used a list of hadron resonances  (QMHRG) that included only established mesons 
and  3- and 4-star baryon resonances listed in the PDG summary tables and had been
augmented with a list of QM states in the strange and non-strange
baryon sectors \cite{Bazavov:2014xya}.  
We now updated this list of hadron resonances by taking into account 
the 1- and 2-star baryon resonances as well as 
mesons not listed as being well established in the
PDG 2020 summary tables \cite{Zyla:2020zbs}. 
We, however, left out the $K_0^*(700)$ (see below).
This defines the list of hadron resonances, QMHRG2020\footnote{The list of hadron resonances, QMHRG2020, and also the PDGHRG list used by us, have been added to this archive publication
as ancillary files. These lists also are given in the data publication that contains all the material needed
to reproduce the figures shown in this paper \cite{bollweg2021dataset}}.

To a large extent the resonances, listed in the PDG tables, have counterparts in the hadron spectrum
calculated in relativistic quark models
\cite{Capstick:1986bmxx,Ebert:2009ub,Faustov:2015eba}.
In order to avoid double counting we used from the QM calculations
only states that have no identified counterparts in the PDG tables.
QMHRG2020 differs from the 
QMHRG2016+ list \cite{Alba:2017mqu} by only a few
resonances. In the strange baryon sector this leads to a few percent differences 
in the HRG model calculation of $\chi_{11}^{BS}$, that are of significance when comparing HRG 
model calculations with QCD results, while in all other cases differences in the two lists
are negligible. For instance, for temperatures below the QCD pseudo-critical temperature,
i.e. for $T\sim (140-155)$~MeV, HRG model results for $\chi_{11}^{BS}$,
obtained from both lists, differ by about 6\%. Eliminating the QM counterparts \cite{Capstick:1986bmxx,Faustov:2015eba} of two 4-star as well as two 3-star resonances  from the QMHRG2016+ list reduces this difference to 3\%. In Table~\ref{tab:HW-Hot-resonances} of Appendix~\ref{HRG-comparison}  we give a list of 8 strange baryon resonances that are listed in the
PDG tables and also have been calculated in relativistic quark models. Eliminating the latter from QMHRG2016+ reproduces results obtained with QMHRG2020
within 1\% accuracy\footnote{Both lists still differ to the extent that masses of strange baryons, obtained in quark model calculations,
are taken from \cite{Faustov:2015eba} in QMHRG2020 whereas QMHRG2016+ uses results from  \cite{Capstick:1986bmxx}.}.

In the following we use the QMHRG2020 list of hadron resonances
as baseline model for comparisons with QCD calculations. 
In the insets of Fig.~\ref{fig:second} we have compared the continuum extrapolated results for four independent second order 
cumulants, obtained in $(2+1)$-flavor QCD, with
HRG model calculations that make use of the QMHRG2020 list. The insets show the ratio of results obtained in HRG
model calculations and
QCD, respectively. As can be seen, at low temperatures the agreement is fairly good. 
In detail, however, the differences seen in QCD and HRG
model calculations are quite 
different for the four second order cumulants and reflect different physics. We will discuss this in 
more detail in the following subsections. We note 
already here that the difference between QCD results and HRG calculations is particularly striking for correlations
between net baryon-number and electric charge, $\chi_{11}^{BQ}$.  

\subsubsection{Net baryon-number fluctuations and correlations}

In Fig.~\ref{fig:HRG} we compare HRG model calculations using different 
hadron resonance spectra with continuum extrapolated lattice QCD results for 
(2+1)-flavor QCD. As can be seen correlations between net baryon-number 
and strangeness fluctuations ($\chi_{11}^{BS}$) are particularly sensitive to 
contributions from the strange baryon resonances, while correlations between net 
baryon-number and electric charge fluctuations ($\chi_{11}^{BQ}$) show only 
little sensitivity to contributions from additional baryon resonances; this cumulant only depends mildly on the strange baryon sector as  only
$|S|=2, 3$ baryons contribute to $\chi_{11}^{BQ}$.
The small differences seen in the magnitude of $\chi_{11}^{BQ}$ when calculated with PDGHRG and QMHRG spectra, respectively, (Fig.~\ref{fig:HRG}~(bottom)) 
mainly arise from additional non-strange baryons obtained in QM calculations.

We note that $|\chi_{11}^{BS}|$, calculated in a HRG model using QMHRG2020,
is larger by about 30\% relative to calculations based
only on 3- and 4-star resonances listed by the PDG. 
HRG model calculations of $\chi_{11}^{BS}$, using QMHRG2020, are consistent
with QCD results at and below the pseudo-critical temperature.
HRG model results for $\chi_{11}^{BQ}$,
on the other hand, are clearly larger than the corresponding QCD results for temperatures 
$T\gsim 145$~MeV. They are about 20\% larger at
$T_{pc,0}$. This is a quite
robust result, as HRG model calculations for $\chi_{11}^{BQ}$ are not very sensitive
to the details of the HRG resonance spectrum used and even the inclusion of 1- and 2-star resonances has little effect, as can be seen in Fig.~\ref{fig:HRG}~(bottom).

In the case of net baryon-number
fluctuations, $\chi_2^B$, which are related to
$\chi_{11}^{BQ}$ and $\chi_{11}^{BS}$ through Eq.~\ref{constraintB}, this still leads to 
10\% larger results in HRG model calculations than 
QCD results obtained at $T_{pc,0}$. 

\begin{figure*}[t]
\hspace{-0.7cm}
\includegraphics[scale=0.47]{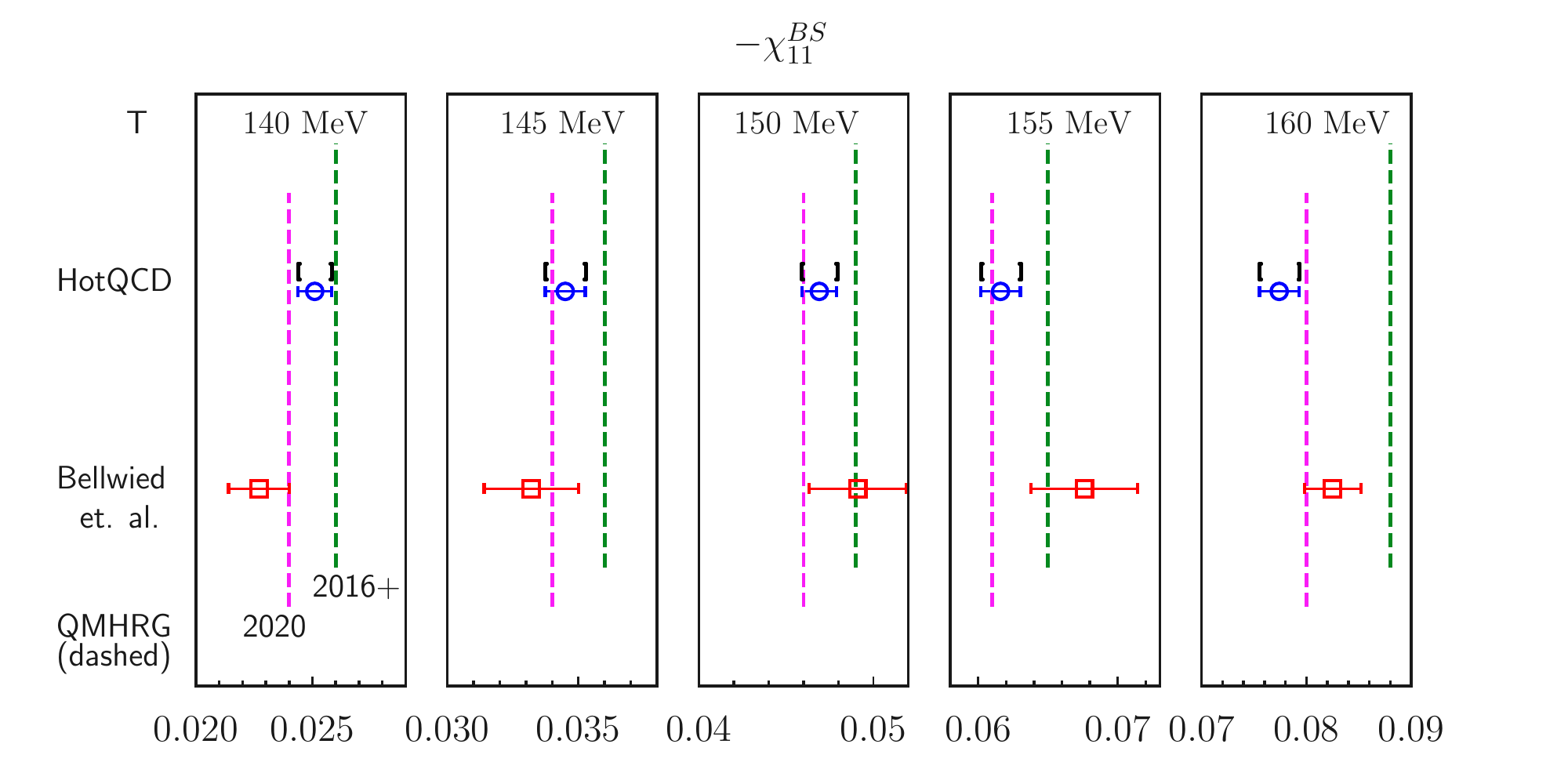}
\hspace{-0.7cm}
\includegraphics[scale=0.47]{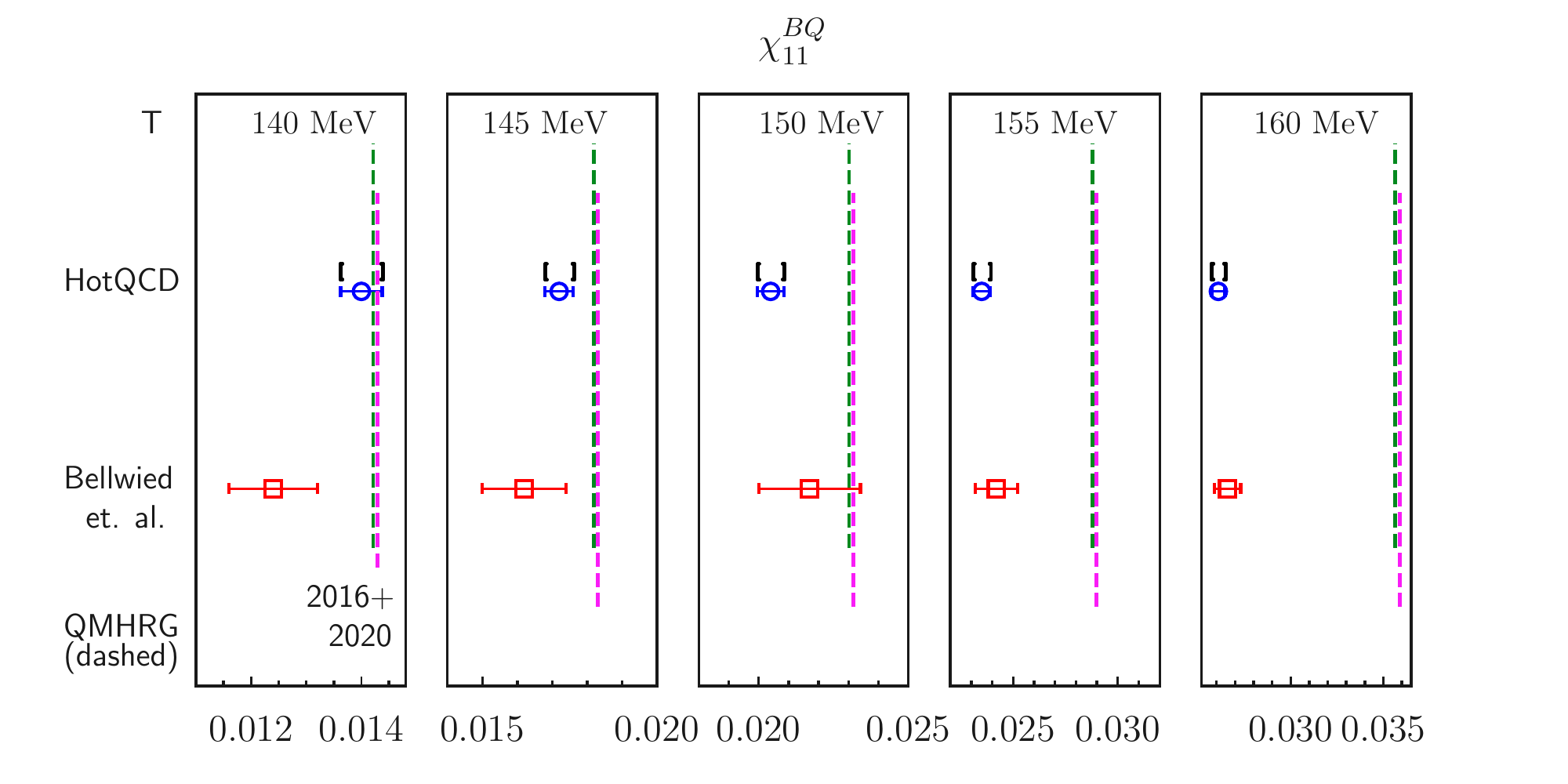}
\hspace{-0.3cm}
\caption{Comparison of continuum extrapolated lattice QCD results for the second order cumulants
$\chi_{11}^{BS}$ (left) and $\chi_{11}^{BQ}$ (right)
with QMHRG model calculations  based on the QMHRG2020 and QMHRG2016+ list of hadron resonances. In addition to the results
from this work (HotQCD) we also show results from
Bellwied et al. \cite{Bellwied:2019pxh}. For our results we show separately statistical errors and 
systematic errors (brackets) arising from the 
systematic error on $r_1$. 
}
\label{fig:cumulants-compare}
\end{figure*}

In Fig.~\ref{fig:cumulants-compare} we show 
for temperatures close to and below $T_{pc,0}$
a comparison between HRG model calculations, performed with the QMHRG2020 and QMHRG2016+ resonance lists, and QCD. In this figure we also
compare our results with results obtained by
Bellwied et al. \cite{Bellwied:2019pxh}.
Within the current statistical and systematic uncertainties the results of both works are consistent with each other.

As outlined above, the larger magnitude obtained for $|\chi_{11}^{BS}|$ when using QMHRG2016+ arises from the fact that this list uses some quark model states that are also listed in the PDG tables. As a consequence the HRG model contains too many strange baryons; QMHRG2020 gives better agreement with QCD results below $T_{pc,0}$. 

The differences between lattice QCD calculations and HRG 
model calculations using point-like, non-interacting hadronic resonances, indicate that these models do not 
correctly reflect interactions in a strongly interacting medium. It has been attempted to improve the model calculations by taking into account additional repulsive interactions in so-called excluded volume HRG models (EVHRG)
\cite{Gorenstein:1981fa,Vovchenko:2016rkn,Taradiy:2019taz}.
Generically the magnitude of second order cumulants involving net baryon-number fluctuations is decreased
in such excluded volume HRG models. 
Indeed, in the case of $\chi_{11}^{BQ}$ this may improve the comparison to QCD. However, at the same time it is obvious that EVHRG model calculations for $\chi_{11}^{BS}$ will spoil the good agreement observed between QCD results and
HRG model calculations with point-like, non-interacting resonances observed below $T_{pc,0}$. While $\chi_{11}^{BS}$
favors only small excluded volumes, $\chi_{11}^{BQ}$
would require large volumes to achieve agreement between
HRG model calculations and QCD results.

In fact, the relative 
change in second order cumulants involving net baryon-number fluctuations, calculated in HRG and EVHRG models, respectively, is identical
\cite{Taradiy:2019taz}
\begin{eqnarray}
R_B^{EV}&=&\frac{(\chi_{11}^{BQ})_{EVHRG}}{(\chi_{11}^{BQ})_{HRG}}
=\frac{(\chi_{11}^{BS})_{EVHRG}}{(\chi_{11}^{BS})_{HRG}}
=\frac{(\chi_{2}^{B})_{EVHRG}}{(\chi_{2}^{B})_{HRG}}
\nonumber \\
&=& 1 - 2 b P_{B}^{HRG}(T)/T + {\cal O}(b^2) \; .
\label{RBEV}
\end{eqnarray}
Here $b$ parametrizes the size of the excluded volume for all
baryons and $P_{B}^{HRG}(T)$ denotes the contribution of baryons and anti-baryons to the pressure in a resonance gas model for non-interacting point-like resonances. It contains all
information on details of the hadron spectrum.  The excluded
volume parameter $b$ is related to the hard-sphere radius ($r$) of a hadron through $b=16\pi r^3/3$.  

The comparison of EVHRG model calculations and QCD results, 
shown in Figs.~\ref{fig:HRG} and \ref{fig:cumulants-compare} for $\chi_{11}^{BS}$ and $\chi_{11}^{BQ}$,  
makes it clear that no unique choice for $b$ exists
that would give better agreement between QCD and HRG 
models for both cumulants simultaneously.
In fact, using Eq.~\ref{RBEV} we may write
\begin{equation}
    \frac{(\chi_{11}^{BX})_{EVHRG}}{(\chi_{11}^{BX})_{QCD}} =
    R_B^{EV} \frac{(\chi_{11}^{BX})_{HRG}}{(\chi_{11}^{BX})_{QCD}} \;\; , \;\; X=Q,\ S .
\end{equation}
Within the errors put on QCD results,
this ratio should be consistent with unity for EVHRG models
in order to be consistent with QCD results. This puts bounds on the
magnitude of the excluded volume parameter $b$. Using the QCD result for $\chi_{11}^{BX}$, $X=S,\ Q$, and including the combined statistical and systematic error $\Delta_X$, we find
the largest ($b^+$) and smallest ($b^-$) excluded volume 
parameters that would yield EVHRG model results consistent
with QCD results,
\begin{equation}
    b^\pm = \frac{1}{2 T^3 (\chi_2^B)_{HRG}}\left( 1-\frac{(\chi_{11}^{BX}\pm\Delta_X)_{QCD}}{(\chi_{11}^{BX})_{HRG}} \right) \;\; . \;\; 
\end{equation}
Here we used the relation between the baryonic part of the pressure and the second order cumulant of net baryon-number
fluctuations in HRG models with point-like hadrons, $P_B/T^4 = \chi_2^B$. At low temperatures these bounds are not very stringent as the hadronic medium is dilute.
However, close to $T_{pc,0}$, i.e. for $T=(150-155)$~MeV, we find that $b\le 0.4$~fm$^3$ is needed in order for EVHRG calculations to be consistent with QCD results for $\chi_{11}^{BS}$, whereas from $\chi_{11}^{BQ}$
we find that $b$ should be significantly larger, i.e.
$1~{\rm fm}^3\le b \le 2~{\rm fm}^3$.  

\begin{figure}[t]
\includegraphics[scale=0.50]{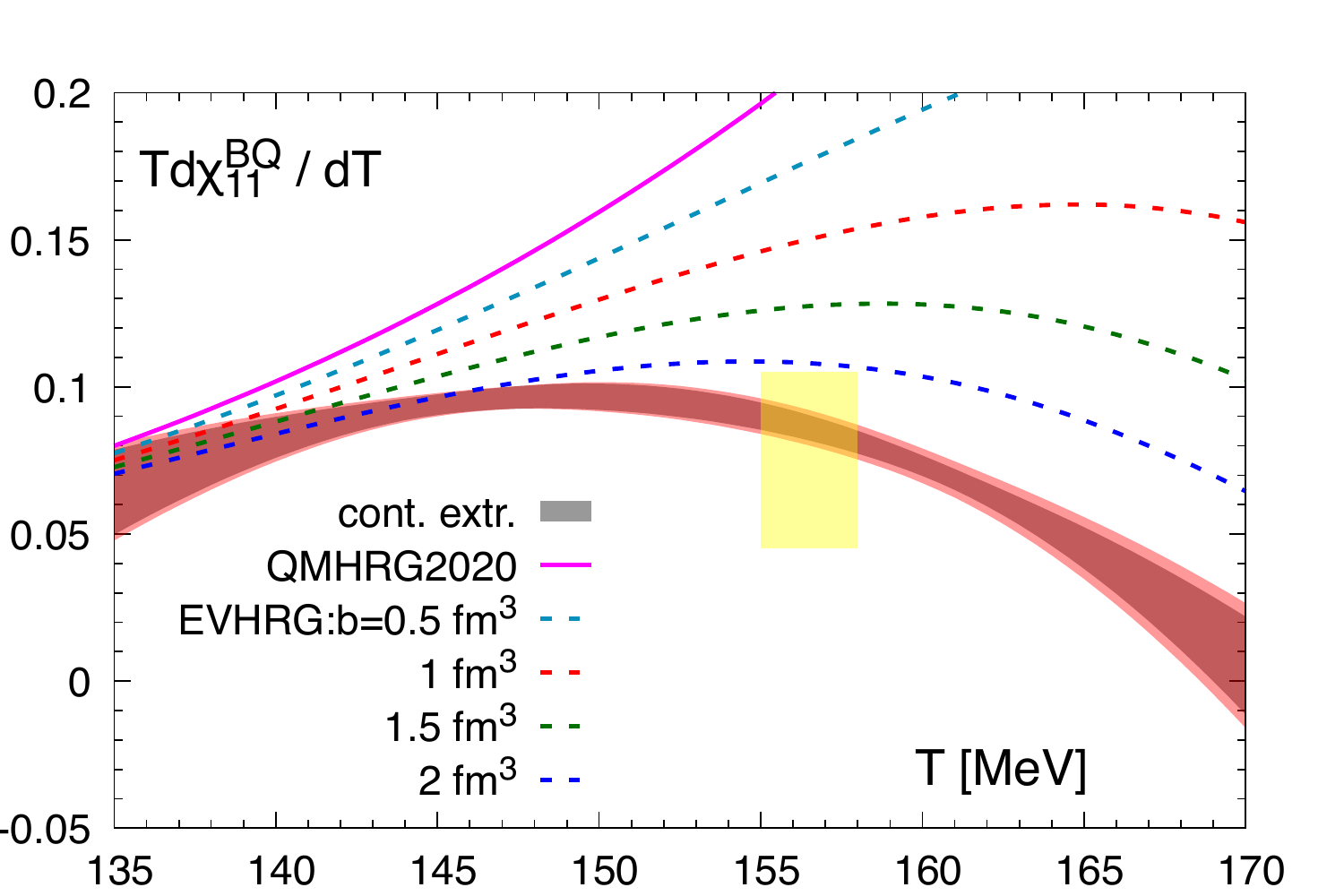}

\caption{Temperature derivative of second order cumulant
$\chi_{11}^{BQ}$ obtained in QCD (band). This result is
compared to EVHRG calculations using the QMHRG2020
and several values of the excluded volume parameter $b$.
}
\label{fig:TderivBQ}
\end{figure}

In fact, the temperature dependence of $\chi_{11}^{BQ}$
favors $b\simeq 2$~fm$^3$. As can be seen in Fig.~\ref{fig:second} (top, left),
in QCD calculations it is found that $\chi_{11}^{BQ}$ rises much more slowly with temperature than in HRG model
calculations using QMHRG2020. A value of $b$, significantly
larger than $1$~fm$^3$ would be needed to account for this difference.
In Fig.~\ref{fig:TderivBQ} we show results for the temperature derivative of $\chi_{11}^{BQ}$ shown in Fig.~\ref{fig:second} (top, left). In the case of QCD these derivatives are obtained from the parametrization given in Eq.~\ref{cumulantfit}, using a bootstrap fit to the 
continuum extrapolated results shown in Fig.~\ref{fig:second}.
As can be seen $b\gsim 2$~fm$^3$ would be needed in an EVHRG calculation to reproduce the QCD results for
the temperature derivative of $\chi_{11}^{BQ}$.

\begin{figure}[t]
\includegraphics[scale=0.70]{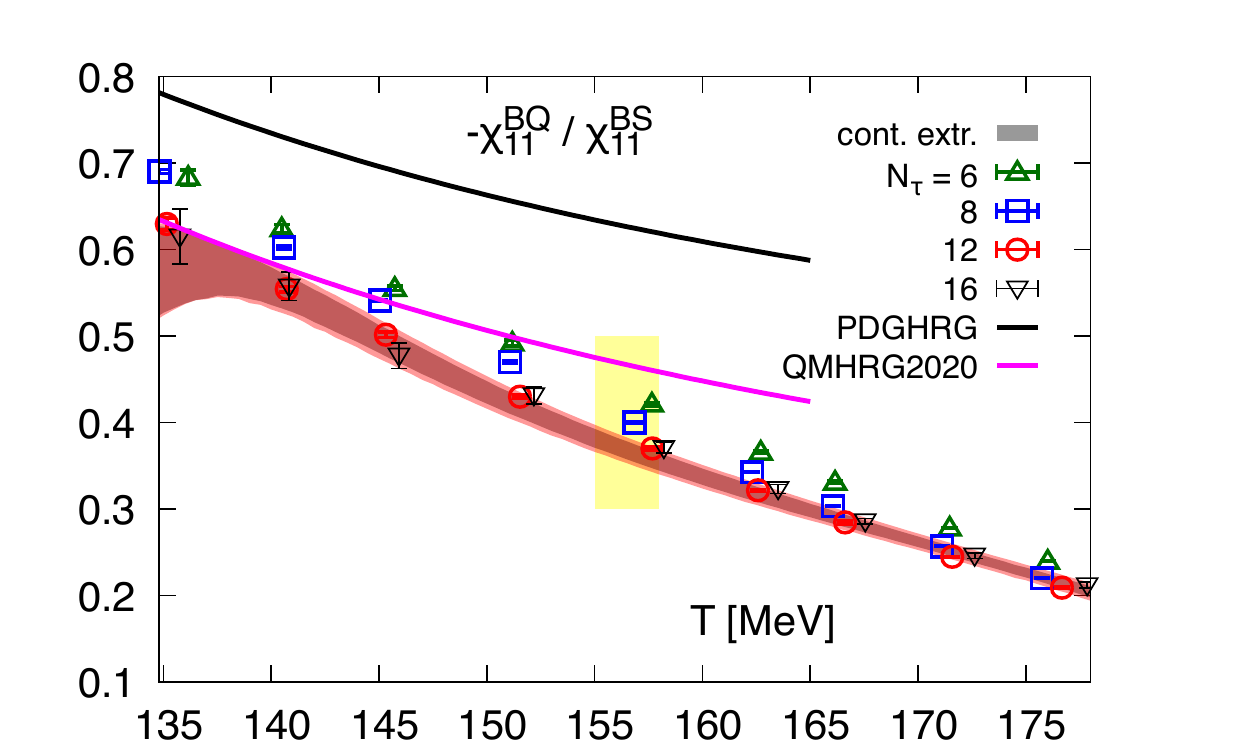}
\caption{Ratio of continuum extrapolated lattice QCD results
	for $\chi_{11}^{BQ}$ and $\chi_{11}^{BS}$ versus 
	temperature.}
\label{fig:BQBS}
\end{figure}

As excluded volume corrections are identical for all
baryon correlations and fluctuations, it is 
instructive to look at ratios of second order cumulants
involving net baryon-number fluctuations. In this case
excluded volume corrections drop out. 
Any difference between HRG and QCD 
cumulants thus is of different origin. 
We also note that
due to the relation given in Eq.~\ref{constraintB}, it suffices to analyze one ratio, e.g. $\chi_{11}^{BQ}/\chi_{11}^{BS}$. Other ratios, are then simply related to this ratio, e.g.,
\begin{eqnarray}
    \frac{\chi_2^B}{\chi_{11}^{BS}} &=& 2 \frac{\chi_{11}^{BQ}}{\chi_{11}^{BS}} -1 \; ,
    \nonumber \\
        \frac{\chi_2^B}{\chi_{11}^{BQ}} &=& 2 - \frac{\chi_{11}^{BS}}{\chi_{11}^{BQ}} \; .
\end{eqnarray}
Differences found 
between QCD and HRG model calculations of $\chi_{11}^{BQ}/\chi_{11}^{BS}$ thus translate into
corresponding differences for the other two ratios.
Results for $\chi_{11}^{BQ}/\chi_{11}^{BS}$ are
presented in Fig.~\ref{fig:BQBS}. This shows that deviations from HRG
model calculations, which cannot be accounted for in 
excluded volume models using a single parameter $b$, 
become significant already at temperatures $T\simeq 145$~MeV. This, of course, is a direct consequence of the
deviation of HRG model calculations from QCD results setting in for $\chi_{11}^{BQ}$ already at $T\simeq 140$~MeV, while at the same time results for $\chi_{11}^{BS}$ obtained in  HRG model calculations are
in good agreement with QCD up to $T_{pc,0}$.

A similar conclusion has been drawn from calculations of the second virial coefficients
using the S-matrix approach
\cite{Lo:2017ldt}, where it has been pointed out that 
the influence
of repulsive interactions, which motivated an excluded volume 
ansatz, is subtle and quite different in various quantum number channels
contributing in the partial wave analysis of the second 
virial expansion coefficient.

\begin{figure}[t]
\includegraphics[scale=0.50]{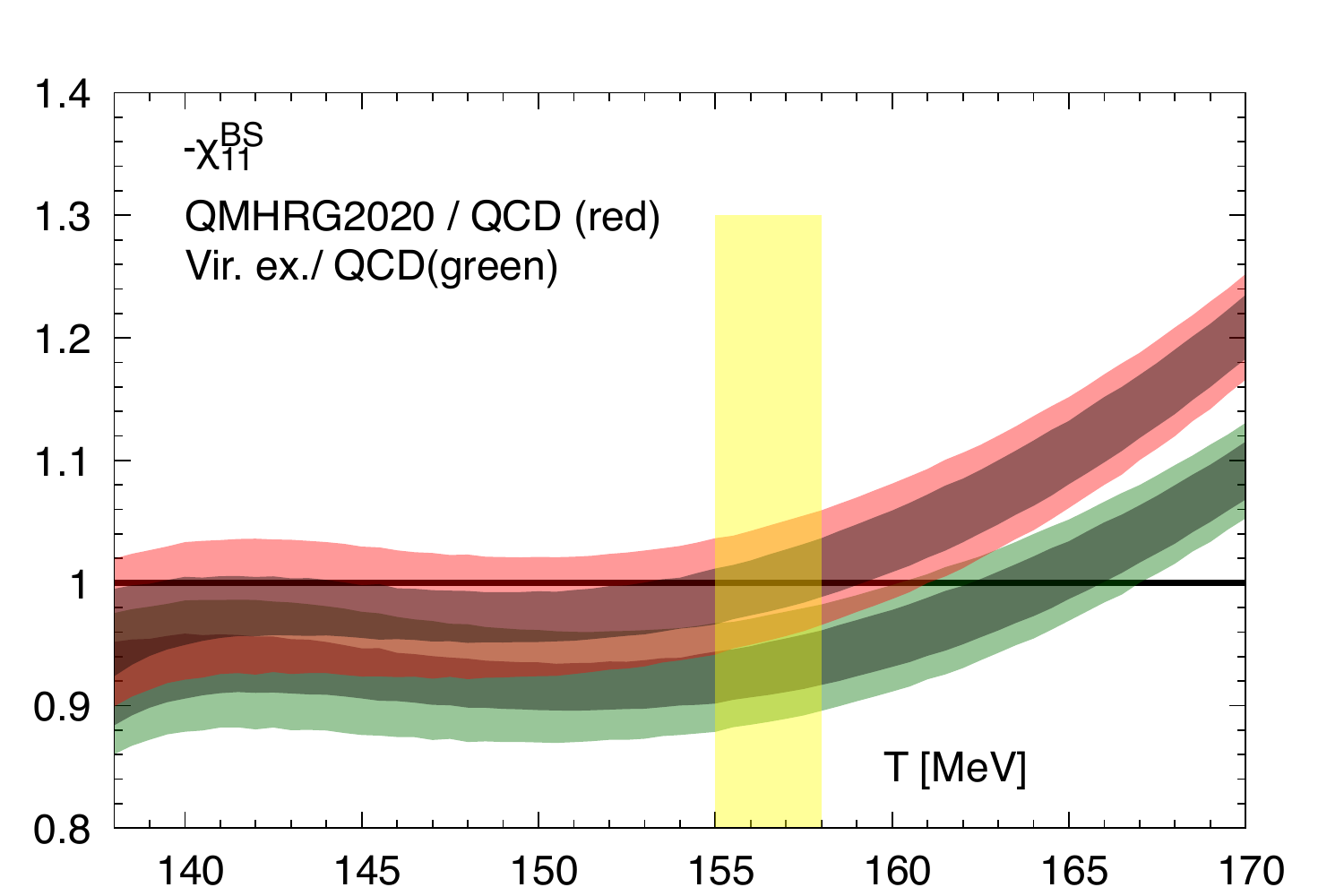}
\includegraphics[scale=0.50]{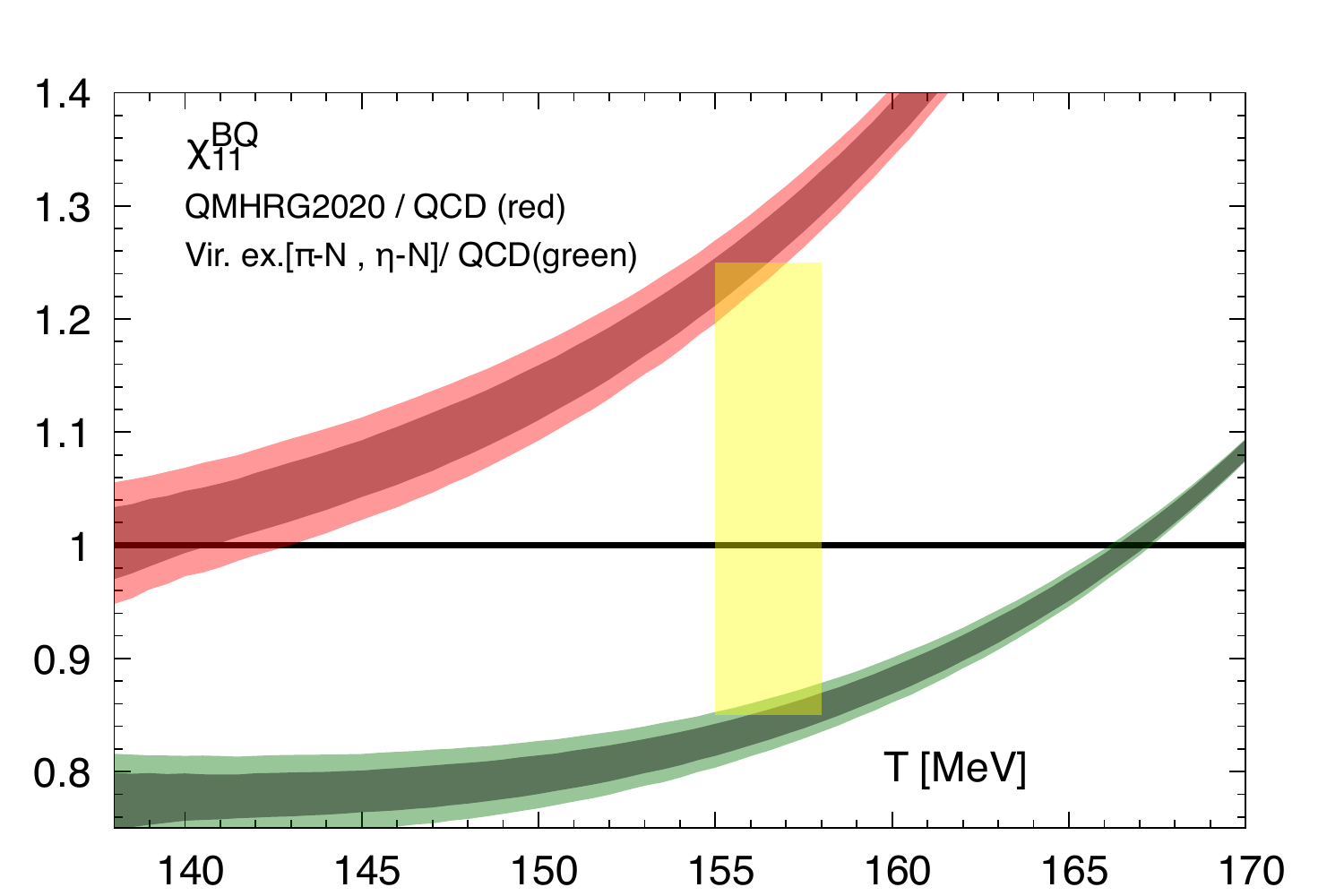}
\caption{Comparison of continuum extrapolated lattice QCD results 
	for $\chi_{11}^{BS}$ (top) and $\chi_{11}^{BQ}$ (bottom) with HRG model results based on QMHRG2020 (red band) and
	second order virial expansion results (green band)
\cite{Lo:2017lym,Fernandez-Ramirez:2018vzu}.
}
\label{fig:Smatrixratio}
\end{figure}

Although the S-matrix approach, which is based on experimental
information on phase shifts contributing to the S-matrix,
provides a rigorous formulation of the thermodynamics
of an interacting hadron gas in the grand canonical ensemble and, at least in principle, does not require
any a-priori information on a particular spectrum of
resonances, it generally is difficult
to arrive at first principle, quantitative results 
on properties of second order cumulants. In a systematic
virial expansion of the partition function, expressed 
in terms of the S-matrix, even the calculation of the
second virial coefficient often suffers from
insufficient experimental information even
for 
two-particle interactions.
Moreover, at higher
densities  multi-particle interactions become of importance
\cite{Dashen:1969ep,Venugopalan:1992hy}, which are poorly known and thus
require approximations when comparing a S-matrix based analysis to e.g.
first principle lattice QCD calculations
\cite{lo_probing_2013,Fernandez-Ramirez:2018vzu,Andronic:2018qqt}.

In Fig.~\ref{fig:Smatrixratio} we compare QCD results for second order cumulants involving
net baryon-number fluctuations, with corresponding
results from calculations of the second virial coefficient.
In Ref.~\cite{Fernandez-Ramirez:2018vzu} the 
second virial coefficient for the description of
correlations between net baryon-number and net strangeness fluctuation has been obtained 
in a unitary, multi-channel analysis \cite{Capstick:1993kb,Hunt:2018wqz}.  
We show results from this analysis in Fig.~\ref{fig:Smatrixratio}~(top). 
As can be seen, agreement between S-matrix calculations and lattice QCD results is within
(10-15)\% below
the pseudo-critical temperature. This result is similar to, but does at present not improve over results that can already be achieved within a HRG model calculation based
on QMHRG2020.

A calculation of the second virial coefficient
for the description of correlations between
net baryon-number and net electric-charge fluctuations is more difficult as less information on the relevant interaction 
channels is known.
In \cite{Lo:2017lym} $\chi_{11}^{BQ}$ has been
analyzed in the S-matrix approach taking into account two body interactions arising from elastic $\pi$-$N$ scattering, and a small contribution from the 
inelastic $\pi$-$N\rightarrow \eta$-$N$ channel. This
partial calculation of the second virial expansion coefficient turns out not to be sufficient to achieve good 
agreement with QCD results \cite{Lo:2017lym,Andronic:2018qqt}.
Although it improves the 
comparison with QCD, deviations 
at low temperatures are about a factor two larger than in the S-matrix analysis of $\chi_{11}^{BS}$.
The inclusion of further interaction channels 
and contributions from higher order corrections clearly is needed. 

At present the insufficient knowledge on scattering process contributing to $\chi_{11}^{BQ}$ 
can only be overcome by modeling contributions from three body and higher order interactions. This has been attempted in \cite{Andronic:2018qqt} 
by approximating $\pi\pi N$ interactions by a 
simple, structure-less vertex.
The strength of the interaction vertex in this model has been tuned to
achieve agreement with lattice QCD results
at $T\simeq 150$~MeV. However, comparing the results obtained in \cite{Andronic:2018qqt} with the precise QCD results presented here also at lower temperatures shows that agreement in a wider temperature range cannot be achieved
with a temperature independent coupling
strength for this interaction vertex.
At lower temperatures, $T\sim 140$~MeV, deviations from QCD are still about 10\%.

\subsubsection{Contribution of $K_0^*(700)$ to $\chi_{11}^{QS}$}

In Fig.~\ref{fig:second}~(bottom, right) we have shown
results for $\chi_{11}^{QS}$ which also are found to agree
well with HRG model calculations using QMHRG2020. As 
mentioned in section~\ref{QMHRG2020} we did not include
the scalar kaon resonance $K_0^*(700)$ (previously $\kappa$ in PDG) \cite{Ablikim:2005ni,Ablikim:2010ab}  in the QMHRG2020
list, although it is listed as an established resonance 
in the PDG tables.
How to best accommodate for this resonance in
thermodynamic calculations is much discussed \cite{Broniowski:2015oha,Friman:2015zua,Giacosa:2018vbw}.

Kaons give the most significant contribution to the correlations between
net electric-charge and strangeness fluctuations.
At low temperatures, $T\sim 130$~MeV, already the ground state kaon and its P-wave
excitation $K^*(892)$ contribute more than 80\% to the $2^{\rm nd}$ order cumulant $\chi_{11}^{QS}$. All heavier strange mesons
and baryons account for the remaining 
contribution to $\chi_{11}^{QS}$.
Adding the
$K_0^*(700)$ to the list of strange meson resonances would change the
HRG model result for $\chi_{11}^{QS}$ by almost 10\%. However, as has been shown in
an analysis of
the strangeness fluctuation cumulant $\chi_2^S$
\cite{Friman:2015zua}, the contribution of $K_0^*(700)$ is largely reduced when
treating its contribution in a virial expansion that makes use of information on scattering amplitudes in 
the S-wave $K$-$\pi$ channel. 

The QCD results for $\chi_{11}^{QS}$ are
shown in Fig.~\ref{fig:second} and compared to our list of hadron resonances
QMHRG2020, in which we do not include the $K_0^*(700)$ resonance.
In Fig.~\ref{fig:QS_Kpi} we compare in more detail the HRG model 
calculations, with and without the $K_0^*(700)$ resonance included, with
QCD results. Here we also show the S-matrix calculation taken from
\cite{Friman:2015zua}, which includes interactions in
the $I=1/2$ and $I=3/2$ S-wave channels. We combined the
result obtained from the virial expansion with those strange hadron resonances from the QMHRG2020 list that are not taken care of in the S-matrix analysis. 
The analysis of interactions in the S-matrix formulation
of an interacting hadron gas thus motivates that 
the $K_0^*(700)$ resonance does not contribute to the
thermodynamics of such a system and thus should be left
out when using a gas of point-like, non-interacting
only.

\subsubsection{Strangeness neutrality and the strangeness chemical potential}
\label{sec:neutral}

The ratio $\chi_{11}^{BS}/\chi_{11}^{QS}$ is directly 
related to the ratio of baryon and strangeness chemical
potentials in a strangeness and electric charge neutral medium. As can be seen in Eq.~\ref{muSmuB} it controls the value of
$\mu_S/\mu_B$ in the isospin symmetric case 
($n_Q/n_B=1/2$),
\begin{equation}
    \frac{\mu_S}{\mu_B} =\left( 1 - 2 \frac{\chi_{11}^{QS}}{\chi_{11}^{BS}}\right)^{-1} + {\cal O}(\mu_B^2)\; ,
    \label{muSmuB2}
\end{equation}
which also gives the dominant contribution
in the case most relevant for comparison with
heavy ion experiments, $n_Q/n_B=0.4$.

\begin{figure}[t]
\includegraphics[scale=0.5]{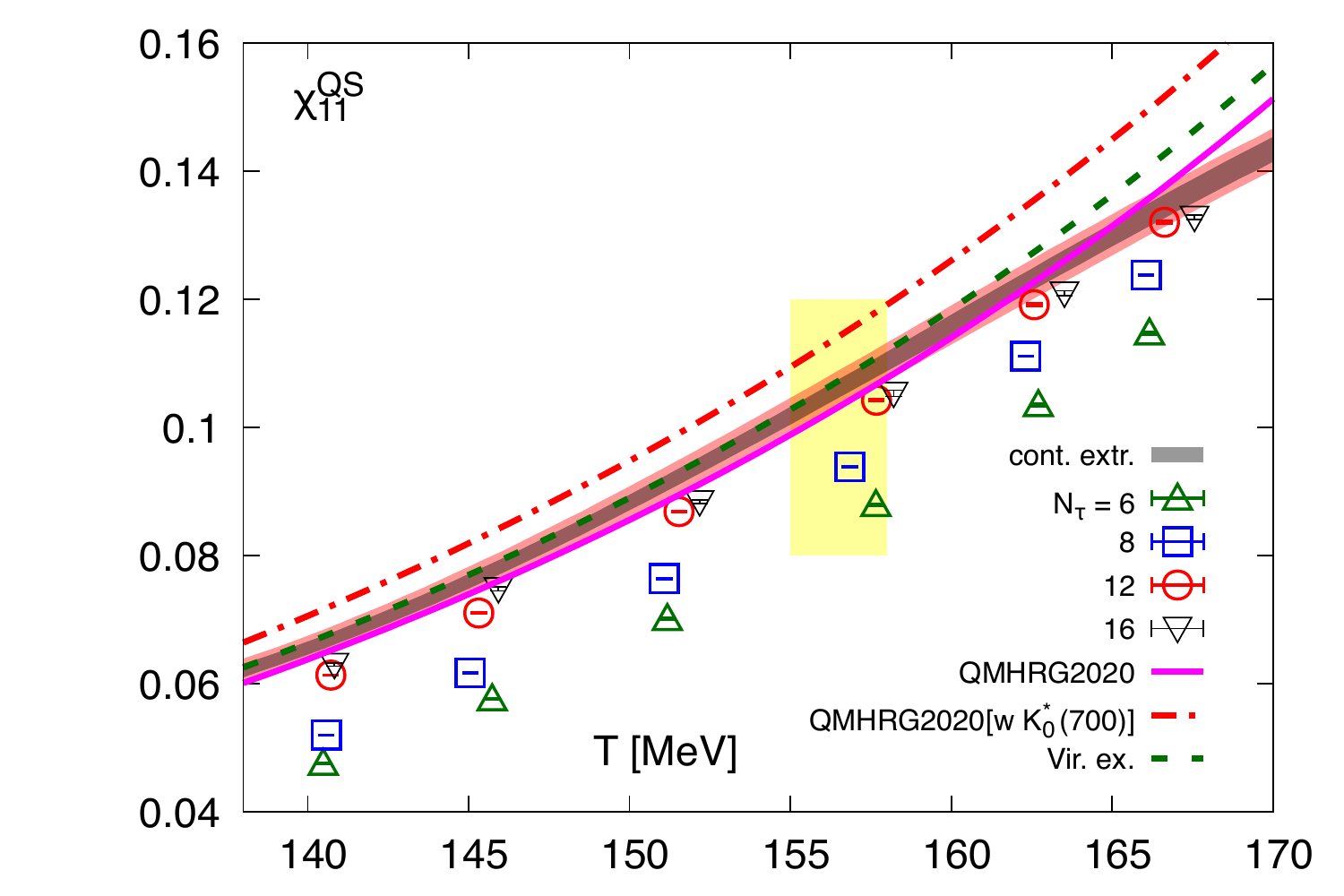}
\includegraphics[scale=0.5]{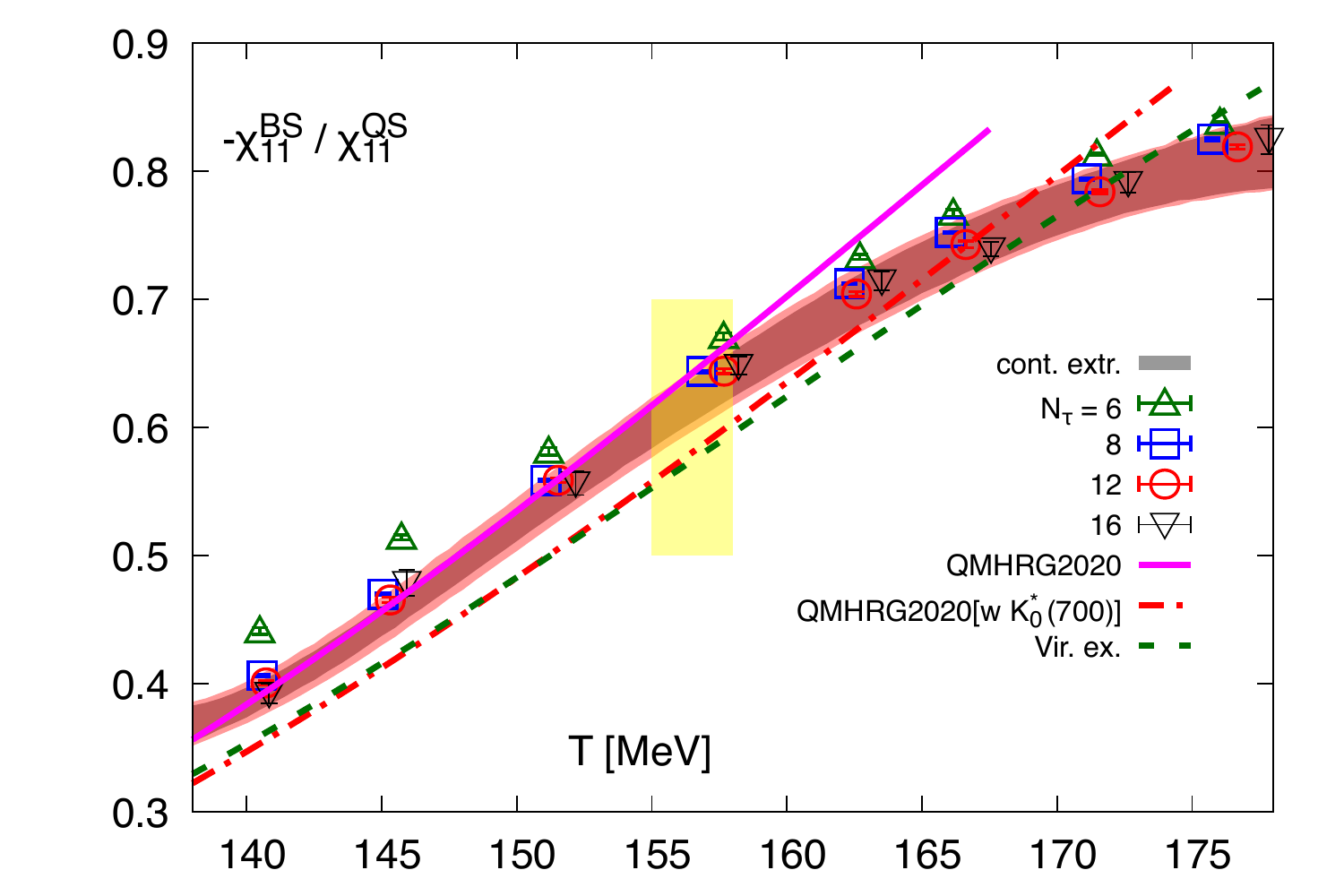}
\caption{Comparison of HRG model calculations with 
QCD results for $\chi_{11}^{QS}$ (top) and the ratio
$\chi_{11}^{BS} / \chi_{11}^{QS}$ (bottom). The dashed-dotted lines show
results from a HRG model calculation where also the contribution of $K_0^*(700)$ is included in the list of
hadron resonances.
In the lower figure we 
show the result of an virial expansion for $\chi_{11}^{BS}$
taken from \cite{Fernandez-Ramirez:2018vzu}.
The upper figure shows the result of a
virial expansion, based on the analysis 
of S-wave scattering contributions in
the $K$-$\pi$ channel to strangeness fluctuations \cite{Friman:2015zua}, as discussed in the text.
}
\label{fig:QS_Kpi}
\end{figure}

The good agreement found for $\chi_{11}^{BS}/\chi_{11}^{QS}$ when calculated in lattice QCD and HRG models using QMHRG2020 (see 
Fig.~\ref{fig:QS_Kpi} (bottom)), suggests that a
determination of the ratio of chemical potentials from experimental data, using as an intermediate
step HRG model based relations, is appropriate at least for small values of the baryon chemical potentials, 
for which the leading order Taylor expansion expressions (Eq.~\ref{densities}) are valid.
The strong sensitivity of  $\chi_{11}^{BS}$ on 
the strange baryon sector, on the one hand, 
and the small sensitivity of $\chi_{11}^{QS}$
on details of 
the spectrum, on the other hand, also suggests that $\mu_S/\mu_B$ provides information on additional strange baryon resonances contributing to the thermodynamics of strongly interacting matter at the freeze-out temperature.

\begin{figure}[t]
   \includegraphics[scale=0.64]{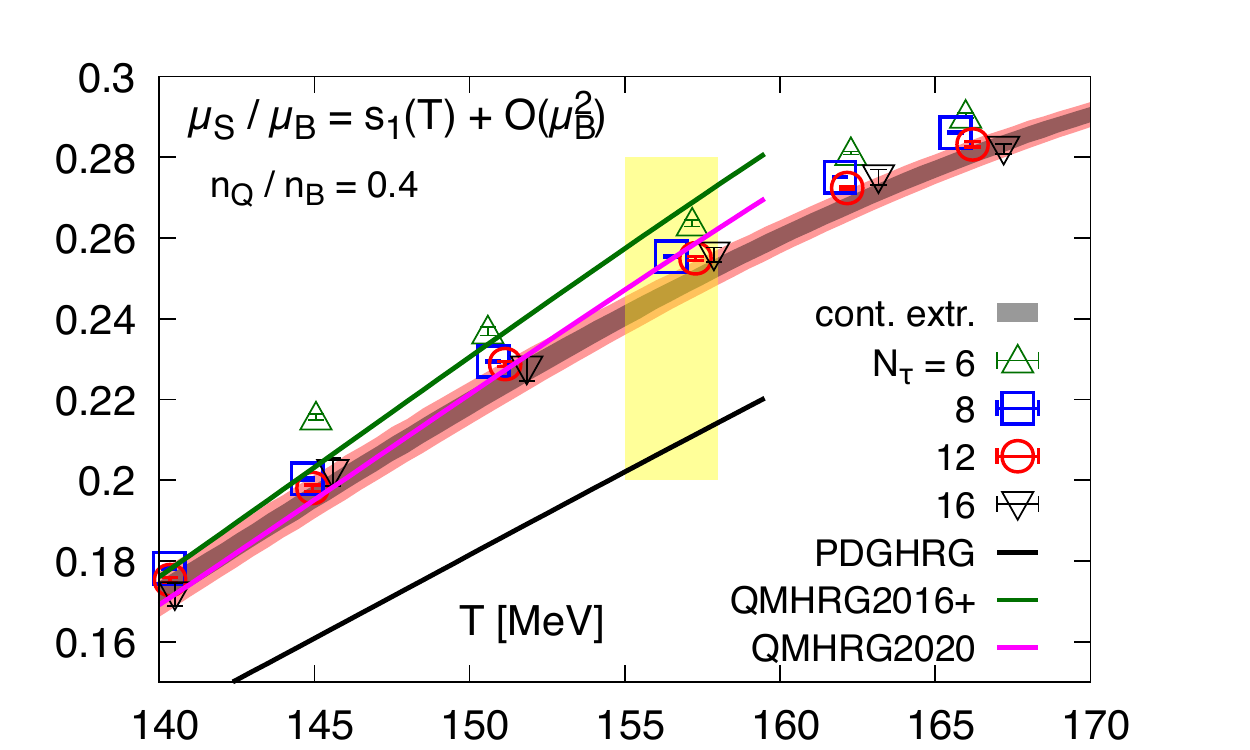}
    \caption{Leading order result for the ratio
    of strangeness and baryon chemical potentials versus temperature for the case 
    $n_Q/n_B=0.4$. QCD results are compared to 
    HRG model calculations using different spectra of hadronic resonances.
}
    \label{fig:muSmuB_STAR}
\end{figure}

In Fig.~\ref{fig:muSmuB_STAR} we
compare results from QCD calculations to HRG
model calculations that use lists of hadron
resonances
with (QMHRG) and without (PDGHRG) additional
strange baryon resonances included. 
As can be seen the ratio
$\mu_S/\mu_B$ obtained in QCD calculations when imposing the strangeness neutrality condition 
$n_S=0$ and $n_Q/n_B=0.4$ differs by about 15\%
from HRG model calculations, that only use the 
PDGHRG states, but is in good agreement with QMHRG model calculations. This shows that the
ratio of strangeness and baryon chemical 
potentials indeed is sensitive to the 
spectrum of strange hadron resonances in a 
strongly interacting medium. We will discuss
this further in a forthcoming publication, where we show that higher order contributions to 
$\mu_S/\mu_B$ are indeed negligible for $\mu_B/T\lsim 1$,
which makes this ratio a good observable for
probing thermal conditions in a strongly-interacting medium.

\subsection{Volume dependence of
the cumulant of net electric-charge fluctuations, \boldmath$\chi_{2}^{Q}$}
\label{chiQ2}

As can be seen in Fig.~\ref{fig:second} (top, right)
even at temperatures as low as $135$~MeV
continuum extrapolated results
for the second order cumulant of net electric-charge fluctuations still differ
from HRG model calculations. At such low temperatures electric charge fluctuations are 
dominated by the contribution from pions. In this case it is well known that
finite volume effects in a pion gas with mass $ m_\pi\lsim T$ can lead to
significant deviations from results obtained in the thermodynamic limit
\cite{Engels:1981ab,Bhattacharyya:2015zka,Karsch:2015zna}.

The lattice QCD calculations presented in Fig.~\ref{fig:second} have been
performed on lattices with a fixed ratio of spatial versus temporal extent,
$N_\sigma/N_\tau =4$. {\it I.e.} the spatial extent of the physical volume,
$L=N_\sigma a$, changes with temperature, $T=1/N_\tau a$, such that $LT=4$
stays constant. In the temperature range shown in Fig.~\ref{fig:second},
we have calculated  $\chi_2^Q$ for a pion gas in a finite volume with $LT=4$. This can be done directly using the partition function of a Bose gas in a finite volume \cite{Bhattacharyya:2015zka,Karsch:2015zna}. In order to mimic periodic boundary conditions and cubic box sizes as they are used in lattice QCD calculations, we instead used a scalar field theory discretized on a lattice as described in \cite{Engels:1981ab}.
For a non-interacting pion and kaon gas we find that deviations from 
infinite volume results are well parametrized by a straight line ansatz
in the temperature interval of interest, $T\in [130~{\rm MeV}:170~{\rm MeV}]$,
\begin{equation}
	\frac{( \chi_2^Q)_{LT=4}}{( \chi_2^Q)_{LT=\infty}}=
\begin{cases}
	0.997-0.126\ T/T_{pc,0} \; , \; {\rm pion}~{\rm gas} \\
	1.002-0.032\ T/T_{pc,0}  \; , \; {\rm kaon}~{\rm gas}
\end{cases}
	\label{bose}
\end{equation}

\begin{table*}[t]
  \centering
\resizebox{\textwidth}{!}{
  \begin{tabular}{l |  ll  ll | c c | c c }
~  & ~~$\chi_{11}^{BQ}$ & ~~~$\chi_{11}^{BS}$  &~~~$\chi_{11}^{QS}$ &$(\chi_{2}^{Q})_{LT=4}$ ~~[$\chi_{2}^{Q}$] &  $\chi_{2}^{S}$ &  $\chi_{2}^{B}$ &
          $\chi_{11}^{BS}/\chi_{2}^{S}$ & $\chi_{11}^{BQ}/\chi_{2}^{B}$
          \\ \hline
QCD[this work] & ~0.0243(7)(9) & -0.066(4)(5) & 0.106(3)(5) & 0.413(8)(9) & 0.279(9)(12) &  0.115(5)(7) &
-0.236(5)(6) & 0.212(4)(5) \\ \hline
QMHRG2020[this work]   &~0.031(3) & -0.066(6)  & 0.103(5) & 0.437(14) [0.466(15)] & 0.272(14) & 0.127(10) & -0.243(8) & 0.244(3) \\
QMHRG2016+ \cite{Alba:2017mqu} & ~0.031(3) & -0.071(7) & 0.104(5) & 0.444(15) [0.472(15)] & 0.277(16) & 0.132(10) & -0.256(7) & 0.235(2) \\
PDGHRG  & ~0.030(2) &-0.046(4) & 0.094(4) & 0.419(12) [0.447(14)] & 0.234(11) & 0.106(8) & -0.197(6) & 0.283(2) \\
\hline
EVHRG2020[$b=1~\rm{fm^3}$]  & ~0.027(2) & -0.059(5) & 0.103(5) & 0.431(13) [0.459(15)] & 0.264(13) & 0.113(8) & -0.223(5) & 0.243(2) \\
S-matrix[\cite{Friman:2015zua},\cite{Fernandez-Ramirez:2018vzu}]  & ~0.020(1) & -0.062(5) & 0.107(4) & -- & -- & -- & -- & --\\
  \end{tabular}
}
  \caption{Continuum-extrapolated values of second order cumulants at
          $T_{pc,0}=156.5(1.5)$~MeV
  and for vanishing values of the chemical potentials. First error in QCD is obtained by linearly combining statistical and systematical error of our continuum extrapolations while the second one error reflects the error on the determination of $T_{pc,0}$. Similarly the error in the HRG is due to the error of $T_{pc,0}$.
  These results agree well with the continuum extrapolated data given in \cite{Bellwied:2019pxh}.
  For $\chi_2^Q$ we give results calculated in a
  finite volume $LT\equiv N_\sigma / N_\tau =4$.
  Numbers in brackets give the corresponding infinite 
  volume result.
          }
\label{tab:2ndmu0}
\end{table*}

At the pseudo-critical temperature, $T_{pc,0}$, the net electric-charge
fluctuations in a pion gas in a volume $LT=4$ thus is about 12\% smaller
than in the infinite volume limit. In the HRG model, calculated
with the resonance spectrum QMHRG2020, this distortion effect gets reduced
by almost a factor two, reflecting the relative contribution of pions
to the entire net electric-charge fluctuations. In the temperature interval 
$T\in [130~{\rm MeV} : 180~{\rm MeV}]$ we find for 
a HRG model using the QMHRG2020 list of hadrons,
\begin{eqnarray}
        \frac{( \chi_2^Q)_{LT=4}}{( \chi_2^Q)_{LT=\infty}} &=& 
	1.324-1.290 T/T_{pc,0} + 1.316 ( T/T_{pc,0})^2 \nonumber \\
	&& -0.411 ( T/T_{pc,0})^3\; , \;
	        \label{QMHRG}
\end{eqnarray}
This amounts to a 6\% finite volume correction for $\chi_2^Q$ at the 
pseudo-critical temperature, $T_{pc,0}$, which only changes slowly as 
function of temperature. We show this finite volume correction to the
QMHRG2020 result for $\chi_2^Q$ in 
Fig.~\ref{fig:second} (top, right).

Using the finite volume corrected QMHRG2020 results for 
a comparison with QCD results in a finite volume, we find
that at $T_{pc,0}$ the latter are still smaller by about 5\%. this is consistent with large
deviations observed in the charged baryon sector
($\chi_{11}^{BQ}$) which contributes only
about 15\% to the total electric charge fluctuations. 

\subsection{Comparison of QCD results with various model
calculations at \boldmath$T_{pc,0}$}

Continuum extrapolated lattice QCD results
for all six second order cumulants
at $T_{pc,0}$ and corresponding results from HRG model calculations, using different hadronic
resonance spectra, as well as results from S-matrix 
calculations are summarized in Table~\ref{tab:2ndmu0}.
Here we also give results for two  of the three independent ratios of second order cumulants. The third
ratio would involve $\chi_2^Q$, for which at present no
infinite volume extrapolated result exists.

\section{Conclusions}

We have presented an update of continuum extrapolated results for second order cumulants of conserved charge
fluctuations and their cross-correlations. These results
are based on high-statistics data obtained in simulations within the HISQ discretization scheme for staggered fermions. Our results are found to be consistent with
previous results obtained by using the stout discretization scheme for staggered fermions
\cite{Bellwied:2019pxh}.

We compiled a new list of hadron resonances, QMHRG2020,
which differs from QMHRG2016+ in particular in the 
strange baryon sector. HRG model calculations with 
QHMHRG2020 provide a good description of
strangeness fluctuations and correlations with baryon-number and
electric charge fluctuations, respectively. Deviations 
are found to be less than 10\% in the temperature range
$135~{\rm MeV}< T < T_{pc,0}$. 
This puts stringent bounds on the magnitude of excluded volume parameters for strange baryon interactions in EVHRG models.

The largest differences between QCD results and HRG model
calculations have been found for correlations between
net baryon-number and electric charge fluctuations. 
They amount to more than 20\% at $T_{pc,0}$ when 
comparing QCD with HRG models based on point-like, 
non-interacting resonances. Modeling these deviations,
and in particular the quite different temperature
dependence of $\chi_{11}^{BQ}$ found in the vicinity of
$T_{pc,0}$, in excluded volume HRG models would require
a large excluded volume parameter $b> 1$~fm$^3$. 

Calculations based on a virial expansion capture
basic features of the interplay between repulsive and 
attractive interactions in the strongly interacting
hadronic medium. They successfully explain why in 
particular the strange meson resonance $K_0^*(700)$ does
not contribute in that medium. They also qualitatively 
describe the smaller magnitude of $\chi_{11}^{BQ}$
found in QCD compared to HRG model calculations with
point-like, non-interacting resonances. However, 
in order to achieve quantitative agreement with QCD
modeling of interactions that are not captured in
calculations of the second virial coefficient is needed.

The results presented here form the basis for
a systematic analysis of second order cumulants
at non-vanishing values of the chemical potentials. This will be discussed in a forthcoming publication. All data from our calculations, presented in the figures of this paper, can be found in \cite{bollweg2021dataset}.

\vspace{0.2cm}
\noindent
\emph{Acknowledgments.---} 
This work was supported by: (i) The U.S. Department of Energy, Office of
Science, Office of Nuclear Physics through the Contract No. DE-SC0012704;
(ii) The U.S. Department of Energy, Office of Science, Office of Nuclear
Physics and Office of Advanced Scientific Computing Research within the
framework of Scientific Discovery through Advance Computing (SciDAC) award
{\it Computing the Properties of Matter with Leadership Computing Resources};
(iii) The Deutsche Forschungsgemeinschaft (DFG, German Research Foundation) - Project number 315477589-TRR 211;
(iv) The grant 05P2018 (ErUM-FSP T01) of the German Bundesministerium f\"ur Bildung und Forschung;
(v) The grant 283286 of the European Union.

This research used awards of computer time provided by:
(i) The INCITE program at Oak Ridge Leadership Computing Facility, a DOE
Office of Science User Facility operated under Contract No. DE-AC05-00OR22725;
(ii) The ALCC program at National Energy Research Scientific Computing Center,
a U.S. Department of Energy Office of Science User Facility operated under
Contract No. DE-AC02-05CH11231;
(iii) The INCITE program at Argonne Leadership Computing Facility, a U.S.
Department of Energy Office of Science User Facility operated under Contract
No. DE-AC02-06CH11357;
(iv) The USQCD resources at the Thomas Jefferson National Accelerator Facility.

This research also used computing resources made available through:
(i) a  PRACE grant at CINECA, Italy;
(ii) the Gauss Center at NIC-J\"ulich, Germany;
(iii) the GPU-cluster at Bielefeld University, Germany.

\appendix

\section{Parametrization of \boldmath$a/r_1$ and $af_K$}
\label{app:scale-setting}
In order to estimate systematic errors in our continuum
extrapolations we used two observables to set the scale
for the lattice spacing at finite values of the gauge
coupling $\beta\equiv 10/g^2$. We use the length scale
$r_1$ which is deduced from the short distance part
of the heavy quark potential and the kaon decay constant, which is obtained from fits to the long 
distance behavior of strange meson correlation functions. Both parametrizations have already been 
introduced and used by us in \cite{Bazavov:2014pvz}.

Using the two-loop beta-function of 3-flavor QCD,
\begin{equation*}
 f(\beta)=\left( \frac{10 b_0}{\beta} \right)^{-b_1/(2 b_0^2)} \exp(-\beta/(20 b_0))
\end{equation*}
with  $b_0=9/(16 \pi^2)$ and $b_1=1/(4 \pi^4)$ we 
parametrize $a/r_1$ and $af_K$ as 
\begin{eqnarray}
 a / r_1(\beta)&=&\frac{c_0f(\beta)+c_2(10/\beta)f^3(\beta)}{1+d_2(10/\beta)f^2(\beta)} \; ,
 \label{eq:r1parametrization} \\
 af_K(\beta)&=&\frac{c_0^Kf(\beta)+c_2^K(10/\beta)f^3(\beta)}{1+d_2^K(10/\beta)f^2(\beta)} \; .
 \label{eq:fKparametrization}
\end{eqnarray}
For $a/r_1$ we use 
\begin{eqnarray}
 c_0= 43.16 (15) \;&,& c_2= 339472(21144) \;\; ,\nonumber \\
 && d_2= 5452(387)\; .
\end{eqnarray}
This parametrization is consistent with that given
in \cite{Bazavov:2014pvz}. The parameters changed slightly which reflects our new bootstrap analysis of
the data for $a/r_1$. For $af_K$ we use
\begin{eqnarray}
 c_0^K= 7.486(25) \;&,& c_2^K= 41935(2247) \;\; ,\nonumber \\
&& d_2^K= 3273(224)\; .
\end{eqnarray}
This parametrization takes into account systematic
errors on the determination of the $af_K$ scale at
non-zero values of the lattices spacing, which arise
from the fact that the original data taken for $af_K$
are not directly taken on the LCP and needed to be 
corrected as discussed in appendix C.3 of \cite{Bazavov:2014pvz}. In the continuum limit
our parametrization gives for the relation of 
$r_1$ and $f_K$ in the continuum limit,
\begin{equation}
    r_1f_K = \frac{c_0^K}{c_0} = 0.1734(9) \; ,
\end{equation}
which for $r_1=0.3106$~fm gives for the kaon decay
constant the FLAG average value 
$f_K= 155.7/\sqrt{2}$~MeV
\cite{Aoki:2019cca}. This value is used in all our figures when showing temperature scales in physical units.

For our spline interpolations at non-zero lattice 
spacing, i.e. at fixed temporal lattice extent $N_\tau$
we add to the statistical error of our data, determined
at a certain value of the coupling $\beta$, a systematic
error arising from the width of the bootstrap band
at this $\beta$-value. This is used as an error on 
the $T$-scale (x-axis) when performing spline interpolations of our data at fixed $N_\tau$. 

\begin{figure}[ht]
\includegraphics[scale=0.60]{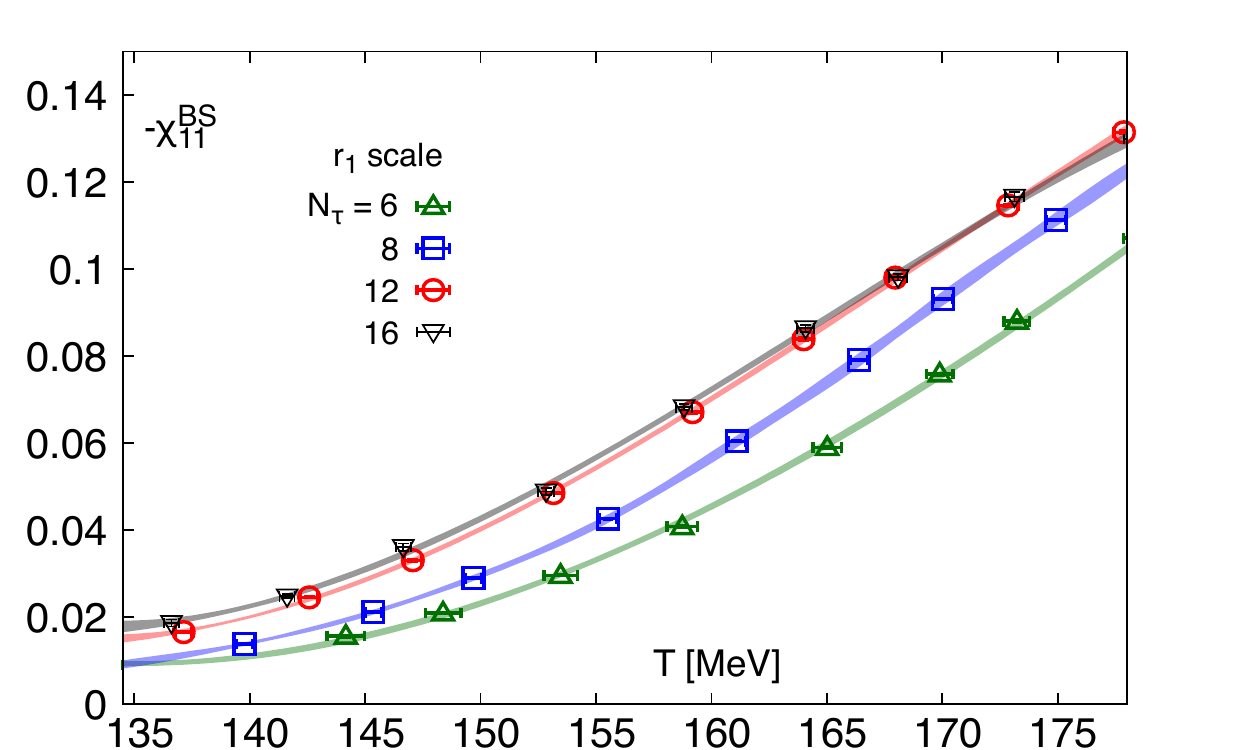}
\includegraphics[scale=0.60]{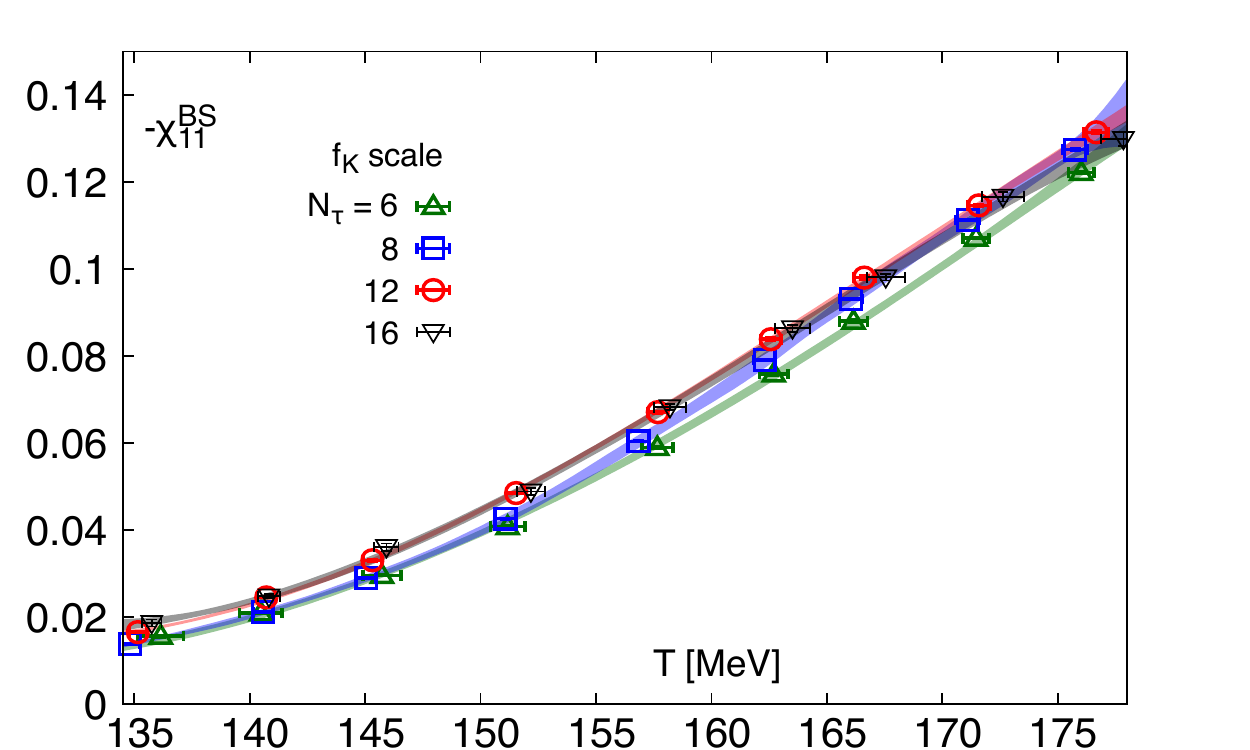}
\caption{Spline interpolations for the cumulant  $\chi_{11}^{BS}$ 
	performed with the temperature scales obtained from $r_1$ (top) and 
	$f_K$ (bottom) for $N_\tau = 6,\ 8,\ 12,\ 16$.
}
\label{fig:BS-data}
\end{figure}

\vspace{0.4cm}
\section{Fits to second order cumulants}
\label{app:fits}
\begin{figure*}[ht]
\includegraphics[scale=0.46]{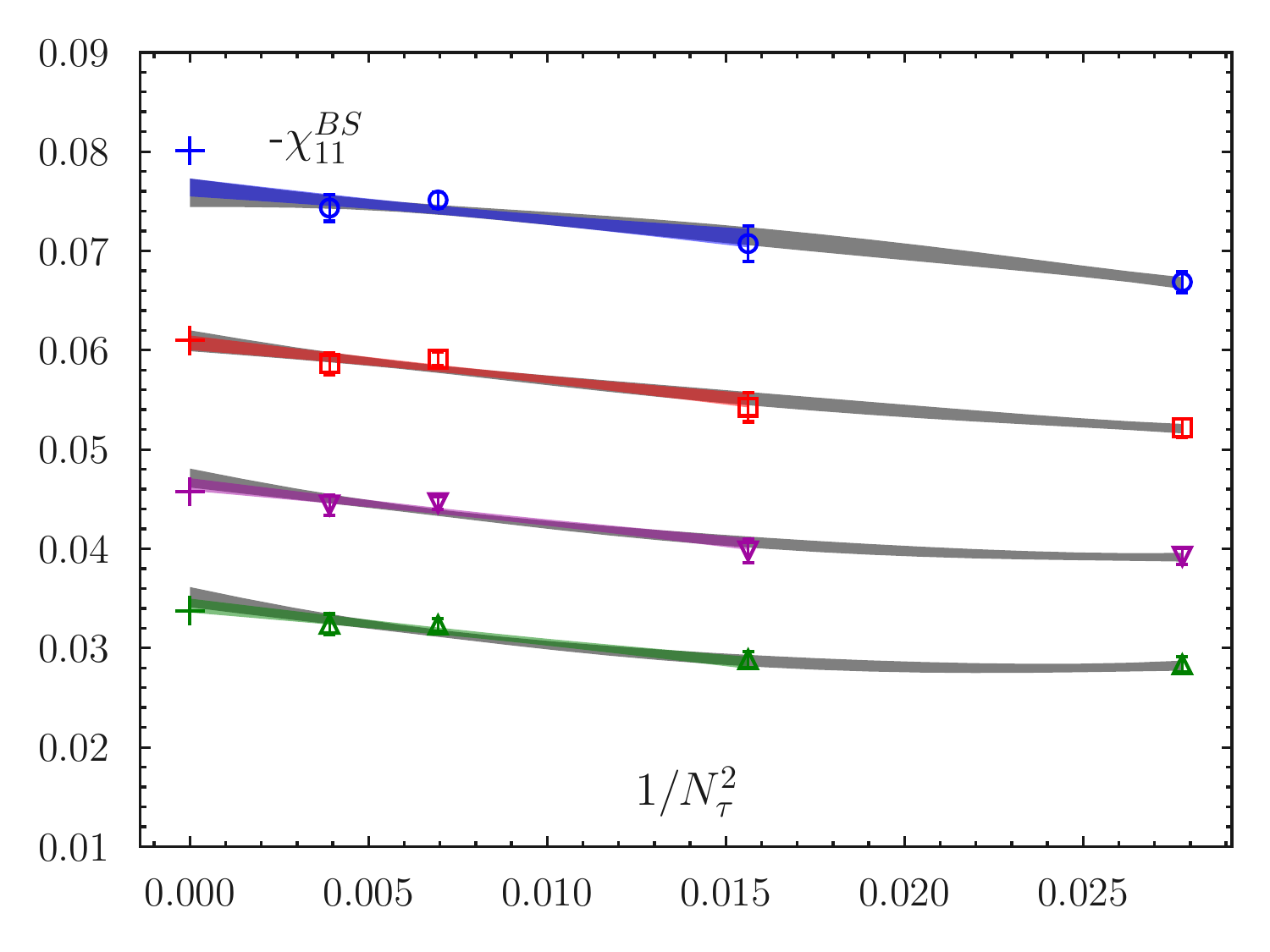}
\includegraphics[scale=0.46]{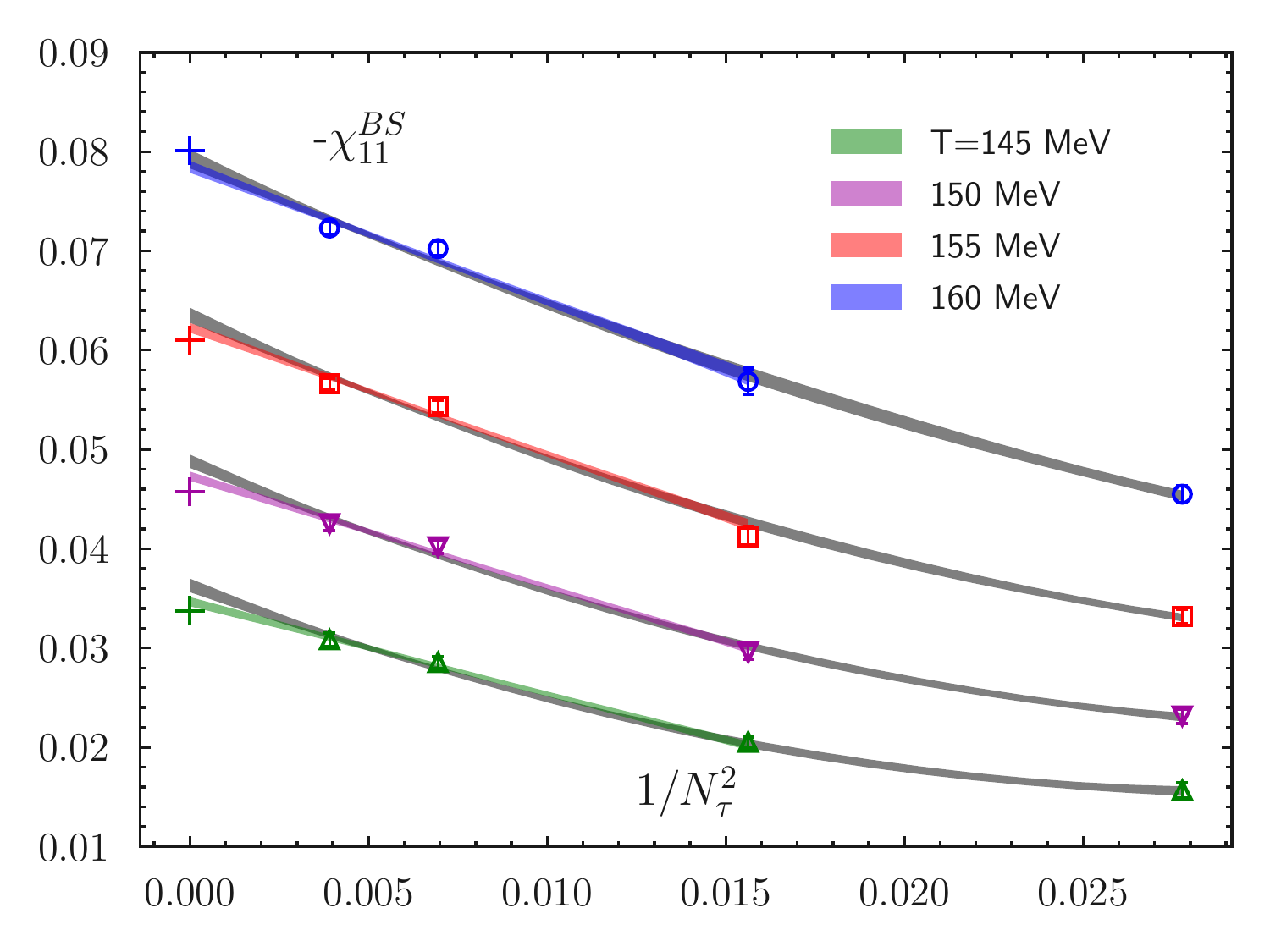}
\includegraphics[scale=0.46]{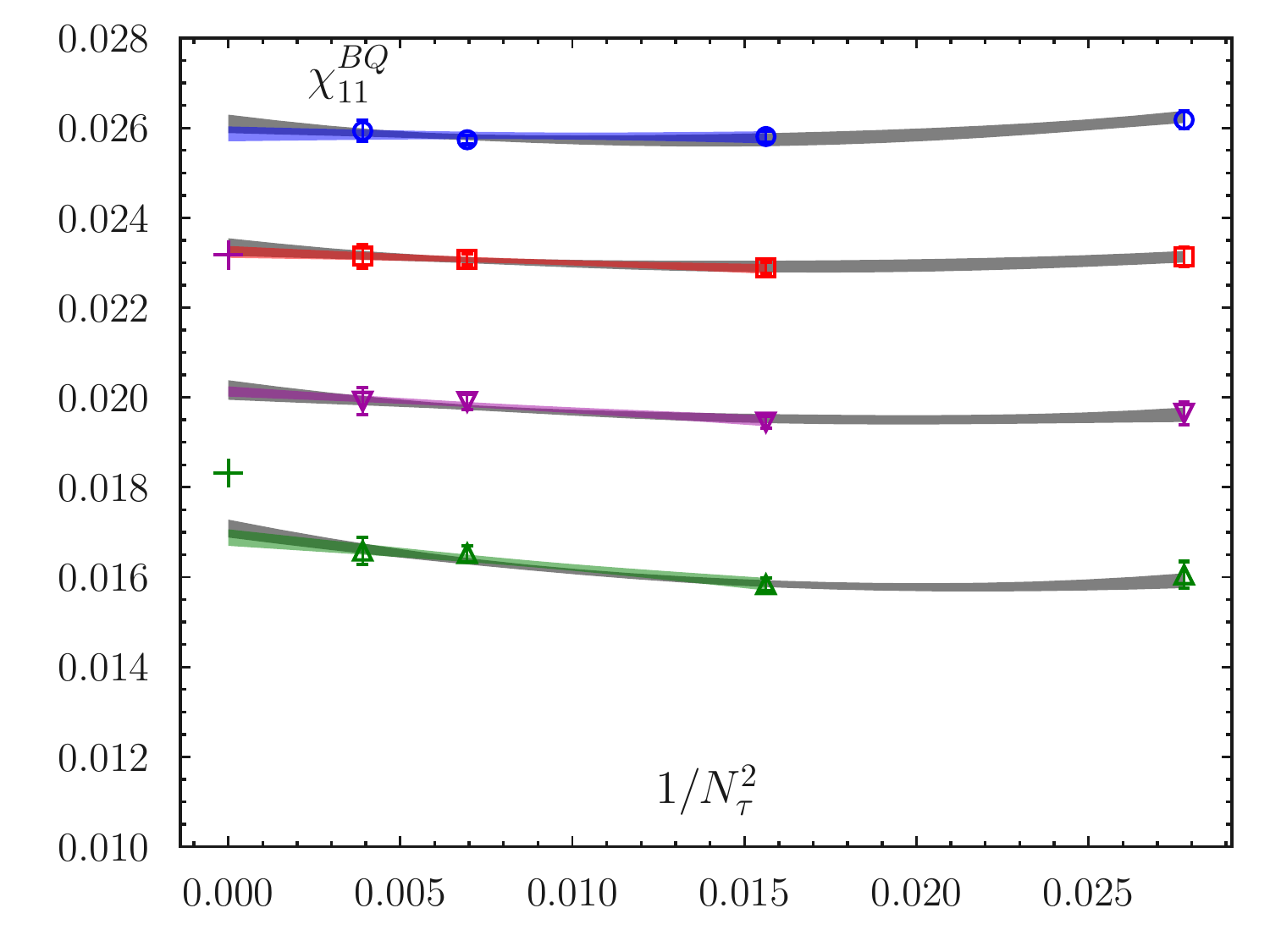}
\includegraphics[scale=0.46]{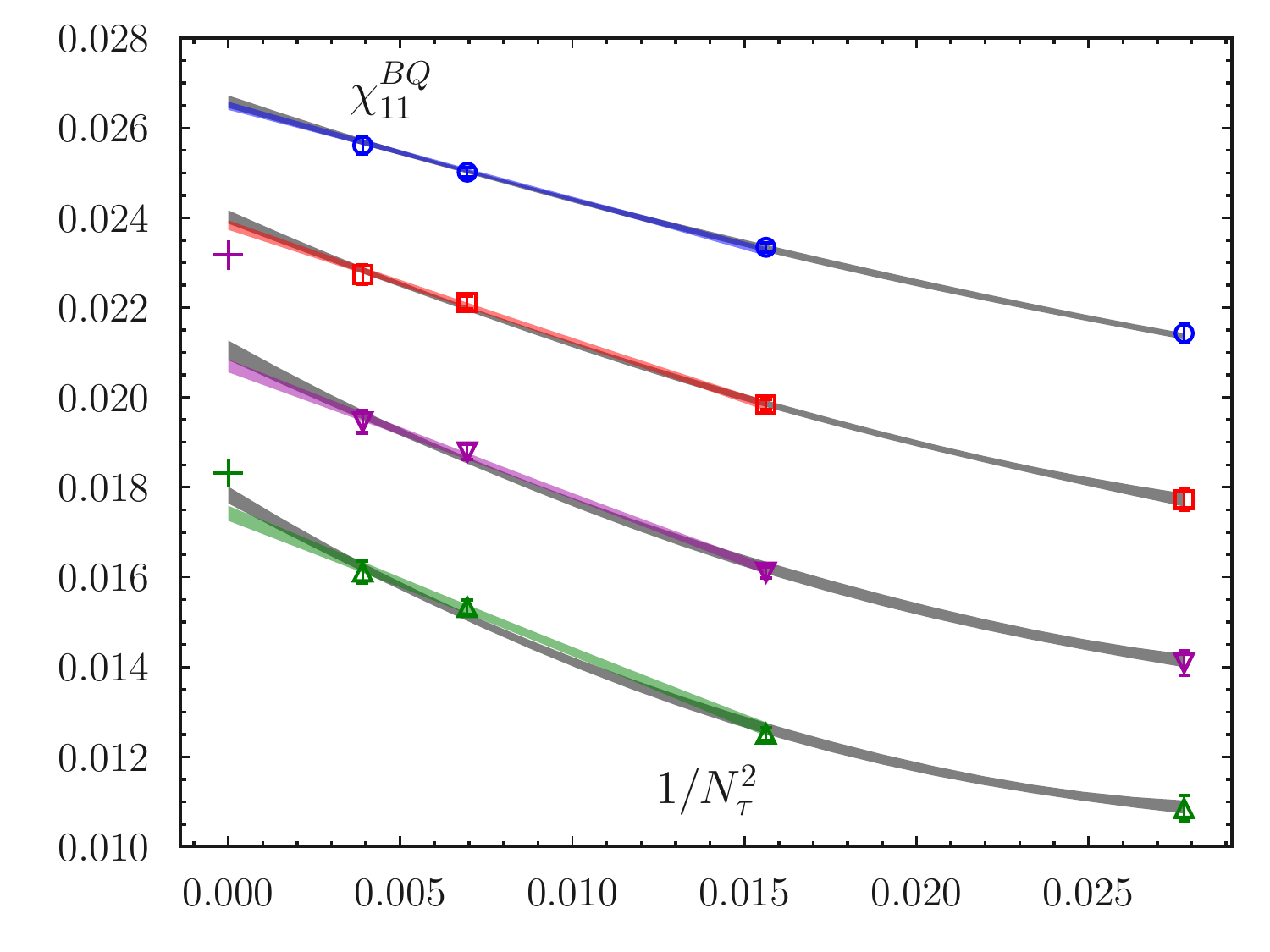}
\includegraphics[scale=0.46]{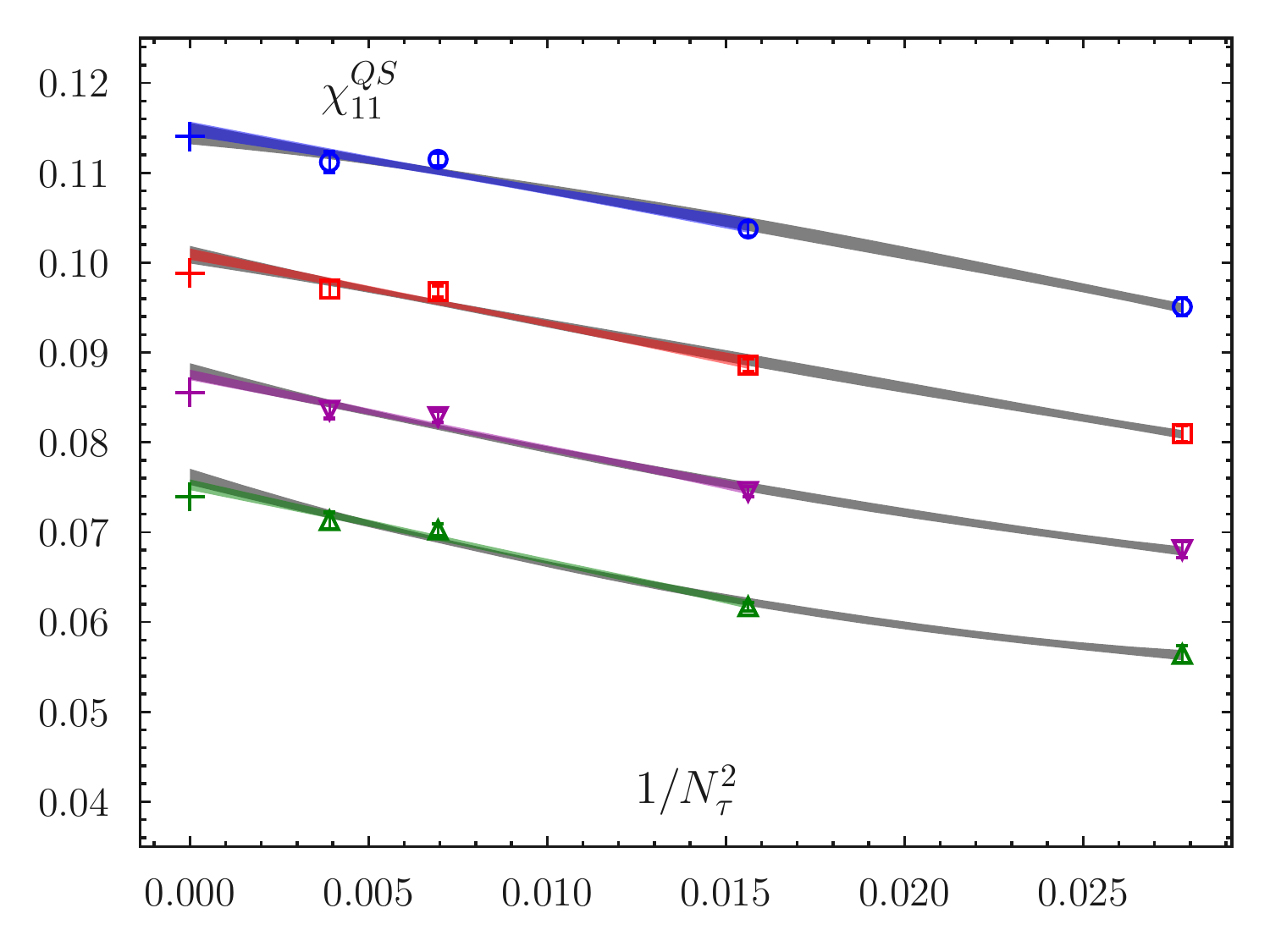}
\includegraphics[scale=0.46]{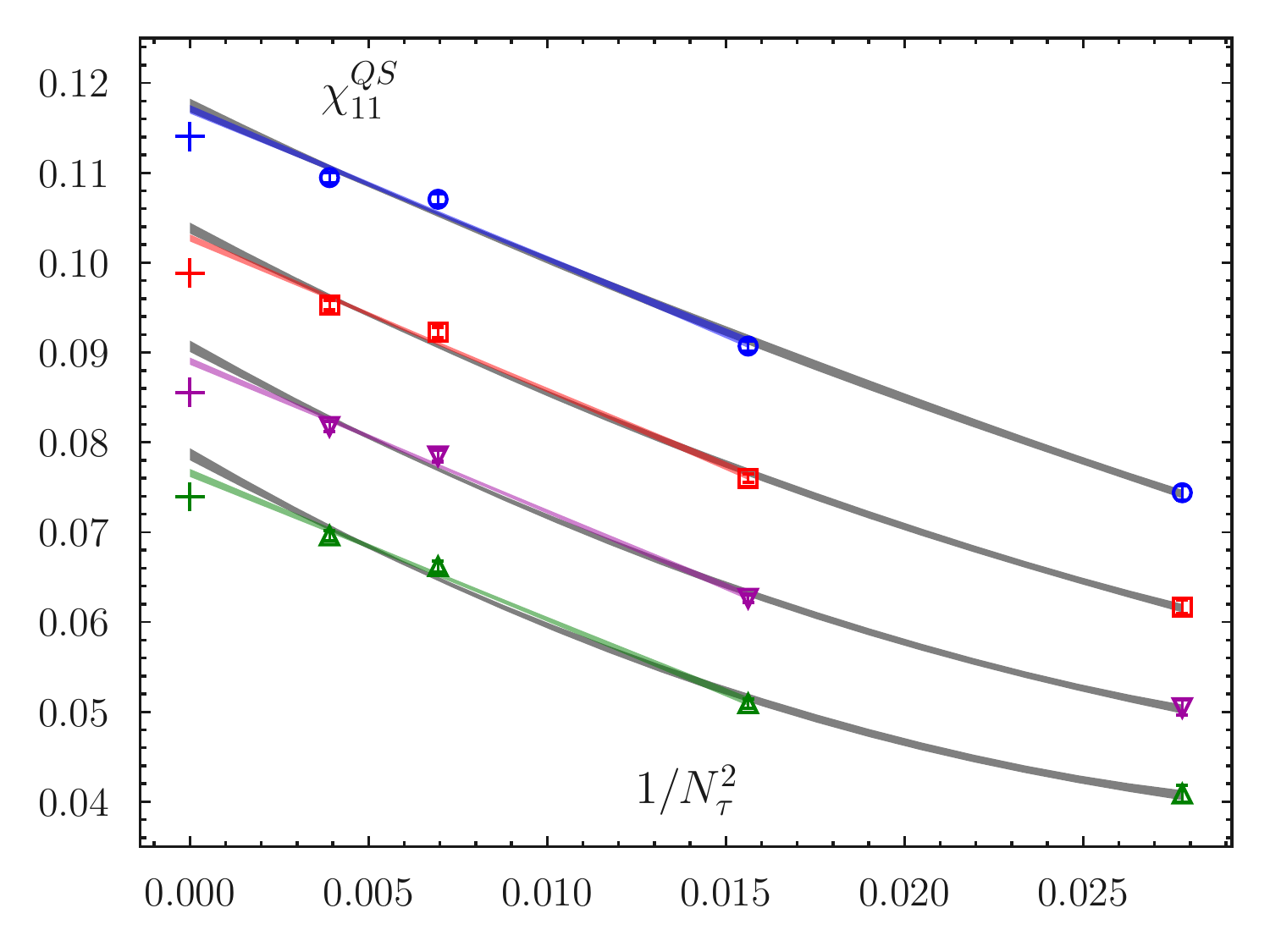}
\includegraphics[scale=0.46]{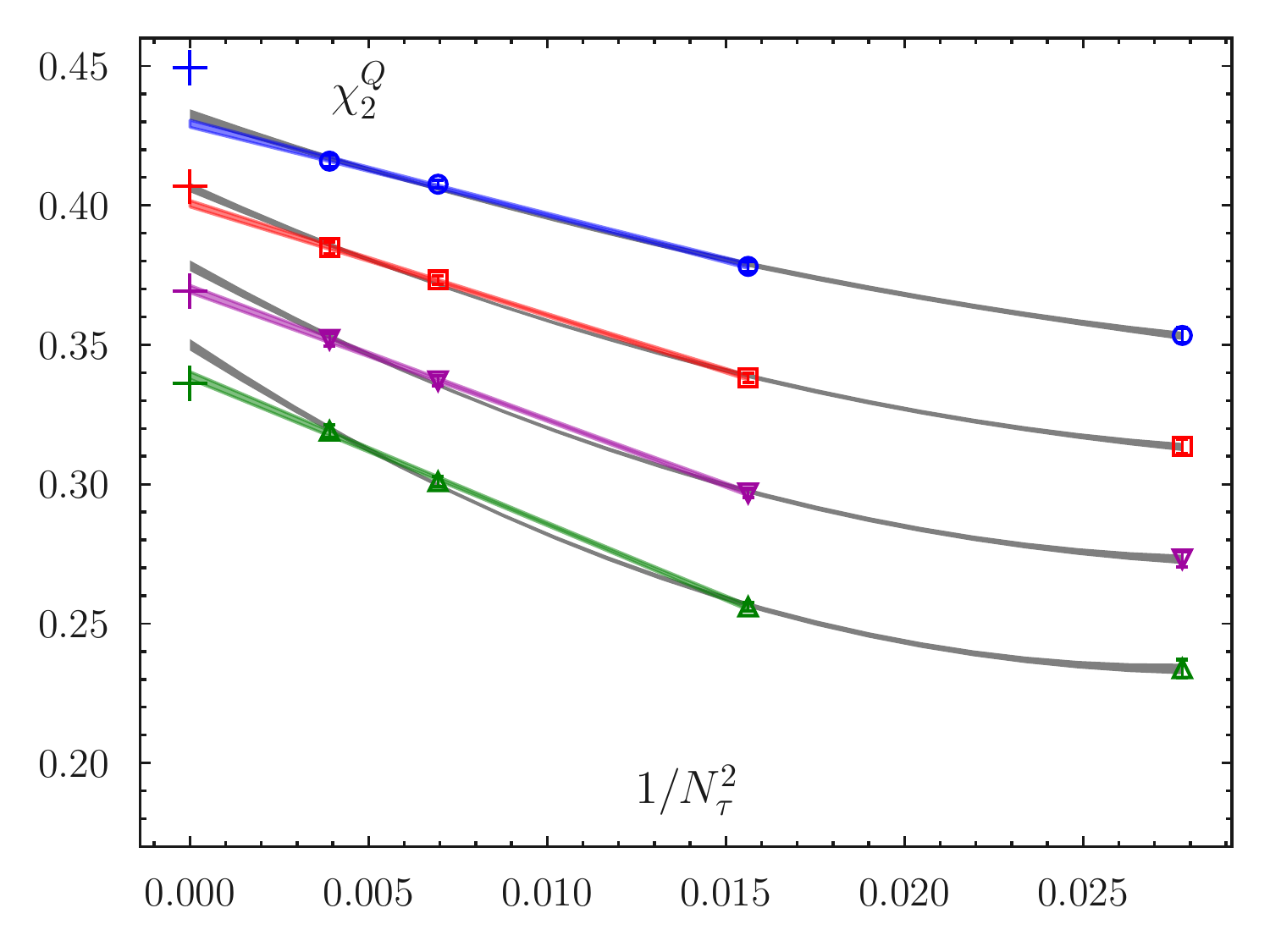}
\includegraphics[scale=0.46]{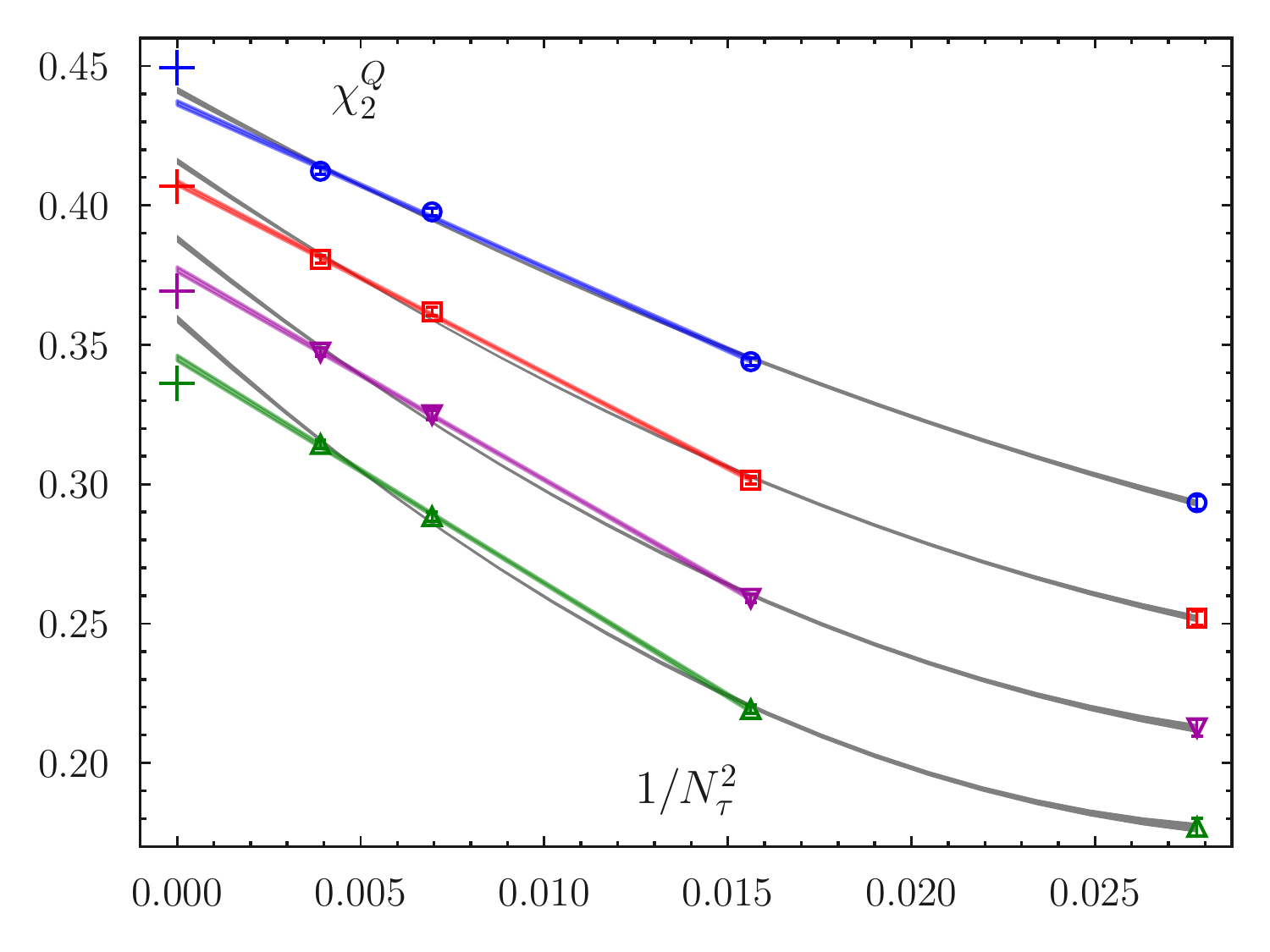}
\caption{Continuum extrapolations for the cumulants  $\chi_{11}^{BS}$, $\chi_{11}^{BQ}$, $\chi_{11}^{QS}$ and 
$\chi_{2}^{Q}$ (top to bottom)
        performed with the temperature scales obtained from $f_K$ (left) and $r_1$ (right) using linear
        and quadratic extrapolations in $1/N_\tau^2$. Shown are results for four values of the temperature below and close to the pseudo-critical temperature of $(2+1)$-flavor QCD. Also shown are results obtained in HRG model calculations using the QMHRG2020 list
        of hadrons (crosses). Note that for
        $\chi_{11}^{BQ}$ and $\chi_2^Q$ these 
        HRG results are not shown for all temperature values as the deviations from the corresponding QCD results are too large. For $\chi_2^Q$  QMHRG2020 with the finite-volume corrected contributions for pions and kaons in a volume
        $LT\equiv N_\sigma / N_\tau =4$ has been used.
}
\label{fig:BS-BQ-cont}
\end{figure*}

Using the temperature scales given in Appendix~\ref{app:scale-setting} we performed 
spline interpolations of our data for each value of $N_\tau$.
These interpolations have been done using 800 bootstrap samples on each of the four different lattice sizes.
These interpolations are shown in Fig.\ref{fig:BS-data} for the case of $\chi_{11}^{BS}$ using $a/r_1$ (left) and $af_K$ (right). Errors on the spline interpolation are obtained from a bootstrap analysis 
where each data point has its statistical error and a systematic error arising from the errors on $a/r_1$
and $af_K$, respectively.
Continuum extrapolations at fixed temperature are
then performed using data at each $N_\tau$ value with errors given by the error band of the spline interpolations. We performed fits linear and quadratic in $1/N_\tau^2$, as introduced in
Eq.~\ref{fit-linear} and Eq.~\ref{fit-quadratic}, respectively. has been changed to Fits to $\chi_{11}^{BS}$, $\chi_{11}^{BQ}$, 
$\chi_{11}^{QS}$ and $\chi_{2}^{Q}$ performed at some temperature values,  are shown in Fig.~\ref{fig:BS-BQ-cont}.  
In the case of quadratic fits we have used the data sets
for all $N_\tau$ and for linear extrapolations
only results for $N_\tau > 6$ have been used. As can be seen the quadratic 
fits generally have larger statistical errors and a $\chi^2/dof$ which is (5-10) times larger.
Linear extrapolations for the $N=\tau = 8,\ 12,\ 16$ data sets
generally yield $\chi^2/dof\simeq 1$. This also is the case for the 
electric charge fluctuations, $\chi_2^Q$, except for the two lowest
temperatures, $T=135$~MeV and $140$~MeV, where the $\chi^2/dof$ is about
2, but becomes even larger when using quadratic fits that include the 
$N_\tau=6$ data sets. We thus used results from the linear extrapolations 
in our final analysis.

\section{Comparison of QMHRG2020 and QMHRG2016+}
\label{HRG-comparison}

We list here eight strange baryon resonances 
that are treated differently in QMHRG2020 and QMHRG2016+ lists of hadrons. When eliminating these
states from the QMHRG2016+ list HRG model calculations
performed with the reduced QMHRG2016+ list agree with 
those performed with the QMHRG2020, although some 
values of the masses differ slightly as QMHRG2020
is based on the masses given in Ref.~\cite{Faustov:2015eba}, while QMHRG2016+ use
masses from Ref.~\cite{Capstick:1986bmxx}.

\begin{table}[ht]
\begin{tabular}{|c|c|c|}
\hline
Particle & PDG [MeV] &
Quark Model [MeV] \\
\hline
& \cite{Zyla:2020zbs} & \cite{Capstick:1986bmxx} \hspace{0.45cm} \cite{Faustov:2015eba} \\
\hline
$\Lambda$(1405) & 1405 & 1550 \;\; 1406  \\
$\Lambda$(1830) & 1830 & 1775 \;\; 1861  \\
$\Lambda$(2050) & 2050 & 2035 \;\; 2030  \\
$\Lambda$(2325) & 2325 & 2185 \;\; 2322  \\
$\Sigma$(1750) & 1750 & 1695  \;\; 1747  \\
$\Sigma$(1910) & 1910 & 1755  \;\; 1856  \\
(was $\Sigma$(1940)) & & \\
$\Sigma$(1940) & 1940 & 2010 \;\; 2025   \\
$\Sigma$(2070) & 2070 & 2030 \;\; 2062   \\
\hline
\end{tabular}
\caption{\label{tab:HW-Hot-resonances}
Strange baryon resonances that are treated differently in the QMHRG2020 and QMHRG2016+ lists. While only states listed by the PDG \cite{Zyla:2020zbs} are kept in the former, also the corresponding quark model states are kept in the latter.
}
\end{table}

\bibliography{bibliography}

\begin{thebibliography}{59}%
\makeatletter
\providecommand \@ifxundefined [1]{%
 \@ifx{#1\undefined}
}%
\providecommand \@ifnum [1]{%
 \ifnum #1\expandafter \@firstoftwo
 \else \expandafter \@secondoftwo
 \fi
}%
\providecommand \@ifx [1]{%
 \ifx #1\expandafter \@firstoftwo
 \else \expandafter \@secondoftwo
 \fi
}%
\providecommand \natexlab [1]{#1}%
\providecommand \enquote  [1]{``#1''}%
\providecommand \bibnamefont  [1]{#1}%
\providecommand \bibfnamefont [1]{#1}%
\providecommand \citenamefont [1]{#1}%
\providecommand \href@noop [0]{\@secondoftwo}%
\providecommand \href [0]{\begingroup \@sanitize@url \@href}%
\providecommand \@href[1]{\@@startlink{#1}\@@href}%
\providecommand \@@href[1]{\endgroup#1\@@endlink}%
\providecommand \@sanitize@url [0]{\catcode `\\12\catcode `\$12\catcode
  `\&12\catcode `\#12\catcode `\^12\catcode `\_12\catcode `\%12\relax}%
\providecommand \@@startlink[1]{}%
\providecommand \@@endlink[0]{}%
\providecommand \url  [0]{\begingroup\@sanitize@url \@url }%
\providecommand \@url [1]{\endgroup\@href {#1}{\urlprefix }}%
\providecommand \urlprefix  [0]{URL }%
\providecommand \Eprint [0]{\href }%
\providecommand \doibase [0]{http://dx.doi.org/}%
\providecommand \selectlanguage [0]{\@gobble}%
\providecommand \bibinfo  [0]{\@secondoftwo}%
\providecommand \bibfield  [0]{\@secondoftwo}%
\providecommand \translation [1]{[#1]}%
\providecommand \BibitemOpen [0]{}%
\providecommand \bibitemStop [0]{}%
\providecommand \bibitemNoStop [0]{.\EOS\space}%
\providecommand \EOS [0]{\spacefactor3000\relax}%
\providecommand \BibitemShut  [1]{\csname bibitem#1\endcsname}%
\let\auto@bib@innerbib\@empty
\bibitem [{\citenamefont {Bazavov}\ \emph
  {et~al.}(2019{\natexlab{a}})\citenamefont {Bazavov} \emph
  {et~al.}}]{Bazavov:2018mes}%
  \BibitemOpen
  \bibfield  {author} {\bibinfo {author} {\bibfnamefont {A.}~\bibnamefont
  {Bazavov}} \emph {et~al.} (\bibinfo {collaboration} {HotQCD}),\ }\href
  {\doibase 10.1016/j.physletb.2019.05.013} {\bibfield  {journal} {\bibinfo
  {journal} {Phys. Lett. B}\ }\textbf {\bibinfo {volume} {795}},\ \bibinfo
  {pages} {15} (\bibinfo {year} {2019}{\natexlab{a}})},\ \Eprint
  {http://arxiv.org/abs/1812.08235} {arXiv:1812.08235 [hep-lat]} \BibitemShut
  {NoStop}%
\bibitem [{\citenamefont {Borsanyi}\ \emph {et~al.}(2020)\citenamefont
  {Borsanyi}, \citenamefont {Fodor}, \citenamefont {Guenther}, \citenamefont
  {Kara}, \citenamefont {Katz}, \citenamefont {Parotto}, \citenamefont
  {Pasztor}, \citenamefont {Ratti},\ and\ \citenamefont
  {Szabo}}]{Borsanyi:2020fev}%
  \BibitemOpen
  \bibfield  {author} {\bibinfo {author} {\bibfnamefont {S.}~\bibnamefont
  {Borsanyi}}, \bibinfo {author} {\bibfnamefont {Z.}~\bibnamefont {Fodor}},
  \bibinfo {author} {\bibfnamefont {J.~N.}\ \bibnamefont {Guenther}}, \bibinfo
  {author} {\bibfnamefont {R.}~\bibnamefont {Kara}}, \bibinfo {author}
  {\bibfnamefont {S.~D.}\ \bibnamefont {Katz}}, \bibinfo {author}
  {\bibfnamefont {P.}~\bibnamefont {Parotto}}, \bibinfo {author} {\bibfnamefont
  {A.}~\bibnamefont {Pasztor}}, \bibinfo {author} {\bibfnamefont
  {C.}~\bibnamefont {Ratti}}, \ and\ \bibinfo {author} {\bibfnamefont {K.~K.}\
  \bibnamefont {Szabo}},\ }\href {\doibase 10.1103/PhysRevLett.125.052001}
  {\bibfield  {journal} {\bibinfo  {journal} {Phys. Rev. Lett.}\ }\textbf
  {\bibinfo {volume} {125}},\ \bibinfo {pages} {052001} (\bibinfo {year}
  {2020})},\ \Eprint {http://arxiv.org/abs/2002.02821} {arXiv:2002.02821
  [hep-lat]} \BibitemShut {NoStop}%
\bibitem [{\citenamefont {Bonati}\ \emph {et~al.}(2015)\citenamefont {Bonati},
  \citenamefont {D'Elia}, \citenamefont {Mariti}, \citenamefont {Mesiti},
  \citenamefont {Negro},\ and\ \citenamefont {Sanfilippo}}]{Bonati:2015bha}%
  \BibitemOpen
  \bibfield  {author} {\bibinfo {author} {\bibfnamefont {C.}~\bibnamefont
  {Bonati}}, \bibinfo {author} {\bibfnamefont {M.}~\bibnamefont {D'Elia}},
  \bibinfo {author} {\bibfnamefont {M.}~\bibnamefont {Mariti}}, \bibinfo
  {author} {\bibfnamefont {M.}~\bibnamefont {Mesiti}}, \bibinfo {author}
  {\bibfnamefont {F.}~\bibnamefont {Negro}}, \ and\ \bibinfo {author}
  {\bibfnamefont {F.}~\bibnamefont {Sanfilippo}},\ }\href {\doibase
  10.1103/PhysRevD.92.054503} {\bibfield  {journal} {\bibinfo  {journal} {Phys.
  Rev.}\ }\textbf {\bibinfo {volume} {D92}},\ \bibinfo {pages} {054503}
  (\bibinfo {year} {2015})},\ \Eprint {http://arxiv.org/abs/1507.03571}
  {arXiv:1507.03571 [hep-lat]} \BibitemShut {NoStop}%
\bibitem [{\citenamefont {Braun-Munzinger}\ \emph {et~al.}()\citenamefont
  {Braun-Munzinger}, \citenamefont {Redlich},\ and\ \citenamefont
  {Stachel}}]{BraunMunzinger:2003zd}%
  \BibitemOpen
  \bibfield  {author} {\bibinfo {author} {\bibfnamefont {P.}~\bibnamefont
  {Braun-Munzinger}}, \bibinfo {author} {\bibfnamefont {K.}~\bibnamefont
  {Redlich}}, \ and\ \bibinfo {author} {\bibfnamefont {J.}~\bibnamefont
  {Stachel}},\ }\bibfield  {booktitle} {\emph {\bibinfo {booktitle} {{Quark
  Gluon Plasma 3, 491 (2003)}}},\ }\href {\doibase 10.1142/9789812795533-0008}
  {\ 10.1142/9789812795533-0008},\ \Eprint
  {http://arxiv.org/abs/nucl-th/0304013} {arXiv:nucl-th/0304013} \BibitemShut
  {NoStop}%
\bibitem [{\citenamefont {Hagedorn}(1965)}]{Hagedorn:1965st}%
  \BibitemOpen
  \bibfield  {author} {\bibinfo {author} {\bibfnamefont {R.}~\bibnamefont
  {Hagedorn}},\ }\href@noop {} {\bibfield  {journal} {\bibinfo  {journal}
  {Nuovo Cim. Suppl.}\ }\textbf {\bibinfo {volume} {3}},\ \bibinfo {pages}
  {147} (\bibinfo {year} {1965})}\BibitemShut {NoStop}%
\bibitem [{\citenamefont {Karsch}\ and\ \citenamefont
  {Redlich}(2011)}]{Karsch:2010ck}%
  \BibitemOpen
  \bibfield  {author} {\bibinfo {author} {\bibfnamefont {F.}~\bibnamefont
  {Karsch}}\ and\ \bibinfo {author} {\bibfnamefont {K.}~\bibnamefont
  {Redlich}},\ }\href {\doibase 10.1016/j.physletb.2010.10.046} {\bibfield
  {journal} {\bibinfo  {journal} {Phys. Lett. B}\ }\textbf {\bibinfo {volume}
  {695}},\ \bibinfo {pages} {136} (\bibinfo {year} {2011})},\ \Eprint
  {http://arxiv.org/abs/1007.2581} {arXiv:1007.2581 [hep-ph]} \BibitemShut
  {NoStop}%
\bibitem [{\citenamefont {Andronic}\ \emph {et~al.}(2018)\citenamefont
  {Andronic}, \citenamefont {Braun-Munzinger}, \citenamefont {Redlich},\ and\
  \citenamefont {Stachel}}]{Andronic:2017pug}%
  \BibitemOpen
  \bibfield  {author} {\bibinfo {author} {\bibfnamefont {A.}~\bibnamefont
  {Andronic}}, \bibinfo {author} {\bibfnamefont {P.}~\bibnamefont
  {Braun-Munzinger}}, \bibinfo {author} {\bibfnamefont {K.}~\bibnamefont
  {Redlich}}, \ and\ \bibinfo {author} {\bibfnamefont {J.}~\bibnamefont
  {Stachel}},\ }\href {\doibase 10.1038/s41586-018-0491-6} {\bibfield
  {journal} {\bibinfo  {journal} {Nature}\ }\textbf {\bibinfo {volume} {561}},\
  \bibinfo {pages} {321} (\bibinfo {year} {2018})},\ \Eprint
  {http://arxiv.org/abs/1710.09425} {arXiv:1710.09425 [nucl-th]} \BibitemShut
  {NoStop}%
\bibitem [{\citenamefont {Dashen}\ \emph {et~al.}(1969)\citenamefont {Dashen},
  \citenamefont {Ma},\ and\ \citenamefont {Bernstein}}]{Dashen:1969ep}%
  \BibitemOpen
  \bibfield  {author} {\bibinfo {author} {\bibfnamefont {R.}~\bibnamefont
  {Dashen}}, \bibinfo {author} {\bibfnamefont {S.-K.}\ \bibnamefont {Ma}}, \
  and\ \bibinfo {author} {\bibfnamefont {H.~J.}\ \bibnamefont {Bernstein}},\
  }\href {\doibase 10.1103/PhysRev.187.345} {\bibfield  {journal} {\bibinfo
  {journal} {Phys. Rev.}\ }\textbf {\bibinfo {volume} {187}},\ \bibinfo {pages}
  {345} (\bibinfo {year} {1969})}\BibitemShut {NoStop}%
\bibitem [{\citenamefont {Venugopalan}\ and\ \citenamefont
  {Prakash}(1992)}]{Venugopalan:1992hy}%
  \BibitemOpen
  \bibfield  {author} {\bibinfo {author} {\bibfnamefont {R.}~\bibnamefont
  {Venugopalan}}\ and\ \bibinfo {author} {\bibfnamefont {M.}~\bibnamefont
  {Prakash}},\ }\href {\doibase 10.1016/0375-9474(92)90005-5} {\bibfield
  {journal} {\bibinfo  {journal} {Nucl. Phys. A}\ }\textbf {\bibinfo {volume}
  {546}},\ \bibinfo {pages} {718} (\bibinfo {year} {1992})}\BibitemShut
  {NoStop}%
\bibitem [{\citenamefont {Lo}\ \emph {et~al.}(2013)\citenamefont {Lo},
  \citenamefont {Friman}, \citenamefont {Kaczmarek}, \citenamefont {Redlich},\
  and\ \citenamefont {Sasaki}}]{lo_probing_2013}%
  \BibitemOpen
  \bibfield  {author} {\bibinfo {author} {\bibfnamefont {P.~M.}\ \bibnamefont
  {Lo}}, \bibinfo {author} {\bibfnamefont {B.}~\bibnamefont {Friman}}, \bibinfo
  {author} {\bibfnamefont {O.}~\bibnamefont {Kaczmarek}}, \bibinfo {author}
  {\bibfnamefont {K.}~\bibnamefont {Redlich}}, \ and\ \bibinfo {author}
  {\bibfnamefont {C.}~\bibnamefont {Sasaki}},\ }\href {\doibase
  10.1103/PhysRevD.88.014506} {\bibfield  {journal} {\bibinfo  {journal} {Phys.
  Rev. D}\ }\textbf {\bibinfo {volume} {88}},\ \bibinfo {pages} {014506}
  (\bibinfo {year} {2013})}\BibitemShut {NoStop}%
\bibitem [{\citenamefont {Fern\'andez-Ram\'\i{}rez}\ \emph
  {et~al.}(2018)\citenamefont {Fern\'andez-Ram\'\i{}rez}, \citenamefont {Lo},\
  and\ \citenamefont {Petreczky}}]{Fernandez-Ramirez:2018vzu}%
  \BibitemOpen
  \bibfield  {author} {\bibinfo {author} {\bibfnamefont {C.}~\bibnamefont
  {Fern\'andez-Ram\'\i{}rez}}, \bibinfo {author} {\bibfnamefont {P.~M.}\
  \bibnamefont {Lo}}, \ and\ \bibinfo {author} {\bibfnamefont {P.}~\bibnamefont
  {Petreczky}},\ }\href {\doibase 10.1103/PhysRevC.98.044910} {\bibfield
  {journal} {\bibinfo  {journal} {Phys. Rev. C}\ }\textbf {\bibinfo {volume}
  {98}},\ \bibinfo {pages} {044910} (\bibinfo {year} {2018})},\ \Eprint
  {http://arxiv.org/abs/1806.02177} {arXiv:1806.02177 [hep-ph]} \BibitemShut
  {NoStop}%
\bibitem [{\citenamefont {Andronic}\ \emph {et~al.}(2019)\citenamefont
  {Andronic}, \citenamefont {Braun-Munzinger}, \citenamefont {Friman},
  \citenamefont {Lo}, \citenamefont {Redlich},\ and\ \citenamefont
  {Stachel}}]{Andronic:2018qqt}%
  \BibitemOpen
  \bibfield  {author} {\bibinfo {author} {\bibfnamefont {A.}~\bibnamefont
  {Andronic}}, \bibinfo {author} {\bibfnamefont {P.}~\bibnamefont
  {Braun-Munzinger}}, \bibinfo {author} {\bibfnamefont {B.}~\bibnamefont
  {Friman}}, \bibinfo {author} {\bibfnamefont {P.~M.}\ \bibnamefont {Lo}},
  \bibinfo {author} {\bibfnamefont {K.}~\bibnamefont {Redlich}}, \ and\
  \bibinfo {author} {\bibfnamefont {J.}~\bibnamefont {Stachel}},\ }\href
  {\doibase 10.1016/j.physletb.2019.03.052} {\bibfield  {journal} {\bibinfo
  {journal} {Phys. Lett. B}\ }\textbf {\bibinfo {volume} {792}},\ \bibinfo
  {pages} {304} (\bibinfo {year} {2019})},\ \Eprint
  {http://arxiv.org/abs/1808.03102} {arXiv:1808.03102 [hep-ph]} \BibitemShut
  {NoStop}%
\bibitem [{\citenamefont {Hagedorn}\ and\ \citenamefont
  {Rafelski}(1980{\natexlab{a}})}]{Hagedorn:1980kb}%
  \BibitemOpen
  \bibfield  {author} {\bibinfo {author} {\bibfnamefont {R.}~\bibnamefont
  {Hagedorn}}\ and\ \bibinfo {author} {\bibfnamefont {J.}~\bibnamefont
  {Rafelski}},\ }\href {\doibase 10.1016/0370-2693(80)90566-3} {\bibfield
  {journal} {\bibinfo  {journal} {Phys. Lett. B}\ }\textbf {\bibinfo {volume}
  {97}},\ \bibinfo {pages} {136} (\bibinfo {year} {1980}{\natexlab{a}})},\
  \Eprint {http://arxiv.org/abs/CERN-TH-2922} {CERN-TH-2922} \BibitemShut
  {NoStop}%
\bibitem [{\citenamefont {Hagedorn}\ and\ \citenamefont
  {Rafelski}(1980{\natexlab{b}})}]{Hagedorn:1980cv}%
  \BibitemOpen
  \bibfield  {author} {\bibinfo {author} {\bibfnamefont {R.}~\bibnamefont
  {Hagedorn}}\ and\ \bibinfo {author} {\bibfnamefont {J.}~\bibnamefont
  {Rafelski}},\ }in\ \href@noop {} {\emph {\bibinfo {booktitle} {{International
  Conference on Nuclear Physics}}}}\ (\bibinfo {year} {1980})\ \Eprint
  {http://arxiv.org/abs/CERN-TH-2947, C80-08-24-1} {CERN-TH-2947, C80-08-24-1}
  \BibitemShut {NoStop}%
\bibitem [{\citenamefont {Gorenstein}\ \emph {et~al.}(1981)\citenamefont
  {Gorenstein}, \citenamefont {Petrov},\ and\ \citenamefont
  {Zinovev}}]{Gorenstein:1981fa}%
  \BibitemOpen
  \bibfield  {author} {\bibinfo {author} {\bibfnamefont {M.~I.}\ \bibnamefont
  {Gorenstein}}, \bibinfo {author} {\bibfnamefont {V.~K.}\ \bibnamefont
  {Petrov}}, \ and\ \bibinfo {author} {\bibfnamefont {G.~M.}\ \bibnamefont
  {Zinovev}},\ }\href {\doibase 10.1016/0370-2693(81)90546-3} {\bibfield
  {journal} {\bibinfo  {journal} {Phys. Lett. B}\ }\textbf {\bibinfo {volume}
  {106}},\ \bibinfo {pages} {327} (\bibinfo {year} {1981})}\BibitemShut
  {NoStop}%
\bibitem [{\citenamefont {Vovchenko}\ \emph
  {et~al.}(2017{\natexlab{a}})\citenamefont {Vovchenko}, \citenamefont
  {Gorenstein},\ and\ \citenamefont {Stoecker}}]{Vovchenko:2016rkn}%
  \BibitemOpen
  \bibfield  {author} {\bibinfo {author} {\bibfnamefont {V.}~\bibnamefont
  {Vovchenko}}, \bibinfo {author} {\bibfnamefont {M.~I.}\ \bibnamefont
  {Gorenstein}}, \ and\ \bibinfo {author} {\bibfnamefont {H.}~\bibnamefont
  {Stoecker}},\ }\href {\doibase 10.1103/PhysRevLett.118.182301} {\bibfield
  {journal} {\bibinfo  {journal} {Phys. Rev. Lett.}\ }\textbf {\bibinfo
  {volume} {118}},\ \bibinfo {pages} {182301} (\bibinfo {year}
  {2017}{\natexlab{a}})},\ \Eprint {http://arxiv.org/abs/1609.03975}
  {arXiv:1609.03975 [hep-ph]} \BibitemShut {NoStop}%
\bibitem [{\citenamefont {Vovchenko}\ \emph
  {et~al.}(2017{\natexlab{b}})\citenamefont {Vovchenko}, \citenamefont
  {Pasztor}, \citenamefont {Fodor}, \citenamefont {Katz},\ and\ \citenamefont
  {Stoecker}}]{Vovchenko:2017xad}%
  \BibitemOpen
  \bibfield  {author} {\bibinfo {author} {\bibfnamefont {V.}~\bibnamefont
  {Vovchenko}}, \bibinfo {author} {\bibfnamefont {A.}~\bibnamefont {Pasztor}},
  \bibinfo {author} {\bibfnamefont {Z.}~\bibnamefont {Fodor}}, \bibinfo
  {author} {\bibfnamefont {S.~D.}\ \bibnamefont {Katz}}, \ and\ \bibinfo
  {author} {\bibfnamefont {H.}~\bibnamefont {Stoecker}},\ }\href {\doibase
  10.1016/j.physletb.2017.10.042} {\bibfield  {journal} {\bibinfo  {journal}
  {Phys. Lett. B}\ }\textbf {\bibinfo {volume} {775}},\ \bibinfo {pages} {71}
  (\bibinfo {year} {2017}{\natexlab{b}})},\ \Eprint
  {http://arxiv.org/abs/1708.02852} {arXiv:1708.02852 [hep-ph]} \BibitemShut
  {NoStop}%
\bibitem [{\citenamefont {Taradiy}\ \emph {et~al.}(2019)\citenamefont
  {Taradiy}, \citenamefont {Motornenko}, \citenamefont {Vovchenko},
  \citenamefont {Gorenstein},\ and\ \citenamefont
  {Stoecker}}]{Taradiy:2019taz}%
  \BibitemOpen
  \bibfield  {author} {\bibinfo {author} {\bibfnamefont {K.}~\bibnamefont
  {Taradiy}}, \bibinfo {author} {\bibfnamefont {A.}~\bibnamefont {Motornenko}},
  \bibinfo {author} {\bibfnamefont {V.}~\bibnamefont {Vovchenko}}, \bibinfo
  {author} {\bibfnamefont {M.~I.}\ \bibnamefont {Gorenstein}}, \ and\ \bibinfo
  {author} {\bibfnamefont {H.}~\bibnamefont {Stoecker}},\ }\href {\doibase
  10.1103/PhysRevC.100.065202} {\bibfield  {journal} {\bibinfo  {journal}
  {Phys. Rev. C}\ }\textbf {\bibinfo {volume} {100}},\ \bibinfo {pages}
  {065202} (\bibinfo {year} {2019})},\ \Eprint
  {http://arxiv.org/abs/1904.08259} {arXiv:1904.08259 [hep-ph]} \BibitemShut
  {NoStop}%
\bibitem [{\citenamefont {Olive}(1981)}]{Olive:1980dy}%
  \BibitemOpen
  \bibfield  {author} {\bibinfo {author} {\bibfnamefont {K.~A.}\ \bibnamefont
  {Olive}},\ }\href {\doibase 10.1016/0550-3213(81)90444-2} {\bibfield
  {journal} {\bibinfo  {journal} {Nucl. Phys. B}\ }\textbf {\bibinfo {volume}
  {190}},\ \bibinfo {pages} {483} (\bibinfo {year} {1981})}\BibitemShut
  {NoStop}%
\bibitem [{\citenamefont {Olive}(1982)}]{Olive:1982we}%
  \BibitemOpen
  \bibfield  {author} {\bibinfo {author} {\bibfnamefont {K.~A.}\ \bibnamefont
  {Olive}},\ }\href {\doibase 10.1016/0550-3213(82)90335-2} {\bibfield
  {journal} {\bibinfo  {journal} {Nucl. Phys. B}\ }\textbf {\bibinfo {volume}
  {198}},\ \bibinfo {pages} {461} (\bibinfo {year} {1982})}\BibitemShut
  {NoStop}%
\bibitem [{\citenamefont {Huovinen}\ and\ \citenamefont
  {Petreczky}(2018)}]{Huovinen:2017ogf}%
  \BibitemOpen
  \bibfield  {author} {\bibinfo {author} {\bibfnamefont {P.}~\bibnamefont
  {Huovinen}}\ and\ \bibinfo {author} {\bibfnamefont {P.}~\bibnamefont
  {Petreczky}},\ }\href {\doibase 10.1016/j.physletb.2017.12.001} {\bibfield
  {journal} {\bibinfo  {journal} {Phys. Lett. B}\ }\textbf {\bibinfo {volume}
  {777}},\ \bibinfo {pages} {125} (\bibinfo {year} {2018})},\ \Eprint
  {http://arxiv.org/abs/1708.00879} {arXiv:1708.00879 [hep-ph]} \BibitemShut
  {NoStop}%
\bibitem [{\citenamefont {Goswami}\ \emph
  {et~al.}(2021{\natexlab{a}})\citenamefont {Goswami}, \citenamefont {Karsch},
  \citenamefont {Schmidt}, \citenamefont {Mukherjee},\ and\ \citenamefont
  {Petreczky}}]{Goswami:2020yez}%
  \BibitemOpen
  \bibfield  {author} {\bibinfo {author} {\bibfnamefont {J.}~\bibnamefont
  {Goswami}}, \bibinfo {author} {\bibfnamefont {F.}~\bibnamefont {Karsch}},
  \bibinfo {author} {\bibfnamefont {C.}~\bibnamefont {Schmidt}}, \bibinfo
  {author} {\bibfnamefont {S.}~\bibnamefont {Mukherjee}}, \ and\ \bibinfo
  {author} {\bibfnamefont {P.}~\bibnamefont {Petreczky}},\ }\href@noop {}
  {\bibfield  {journal} {\bibinfo  {journal} {Acta Phys. Polon. Supp.}\
  }\textbf {\bibinfo {volume} {14}},\ \bibinfo {pages} {251} (\bibinfo {year}
  {2021}{\natexlab{a}})},\ \Eprint {http://arxiv.org/abs/2011.02812}
  {arXiv:2011.02812 [hep-lat]} \BibitemShut {NoStop}%
\bibitem [{\citenamefont {Goswami}\ \emph
  {et~al.}(2021{\natexlab{b}})\citenamefont {Goswami}, \citenamefont {Karsch},
  \citenamefont {Mukherjee}, \citenamefont {Petreczky},\ and\ \citenamefont
  {Schmidt}}]{Goswami:2021opr}%
  \BibitemOpen
  \bibfield  {author} {\bibinfo {author} {\bibfnamefont {J.}~\bibnamefont
  {Goswami}}, \bibinfo {author} {\bibfnamefont {F.}~\bibnamefont {Karsch}},
  \bibinfo {author} {\bibfnamefont {S.}~\bibnamefont {Mukherjee}}, \bibinfo
  {author} {\bibfnamefont {P.}~\bibnamefont {Petreczky}}, \ and\ \bibinfo
  {author} {\bibfnamefont {C.}~\bibnamefont {Schmidt}},\ }\href@noop {} {\
  (\bibinfo {year} {2021}{\natexlab{b}})},\ \Eprint
  {http://arxiv.org/abs/2109.00268} {arXiv:2109.00268 [hep-lat]} \BibitemShut
  {NoStop}%
\bibitem [{\citenamefont {Karthein}\ \emph {et~al.}(2021)\citenamefont
  {Karthein}, \citenamefont {Koch}, \citenamefont {Ratti},\ and\ \citenamefont
  {Vovchenko}}]{Karthein:2021cmb}%
  \BibitemOpen
  \bibfield  {author} {\bibinfo {author} {\bibfnamefont {J.~M.}\ \bibnamefont
  {Karthein}}, \bibinfo {author} {\bibfnamefont {V.}~\bibnamefont {Koch}},
  \bibinfo {author} {\bibfnamefont {C.}~\bibnamefont {Ratti}}, \ and\ \bibinfo
  {author} {\bibfnamefont {V.}~\bibnamefont {Vovchenko}},\ }\href@noop {} {\
  (\bibinfo {year} {2021})},\ \Eprint {http://arxiv.org/abs/2107.00588}
  {arXiv:2107.00588 [nucl-th]} \BibitemShut {NoStop}%
\bibitem [{\citenamefont {Follana}\ \emph {et~al.}(2007)\citenamefont
  {Follana}, \citenamefont {Mason}, \citenamefont {Davies}, \citenamefont
  {Hornbostel}, \citenamefont {Lepage}, \citenamefont {Shigemitsu},
  \citenamefont {Trottier},\ and\ \citenamefont {Wong}}]{Follana:2006rc}%
  \BibitemOpen
  \bibfield  {author} {\bibinfo {author} {\bibfnamefont {E.}~\bibnamefont
  {Follana}}, \bibinfo {author} {\bibfnamefont {Q.}~\bibnamefont {Mason}},
  \bibinfo {author} {\bibfnamefont {C.}~\bibnamefont {Davies}}, \bibinfo
  {author} {\bibfnamefont {K.}~\bibnamefont {Hornbostel}}, \bibinfo {author}
  {\bibfnamefont {G.}~\bibnamefont {Lepage}}, \bibinfo {author} {\bibfnamefont
  {J.}~\bibnamefont {Shigemitsu}}, \bibinfo {author} {\bibfnamefont
  {H.}~\bibnamefont {Trottier}}, \ and\ \bibinfo {author} {\bibfnamefont
  {K.}~\bibnamefont {Wong}} (\bibinfo {collaboration} {HPQCD, UKQCD}),\ }\href
  {\doibase 10.1103/PhysRevD.75.054502} {\bibfield  {journal} {\bibinfo
  {journal} {Phys. Rev. D}\ }\textbf {\bibinfo {volume} {75}},\ \bibinfo
  {pages} {054502} (\bibinfo {year} {2007})},\ \Eprint
  {http://arxiv.org/abs/hep-lat/0610092} {arXiv:hep-lat/0610092} \BibitemShut
  {NoStop}%
\bibitem [{\citenamefont {Bazavov}\ \emph {et~al.}(2020)\citenamefont {Bazavov}
  \emph {et~al.}}]{Bazavov:2020bjn}%
  \BibitemOpen
  \bibfield  {author} {\bibinfo {author} {\bibfnamefont {A.}~\bibnamefont
  {Bazavov}} \emph {et~al.},\ }\href {\doibase 10.1103/PhysRevD.101.074502}
  {\bibfield  {journal} {\bibinfo  {journal} {Phys. Rev. D}\ }\textbf {\bibinfo
  {volume} {101}},\ \bibinfo {pages} {074502} (\bibinfo {year} {2020})},\
  \Eprint {http://arxiv.org/abs/2001.08530} {arXiv:2001.08530 [hep-lat]}
  \BibitemShut {NoStop}%
\bibitem [{\citenamefont {Bazavov}\ \emph
  {et~al.}(2017{\natexlab{a}})\citenamefont {Bazavov} \emph
  {et~al.}}]{Bazavov:2017dus}%
  \BibitemOpen
  \bibfield  {author} {\bibinfo {author} {\bibfnamefont {A.}~\bibnamefont
  {Bazavov}} \emph {et~al.},\ }\href {\doibase 10.1103/PhysRevD.95.054504}
  {\bibfield  {journal} {\bibinfo  {journal} {Phys. Rev. D}\ }\textbf {\bibinfo
  {volume} {95}},\ \bibinfo {pages} {054504} (\bibinfo {year}
  {2017}{\natexlab{a}})},\ \Eprint {http://arxiv.org/abs/1701.04325}
  {arXiv:1701.04325 [hep-lat]} \BibitemShut {NoStop}%
\bibitem [{\citenamefont {Bazavov}\ \emph
  {et~al.}(2014{\natexlab{a}})\citenamefont {Bazavov} \emph
  {et~al.}}]{Bazavov:2014pvz}%
  \BibitemOpen
  \bibfield  {author} {\bibinfo {author} {\bibfnamefont {A.}~\bibnamefont
  {Bazavov}} \emph {et~al.} (\bibinfo {collaboration} {HotQCD}),\ }\href
  {\doibase 10.1103/PhysRevD.90.094503} {\bibfield  {journal} {\bibinfo
  {journal} {Phys.\ Rev.\ D}\ }\textbf {\bibinfo {volume} {90}},\ \bibinfo
  {pages} {094503} (\bibinfo {year} {2014}{\natexlab{a}})},\ \Eprint
  {http://arxiv.org/abs/1407.6387} {arXiv:1407.6387 [hep-lat]} \BibitemShut
  {NoStop}%
\bibitem [{\citenamefont {Bazavov}\ \emph
  {et~al.}(2019{\natexlab{b}})\citenamefont {Bazavov} \emph
  {et~al.}}]{Bazavov:2019www}%
  \BibitemOpen
  \bibfield  {author} {\bibinfo {author} {\bibfnamefont {A.}~\bibnamefont
  {Bazavov}} \emph {et~al.},\ }\href {\doibase 10.1103/PhysRevD.100.094510}
  {\bibfield  {journal} {\bibinfo  {journal} {Phys. Rev. D}\ }\textbf {\bibinfo
  {volume} {100}},\ \bibinfo {pages} {094510} (\bibinfo {year}
  {2019}{\natexlab{b}})},\ \Eprint {http://arxiv.org/abs/1908.09552}
  {arXiv:1908.09552 [hep-lat]} \BibitemShut {NoStop}%
\bibitem [{\citenamefont {Bazavov}\ \emph
  {et~al.}(2012{\natexlab{a}})\citenamefont {Bazavov} \emph
  {et~al.}}]{Bazavov:2012vg}%
  \BibitemOpen
  \bibfield  {author} {\bibinfo {author} {\bibfnamefont {A.}~\bibnamefont
  {Bazavov}} \emph {et~al.},\ }\href {\doibase 10.1103/PhysRevLett.109.192302}
  {\bibfield  {journal} {\bibinfo  {journal} {Phys. Rev. Lett.}\ }\textbf
  {\bibinfo {volume} {109}},\ \bibinfo {pages} {192302} (\bibinfo {year}
  {2012}{\natexlab{a}})},\ \Eprint {http://arxiv.org/abs/1208.1220}
  {arXiv:1208.1220 [hep-lat]} \BibitemShut {NoStop}%
\bibitem [{\citenamefont {Bazavov}\ \emph
  {et~al.}(2012{\natexlab{b}})\citenamefont {Bazavov} \emph
  {et~al.}}]{Bazavov:2012jq}%
  \BibitemOpen
  \bibfield  {author} {\bibinfo {author} {\bibfnamefont {A.}~\bibnamefont
  {Bazavov}} \emph {et~al.} (\bibinfo {collaboration} {HotQCD}),\ }\href
  {\doibase 10.1103/PhysRevD.86.034509} {\bibfield  {journal} {\bibinfo
  {journal} {Phys. Rev. D}\ }\textbf {\bibinfo {volume} {86}},\ \bibinfo
  {pages} {034509} (\bibinfo {year} {2012}{\natexlab{b}})},\ \Eprint
  {http://arxiv.org/abs/1203.0784} {arXiv:1203.0784 [hep-lat]} \BibitemShut
  {NoStop}%
\bibitem [{\citenamefont {Bellwied}\ \emph {et~al.}(2020)\citenamefont
  {Bellwied}, \citenamefont {Borsanyi}, \citenamefont {Fodor}, \citenamefont
  {Guenther}, \citenamefont {Noronha-Hostler}, \citenamefont {Parotto},
  \citenamefont {Pasztor}, \citenamefont {Ratti},\ and\ \citenamefont
  {Stafford}}]{Bellwied:2019pxh}%
  \BibitemOpen
  \bibfield  {author} {\bibinfo {author} {\bibfnamefont {R.}~\bibnamefont
  {Bellwied}}, \bibinfo {author} {\bibfnamefont {S.}~\bibnamefont {Borsanyi}},
  \bibinfo {author} {\bibfnamefont {Z.}~\bibnamefont {Fodor}}, \bibinfo
  {author} {\bibfnamefont {J.~N.}\ \bibnamefont {Guenther}}, \bibinfo {author}
  {\bibfnamefont {J.}~\bibnamefont {Noronha-Hostler}}, \bibinfo {author}
  {\bibfnamefont {P.}~\bibnamefont {Parotto}}, \bibinfo {author} {\bibfnamefont
  {A.}~\bibnamefont {Pasztor}}, \bibinfo {author} {\bibfnamefont
  {C.}~\bibnamefont {Ratti}}, \ and\ \bibinfo {author} {\bibfnamefont {J.~M.}\
  \bibnamefont {Stafford}},\ }\href {\doibase 10.1103/PhysRevD.101.034506}
  {\bibfield  {journal} {\bibinfo  {journal} {Phys. Rev. D}\ }\textbf {\bibinfo
  {volume} {101}},\ \bibinfo {pages} {034506} (\bibinfo {year} {2020})},\
  \Eprint {http://arxiv.org/abs/1910.14592} {arXiv:1910.14592 [hep-lat]}
  \BibitemShut {NoStop}%
\bibitem [{\citenamefont {Zyla}\ \emph {et~al.}(2020)\citenamefont {Zyla} \emph
  {et~al.}}]{Zyla:2020zbs}%
  \BibitemOpen
  \bibfield  {author} {\bibinfo {author} {\bibfnamefont {P.}~\bibnamefont
  {Zyla}} \emph {et~al.} (\bibinfo {collaboration} {Particle Data Group}),\
  }\href {\doibase 10.1093/ptep/ptaa104} {\bibfield  {journal} {\bibinfo
  {journal} {PTEP}\ }\textbf {\bibinfo {volume} {2020}},\ \bibinfo {pages}
  {083C01} (\bibinfo {year} {2020})}\BibitemShut {NoStop}%
\bibitem [{\citenamefont {Capstick}\ and\ \citenamefont
  {Isgur}(1986)}]{Capstick:1986bmxx}%
  \BibitemOpen
  \bibfield  {author} {\bibinfo {author} {\bibfnamefont {S.}~\bibnamefont
  {Capstick}}\ and\ \bibinfo {author} {\bibfnamefont {N.}~\bibnamefont
  {Isgur}},\ }\href {\doibase 10.1103/PhysRevD.34.2809} {\bibfield  {journal}
  {\bibinfo  {journal} {Phys. Rev. D}\ }\textbf {\bibinfo {volume} {34}},\
  \bibinfo {pages} {2809} (\bibinfo {year} {1986})}\BibitemShut {NoStop}%
\bibitem [{\citenamefont {Ebert}\ \emph {et~al.}(2009)\citenamefont {Ebert},
  \citenamefont {Faustov},\ and\ \citenamefont {Galkin}}]{Ebert:2009ub}%
  \BibitemOpen
  \bibfield  {author} {\bibinfo {author} {\bibfnamefont {D.}~\bibnamefont
  {Ebert}}, \bibinfo {author} {\bibfnamefont {R.}~\bibnamefont {Faustov}}, \
  and\ \bibinfo {author} {\bibfnamefont {V.}~\bibnamefont {Galkin}},\ }\href
  {\doibase 10.1103/PhysRevD.79.114029} {\bibfield  {journal} {\bibinfo
  {journal} {Phys. Rev. D}\ }\textbf {\bibinfo {volume} {79}},\ \bibinfo
  {pages} {114029} (\bibinfo {year} {2009})},\ \Eprint
  {http://arxiv.org/abs/0903.5183} {arXiv:0903.5183 [hep-ph]} \BibitemShut
  {NoStop}%
\bibitem [{\citenamefont {Faustov}\ and\ \citenamefont
  {Galkin}(2015)}]{Faustov:2015eba}%
  \BibitemOpen
  \bibfield  {author} {\bibinfo {author} {\bibfnamefont {R.}~\bibnamefont
  {Faustov}}\ and\ \bibinfo {author} {\bibfnamefont {V.}~\bibnamefont
  {Galkin}},\ }\href {\doibase 10.1103/PhysRevD.92.054005} {\bibfield
  {journal} {\bibinfo  {journal} {Phys. Rev. D}\ }\textbf {\bibinfo {volume}
  {92}},\ \bibinfo {pages} {054005} (\bibinfo {year} {2015})},\ \Eprint
  {http://arxiv.org/abs/1507.04530} {arXiv:1507.04530 [hep-ph]} \BibitemShut
  {NoStop}%
\bibitem [{\citenamefont {Edwards}\ \emph {et~al.}(2011)\citenamefont
  {Edwards}, \citenamefont {Dudek}, \citenamefont {Richards},\ and\
  \citenamefont {Wallace}}]{Edwards:2011jj}%
  \BibitemOpen
  \bibfield  {author} {\bibinfo {author} {\bibfnamefont {R.~G.}\ \bibnamefont
  {Edwards}}, \bibinfo {author} {\bibfnamefont {J.~J.}\ \bibnamefont {Dudek}},
  \bibinfo {author} {\bibfnamefont {D.~G.}\ \bibnamefont {Richards}}, \ and\
  \bibinfo {author} {\bibfnamefont {S.~J.}\ \bibnamefont {Wallace}},\ }\href
  {\doibase 10.1103/PhysRevD.84.074508} {\bibfield  {journal} {\bibinfo
  {journal} {Phys. Rev. D}\ }\textbf {\bibinfo {volume} {84}},\ \bibinfo
  {pages} {074508} (\bibinfo {year} {2011})},\ \Eprint
  {http://arxiv.org/abs/1104.5152} {arXiv:1104.5152 [hep-ph]} \BibitemShut
  {NoStop}%
\bibitem [{\citenamefont {Edwards}(2020)}]{Edwards:2020rbo}%
  \BibitemOpen
  \bibfield  {author} {\bibinfo {author} {\bibfnamefont {R.~G.}\ \bibnamefont
  {Edwards}},\ }\href {\doibase 10.22323/1.363.0253} {\bibfield  {journal}
  {\bibinfo  {journal} {PoS}\ }\textbf {\bibinfo {volume} {LATTICE2019}},\
  \bibinfo {pages} {253} (\bibinfo {year} {2020})}\BibitemShut {NoStop}%
\bibitem [{\citenamefont {Allton}\ \emph {et~al.}(2005)\citenamefont {Allton},
  \citenamefont {Doring}, \citenamefont {Ejiri}, \citenamefont {Hands},
  \citenamefont {Kaczmarek}, \citenamefont {Karsch}, \citenamefont {Laermann},\
  and\ \citenamefont {Redlich}}]{Allton:2005gk}%
  \BibitemOpen
  \bibfield  {author} {\bibinfo {author} {\bibfnamefont {C.~R.}\ \bibnamefont
  {Allton}}, \bibinfo {author} {\bibfnamefont {M.}~\bibnamefont {Doring}},
  \bibinfo {author} {\bibfnamefont {S.}~\bibnamefont {Ejiri}}, \bibinfo
  {author} {\bibfnamefont {S.~J.}\ \bibnamefont {Hands}}, \bibinfo {author}
  {\bibfnamefont {O.}~\bibnamefont {Kaczmarek}}, \bibinfo {author}
  {\bibfnamefont {F.}~\bibnamefont {Karsch}}, \bibinfo {author} {\bibfnamefont
  {E.}~\bibnamefont {Laermann}}, \ and\ \bibinfo {author} {\bibfnamefont
  {K.}~\bibnamefont {Redlich}},\ }\href {\doibase 10.1103/PhysRevD.71.054508}
  {\bibfield  {journal} {\bibinfo  {journal} {Phys. Rev. D}\ }\textbf {\bibinfo
  {volume} {71}},\ \bibinfo {pages} {054508} (\bibinfo {year} {2005})},\
  \Eprint {http://arxiv.org/abs/hep-lat/0501030} {arXiv:hep-lat/0501030}
  \BibitemShut {NoStop}%
\bibitem [{\citenamefont {Bazavov}\ \emph
  {et~al.}(2017{\natexlab{b}})\citenamefont {Bazavov} \emph
  {et~al.}}]{HotQCD:2017qwq}%
  \BibitemOpen
  \bibfield  {author} {\bibinfo {author} {\bibfnamefont {A.}~\bibnamefont
  {Bazavov}} \emph {et~al.} (\bibinfo {collaboration} {HotQCD}),\ }\href
  {\doibase 10.1103/PhysRevD.96.074510} {\bibfield  {journal} {\bibinfo
  {journal} {Phys. Rev. D}\ }\textbf {\bibinfo {volume} {96}},\ \bibinfo
  {pages} {074510} (\bibinfo {year} {2017}{\natexlab{b}})},\ \Eprint
  {http://arxiv.org/abs/1708.04897} {arXiv:1708.04897 [hep-lat]} \BibitemShut
  {NoStop}%
\bibitem [{\citenamefont {Bazavov}\ \emph {et~al.}(2010)\citenamefont {Bazavov}
  \emph {et~al.}}]{MILC:2010hzw}%
  \BibitemOpen
  \bibfield  {author} {\bibinfo {author} {\bibfnamefont {A.}~\bibnamefont
  {Bazavov}} \emph {et~al.} (\bibinfo {collaboration} {MILC}),\ }\href
  {\doibase 10.22323/1.105.0074} {\bibfield  {journal} {\bibinfo  {journal}
  {PoS}\ }\textbf {\bibinfo {volume} {LATTICE2010}},\ \bibinfo {pages} {074}
  (\bibinfo {year} {2010})},\ \Eprint {http://arxiv.org/abs/1012.0868}
  {arXiv:1012.0868 [hep-lat]} \BibitemShut {NoStop}%
\bibitem [{\citenamefont {Aoki}\ \emph {et~al.}(2020)\citenamefont {Aoki} \emph
  {et~al.}}]{Aoki:2019cca}%
  \BibitemOpen
  \bibfield  {author} {\bibinfo {author} {\bibfnamefont {S.}~\bibnamefont
  {Aoki}} \emph {et~al.} (\bibinfo {collaboration} {Flavour Lattice Averaging
  Group}),\ }\href {\doibase 10.1140/epjc/s10052-019-7354-7} {\bibfield
  {journal} {\bibinfo  {journal} {Eur. Phys. J. C}\ }\textbf {\bibinfo {volume}
  {80}},\ \bibinfo {pages} {113} (\bibinfo {year} {2020})},\ \Eprint
  {http://arxiv.org/abs/1902.08191} {arXiv:1902.08191 [hep-lat]} \BibitemShut
  {NoStop}%
\bibitem [{\citenamefont {Bazavov}\ \emph
  {et~al.}(2014{\natexlab{b}})\citenamefont {Bazavov} \emph
  {et~al.}}]{Bazavov:2014xya}%
  \BibitemOpen
  \bibfield  {author} {\bibinfo {author} {\bibfnamefont {A.}~\bibnamefont
  {Bazavov}} \emph {et~al.},\ }\href {\doibase 10.1103/PhysRevLett.113.072001}
  {\bibfield  {journal} {\bibinfo  {journal} {Phys. Rev. Lett.}\ }\textbf
  {\bibinfo {volume} {113}},\ \bibinfo {pages} {072001} (\bibinfo {year}
  {2014}{\natexlab{b}})},\ \Eprint {http://arxiv.org/abs/1404.6511}
  {arXiv:1404.6511 [hep-lat]} \BibitemShut {NoStop}%
\bibitem [{\citenamefont {Menapara}\ and\ \citenamefont
  {Rai}(2021)}]{Menapara:2021dzi}%
  \BibitemOpen
  \bibfield  {author} {\bibinfo {author} {\bibfnamefont {C.}~\bibnamefont
  {Menapara}}\ and\ \bibinfo {author} {\bibfnamefont {A.~K.}\ \bibnamefont
  {Rai}},\ }\href {\doibase 10.1088/1674-1137/abf4f4} {\bibfield  {journal}
  {\bibinfo  {journal} {Chin. Phys. C}\ }\textbf {\bibinfo {volume} {45}},\
  \bibinfo {pages} {063108} (\bibinfo {year} {2021})},\ \Eprint
  {http://arxiv.org/abs/2104.00439} {arXiv:2104.00439 [hep-ph]} \BibitemShut
  {NoStop}%
\bibitem [{\citenamefont {Fiore}\ \emph {et~al.}(1977)\citenamefont {Fiore},
  \citenamefont {Page},\ and\ \citenamefont {Sertorio}}]{Fiore:1977pb}%
  \BibitemOpen
  \bibfield  {author} {\bibinfo {author} {\bibfnamefont {R.}~\bibnamefont
  {Fiore}}, \bibinfo {author} {\bibfnamefont {R.}~\bibnamefont {Page}}, \ and\
  \bibinfo {author} {\bibfnamefont {L.}~\bibnamefont {Sertorio}},\ }\href
  {\doibase 10.1007/BF02790581} {\bibfield  {journal} {\bibinfo  {journal}
  {Nuovo Cim. A}\ }\textbf {\bibinfo {volume} {37}},\ \bibinfo {pages} {45}
  (\bibinfo {year} {1977})}\BibitemShut {NoStop}%
\bibitem [{\citenamefont {Broniowski}\ \emph {et~al.}(2015)\citenamefont
  {Broniowski}, \citenamefont {Giacosa},\ and\ \citenamefont
  {Begun}}]{Broniowski:2015oha}%
  \BibitemOpen
  \bibfield  {author} {\bibinfo {author} {\bibfnamefont {W.}~\bibnamefont
  {Broniowski}}, \bibinfo {author} {\bibfnamefont {F.}~\bibnamefont {Giacosa}},
  \ and\ \bibinfo {author} {\bibfnamefont {V.}~\bibnamefont {Begun}},\ }\href
  {\doibase 10.1103/PhysRevC.92.034905} {\bibfield  {journal} {\bibinfo
  {journal} {Phys. Rev. C}\ }\textbf {\bibinfo {volume} {92}},\ \bibinfo
  {pages} {034905} (\bibinfo {year} {2015})},\ \Eprint
  {http://arxiv.org/abs/1506.01260} {arXiv:1506.01260 [nucl-th]} \BibitemShut
  {NoStop}%
\bibitem [{\citenamefont {Friman}\ \emph {et~al.}(2015)\citenamefont {Friman},
  \citenamefont {Lo}, \citenamefont {Marczenko}, \citenamefont {Redlich},\ and\
  \citenamefont {Sasaki}}]{Friman:2015zua}%
  \BibitemOpen
  \bibfield  {author} {\bibinfo {author} {\bibfnamefont {B.}~\bibnamefont
  {Friman}}, \bibinfo {author} {\bibfnamefont {P.~M.}\ \bibnamefont {Lo}},
  \bibinfo {author} {\bibfnamefont {M.}~\bibnamefont {Marczenko}}, \bibinfo
  {author} {\bibfnamefont {K.}~\bibnamefont {Redlich}}, \ and\ \bibinfo
  {author} {\bibfnamefont {C.}~\bibnamefont {Sasaki}},\ }\href {\doibase
  10.1103/PhysRevD.92.074003} {\bibfield  {journal} {\bibinfo  {journal} {Phys.
  Rev. D}\ }\textbf {\bibinfo {volume} {92}},\ \bibinfo {pages} {074003}
  (\bibinfo {year} {2015})},\ \Eprint {http://arxiv.org/abs/1507.04183}
  {arXiv:1507.04183 [hep-ph]} \BibitemShut {NoStop}%
\bibitem [{\citenamefont {Bollweg}\ \emph {et~al.}(2021)\citenamefont
  {Bollweg}, \citenamefont {Goswami}, \citenamefont {Kaczmarek}, \citenamefont
  {Karsch}, \citenamefont {Mukherjee}, \citenamefont {Petreczky}, \citenamefont
  {Schmidt},\ and\ \citenamefont {Scior}}]{bollweg2021dataset}%
  \BibitemOpen
  \bibfield  {author} {\bibinfo {author} {\bibfnamefont {D.}~\bibnamefont
  {Bollweg}}, \bibinfo {author} {\bibfnamefont {J.}~\bibnamefont {Goswami}},
  \bibinfo {author} {\bibfnamefont {O.}~\bibnamefont {Kaczmarek}}, \bibinfo
  {author} {\bibfnamefont {F.}~\bibnamefont {Karsch}}, \bibinfo {author}
  {\bibfnamefont {S.}~\bibnamefont {Mukherjee}}, \bibinfo {author}
  {\bibfnamefont {P.}~\bibnamefont {Petreczky}}, \bibinfo {author}
  {\bibfnamefont {C.}~\bibnamefont {Schmidt}}, \ and\ \bibinfo {author}
  {\bibfnamefont {P.}~\bibnamefont {Scior}},\ }\href {\doibase
  https://doi.org/10.4119/unibi/2957724} {\  (\bibinfo {year} {2021}),\
  https://doi.org/10.4119/unibi/2957724}\BibitemShut {NoStop}%
\bibitem [{\citenamefont {Alba}\ \emph {et~al.}(2017)\citenamefont {Alba} \emph
  {et~al.}}]{Alba:2017mqu}%
  \BibitemOpen
  \bibfield  {author} {\bibinfo {author} {\bibfnamefont {P.}~\bibnamefont
  {Alba}} \emph {et~al.},\ }\href {\doibase 10.1103/PhysRevD.96.034517}
  {\bibfield  {journal} {\bibinfo  {journal} {Phys. Rev. D}\ }\textbf {\bibinfo
  {volume} {96}},\ \bibinfo {pages} {034517} (\bibinfo {year} {2017})},\
  \Eprint {http://arxiv.org/abs/1702.01113} {arXiv:1702.01113 [hep-lat]}
  \BibitemShut {NoStop}%
\bibitem [{\citenamefont {Lo}\ \emph {et~al.}(2018)\citenamefont {Lo},
  \citenamefont {Friman}, \citenamefont {Redlich},\ and\ \citenamefont
  {Sasaki}}]{Lo:2017lym}%
  \BibitemOpen
  \bibfield  {author} {\bibinfo {author} {\bibfnamefont {P.~M.}\ \bibnamefont
  {Lo}}, \bibinfo {author} {\bibfnamefont {B.}~\bibnamefont {Friman}}, \bibinfo
  {author} {\bibfnamefont {K.}~\bibnamefont {Redlich}}, \ and\ \bibinfo
  {author} {\bibfnamefont {C.}~\bibnamefont {Sasaki}},\ }\href {\doibase
  10.1016/j.physletb.2018.01.016} {\bibfield  {journal} {\bibinfo  {journal}
  {Phys. Lett. B}\ }\textbf {\bibinfo {volume} {778}},\ \bibinfo {pages} {454}
  (\bibinfo {year} {2018})},\ \Eprint {http://arxiv.org/abs/1710.02711}
  {arXiv:1710.02711 [hep-ph]} \BibitemShut {NoStop}%
\bibitem [{\citenamefont {Lo}\ \emph {et~al.}(2017)\citenamefont {Lo},
  \citenamefont {Friman}, \citenamefont {Marczenko}, \citenamefont {Redlich},\
  and\ \citenamefont {Sasaki}}]{Lo:2017ldt}%
  \BibitemOpen
  \bibfield  {author} {\bibinfo {author} {\bibfnamefont {P.~M.}\ \bibnamefont
  {Lo}}, \bibinfo {author} {\bibfnamefont {B.}~\bibnamefont {Friman}}, \bibinfo
  {author} {\bibfnamefont {M.}~\bibnamefont {Marczenko}}, \bibinfo {author}
  {\bibfnamefont {K.}~\bibnamefont {Redlich}}, \ and\ \bibinfo {author}
  {\bibfnamefont {C.}~\bibnamefont {Sasaki}},\ }\href {\doibase
  10.1103/PhysRevC.96.015207} {\bibfield  {journal} {\bibinfo  {journal} {Phys.
  Rev. C}\ }\textbf {\bibinfo {volume} {96}},\ \bibinfo {pages} {015207}
  (\bibinfo {year} {2017})},\ \Eprint {http://arxiv.org/abs/1703.00306}
  {arXiv:1703.00306 [nucl-th]} \BibitemShut {NoStop}%
\bibitem [{\citenamefont {Capstick}\ and\ \citenamefont
  {Roberts}(1994)}]{Capstick:1993kb}%
  \BibitemOpen
  \bibfield  {author} {\bibinfo {author} {\bibfnamefont {S.}~\bibnamefont
  {Capstick}}\ and\ \bibinfo {author} {\bibfnamefont {W.}~\bibnamefont
  {Roberts}},\ }\href {\doibase 10.1103/PhysRevD.49.4570} {\bibfield  {journal}
  {\bibinfo  {journal} {Phys. Rev. D}\ }\textbf {\bibinfo {volume} {49}},\
  \bibinfo {pages} {4570} (\bibinfo {year} {1994})},\ \Eprint
  {http://arxiv.org/abs/nucl-th/9310030} {arXiv:nucl-th/9310030} \BibitemShut
  {NoStop}%
\bibitem [{\citenamefont {Hunt}\ and\ \citenamefont
  {Manley}(2019)}]{Hunt:2018wqz}%
  \BibitemOpen
  \bibfield  {author} {\bibinfo {author} {\bibfnamefont {B.~C.}\ \bibnamefont
  {Hunt}}\ and\ \bibinfo {author} {\bibfnamefont {D.~M.}\ \bibnamefont
  {Manley}},\ }\href {\doibase 10.1103/PhysRevC.99.055205} {\bibfield
  {journal} {\bibinfo  {journal} {Phys. Rev. C}\ }\textbf {\bibinfo {volume}
  {99}},\ \bibinfo {pages} {055205} (\bibinfo {year} {2019})},\ \Eprint
  {http://arxiv.org/abs/1810.13086} {arXiv:1810.13086 [nucl-ex]} \BibitemShut
  {NoStop}%
\bibitem [{\citenamefont {Ablikim}\ \emph {et~al.}(2006)\citenamefont {Ablikim}
  \emph {et~al.}}]{Ablikim:2005ni}%
  \BibitemOpen
  \bibfield  {author} {\bibinfo {author} {\bibfnamefont {M.}~\bibnamefont
  {Ablikim}} \emph {et~al.} (\bibinfo {collaboration} {BES}),\ }\href {\doibase
  10.1016/j.physletb.2005.12.062} {\bibfield  {journal} {\bibinfo  {journal}
  {Phys. Lett. B}\ }\textbf {\bibinfo {volume} {633}},\ \bibinfo {pages} {681}
  (\bibinfo {year} {2006})},\ \Eprint {http://arxiv.org/abs/hep-ex/0506055}
  {arXiv:hep-ex/0506055} \BibitemShut {NoStop}%
\bibitem [{\citenamefont {Ablikim}\ \emph {et~al.}(2011)\citenamefont {Ablikim}
  \emph {et~al.}}]{Ablikim:2010ab}%
  \BibitemOpen
  \bibfield  {author} {\bibinfo {author} {\bibfnamefont {M.}~\bibnamefont
  {Ablikim}} \emph {et~al.} (\bibinfo {collaboration} {BES}),\ }\href {\doibase
  10.1016/j.physletb.2011.03.011} {\bibfield  {journal} {\bibinfo  {journal}
  {Phys. Lett. B}\ }\textbf {\bibinfo {volume} {698}},\ \bibinfo {pages} {183}
  (\bibinfo {year} {2011})},\ \Eprint {http://arxiv.org/abs/1008.4489}
  {arXiv:1008.4489 [hep-ex]} \BibitemShut {NoStop}%
\bibitem [{\citenamefont {Giacosa}(2019)}]{Giacosa:2018vbw}%
  \BibitemOpen
  \bibfield  {author} {\bibinfo {author} {\bibfnamefont {F.}~\bibnamefont
  {Giacosa}},\ }\href@noop {} {\bibfield  {journal} {\bibinfo  {journal} {Acta
  Phys. Pol. B Proc. Suppl.}\ }\textbf {\bibinfo {volume} {12}},\ \bibinfo
  {pages} {283} (\bibinfo {year} {2019})},\ \Eprint
  {http://arxiv.org/abs/1811.00298} {arXiv:1811.00298 [hep-ph]} \BibitemShut
  {NoStop}%
\bibitem [{\citenamefont {Engels}\ \emph {et~al.}(1982)\citenamefont {Engels},
  \citenamefont {Karsch},\ and\ \citenamefont {Satz}}]{Engels:1981ab}%
  \BibitemOpen
  \bibfield  {author} {\bibinfo {author} {\bibfnamefont {J.}~\bibnamefont
  {Engels}}, \bibinfo {author} {\bibfnamefont {F.}~\bibnamefont {Karsch}}, \
  and\ \bibinfo {author} {\bibfnamefont {H.}~\bibnamefont {Satz}},\ }\href
  {\doibase 10.1016/0550-3213(82)90387-X} {\bibfield  {journal} {\bibinfo
  {journal} {Nucl. Phys. B}\ }\textbf {\bibinfo {volume} {205}},\ \bibinfo
  {pages} {239} (\bibinfo {year} {1982})}\BibitemShut {NoStop}%
\bibitem [{\citenamefont {Bhattacharyya}\ \emph {et~al.}(2015)\citenamefont
  {Bhattacharyya}, \citenamefont {Ray}, \citenamefont {Samanta},\ and\
  \citenamefont {Sur}}]{Bhattacharyya:2015zka}%
  \BibitemOpen
  \bibfield  {author} {\bibinfo {author} {\bibfnamefont {A.}~\bibnamefont
  {Bhattacharyya}}, \bibinfo {author} {\bibfnamefont {R.}~\bibnamefont {Ray}},
  \bibinfo {author} {\bibfnamefont {S.}~\bibnamefont {Samanta}}, \ and\
  \bibinfo {author} {\bibfnamefont {S.}~\bibnamefont {Sur}},\ }\href {\doibase
  10.1103/PhysRevC.91.041901} {\bibfield  {journal} {\bibinfo  {journal} {Phys.
  Rev. C}\ }\textbf {\bibinfo {volume} {91}},\ \bibinfo {pages} {041901}
  (\bibinfo {year} {2015})},\ \Eprint {http://arxiv.org/abs/1502.00889}
  {arXiv:1502.00889 [hep-ph]} \BibitemShut {NoStop}%
\bibitem [{\citenamefont {Karsch}\ \emph {et~al.}(2016)\citenamefont {Karsch},
  \citenamefont {Morita},\ and\ \citenamefont {Redlich}}]{Karsch:2015zna}%
  \BibitemOpen
  \bibfield  {author} {\bibinfo {author} {\bibfnamefont {F.}~\bibnamefont
  {Karsch}}, \bibinfo {author} {\bibfnamefont {K.}~\bibnamefont {Morita}}, \
  and\ \bibinfo {author} {\bibfnamefont {K.}~\bibnamefont {Redlich}},\ }\href
  {\doibase 10.1103/PhysRevC.93.034907} {\bibfield  {journal} {\bibinfo
  {journal} {Phys. Rev. C}\ }\textbf {\bibinfo {volume} {93}},\ \bibinfo
  {pages} {034907} (\bibinfo {year} {2016})},\ \Eprint
  {http://arxiv.org/abs/1508.02614} {arXiv:1508.02614 [hep-ph]} \BibitemShut
  {NoStop}%
\end{thebibliography}%
\end{document}